%% file: main.tex
\documentclass[manuscript,screen]{acmart}

\usepackage{multirow}
\usepackage{float}
\usepackage{url}
\usepackage{algorithm}
\usepackage{algpseudocode}
\usepackage{bm}
\usepackage{graphicx}
\usepackage{caption}
\usepackage[skip=0cm,list=true,labelfont=it]{subcaption}
\usepackage{enumitem}
\usepackage{array}
\theoremstyle{definition}
\newtheorem{definition}{DEFINITION}[]

\AtBeginDocument{%
  \providecommand\BibTeX{{%
    \normalfont B\kern-0.5em{\scshape i\kern-0.25em b}\kern-0.8em\TeX}}}

\copyrightyear{2025}
\acmYear{2025}
\setcopyright{acmcopyright}

\setcopyright{cc}
\setcctype{by}
\acmJournal{TOIS}
\acmYear{2025} \acmVolume{1} \acmNumber{1} \acmArticle{1} \acmMonth{1} \acmPrice{}\acmDOI{10.1145/3744240}

\setcopyright{none}

\begin{document}


\title[Causality-Inspired Fair Representation Learning for Multimodal Recommendation]{Causality-Inspired Fair Representation Learning for Multimodal Recommendation}


    \author{Weixin Chen}
    \affiliation{%
		\institution{Hong Kong Baptist University}
        \country{Hong Kong, China}
    }
    \email{cswxchen@comp.hkbu.edu.hk}

    \author{Li Chen}
	\affiliation{%
		\institution{Hong Kong Baptist University}
        \country{Hong Kong, China}
    }
    \email{lichen@comp.hkbu.edu.hk}

    \author{Yongxin Ni}
	\affiliation{%
		\institution{Westlake University}
        \country{China}
    }
    \email{niyongxin@westlake.edu.cn}

    \author{Yuhan Zhao}
    \affiliation{%
      \institution{Hong Kong Baptist University}
      \country{Hong Kong, China}
    }
    \email{csyhzhao@comp.hkbu.edu.hk}


	
\renewcommand{\shortauthors}{Chen et al.}

\input{0.abstract}

\maketitle


\section{INTRODUCTION}
\label{introduction}

\input{1.introduction}

\section{RELATED WORK}
\label{related_work}

\input{2.related_work}

\section{PRELIMINARIES}
\label{preliminary}

\input{3.preliminary}

\section{METHODOLOGY}
\label{methodology}

\input{4.methodology}

\section{EXPERIMENTs}
\label{experiment}
\input{5.experiment}
\label{result}

\input{6.result}

\section{CONCLUSION}
\label{conclusion}

\input{7.conclusion}

\input{acks}


\bibliographystyle{ACM-Reference-Format}
\bibliography{paper}


\end{document}

%% file: 0.abstract.tex
\begin{abstract}
Recently, multimodal recommendations (MMR) have gained increasing attention for alleviating the data sparsity problem of traditional recommender systems by incorporating modality-based representations. 
Although MMR {exhibits} notable improvement in recommendation accuracy, we empirically validate that an increase in the quantity or variety of modalities leads to a higher degree of users’ sensitive information leakage due to entangled causal relationships, risking fair representation learning.
On the other hand, existing fair representation learning approaches are mostly based on the assumption that sensitive information is solely leaked from users’ interaction data and do not explicitly model the causal relationships introduced by multimodal data, which limits their applicability in multimodal scenarios.
To address this limitation, we propose a novel fair multimodal recommendation approach (dubbed FMMRec) through causality-inspired fairness-oriented modal disentanglement and relation-aware fairness learning. 
Particularly, we disentangle biased and filtered modal embeddings inspired by causal inference techniques, enabling the mining of modality-based unfair and fair user-user relations, thereby enhancing the fairness and informativeness of user representations. 
By addressing the causal effects of sensitive attributes on user preferences, our approach aims to achieve counterfactual fairness in multimodal recommendations.
Experiments on two public datasets demonstrate the superiority of our FMMRec relative to the state-of-the-art baselines.
Our source code is available at \url{https://github.com/WeixinChen98/FMMRec}.
\end{abstract}

%% file: 1.introduction.tex

It has been widely recognized that recommender systems (RS) are useful in addressing information overload problems by providing personalized information or service, but their effectiveness is usually restricted by the density of user-item interactions~\cite{zhou2023comprehensive,ZCH24,ZCC25}.
{In recent years, multimodal recommendations (MMR), which exploit the rich multimodal content of items such as images, text, and audio, have been proposed to alleviate the data sparsity problem with remarkable performance~\cite{yang2018mmcf, GRCN, LightGT, li2023exploring}}.
{In particular, benefiting from the technical advances in other fields like natural language processing (NLP) and computer vision (CV), items' multimodal information can be encoded into high-level representations and further incorporated into RS to establish in-depth modeling of user preferences for different modalities~\cite{MoRec, DBLP:conf/www/LiuCLH19, DBLP:conf/sigir/ChenCXZ0QZ19, zhang2017joint}.} 

\begin{figure}[htbp]
  \centering

   \captionsetup[subfigure]{justification=raggedleft}

    \hspace{\fill}
  \begin{subfigure}[b]{0.42\textwidth}
    \includegraphics[width=1\textwidth]{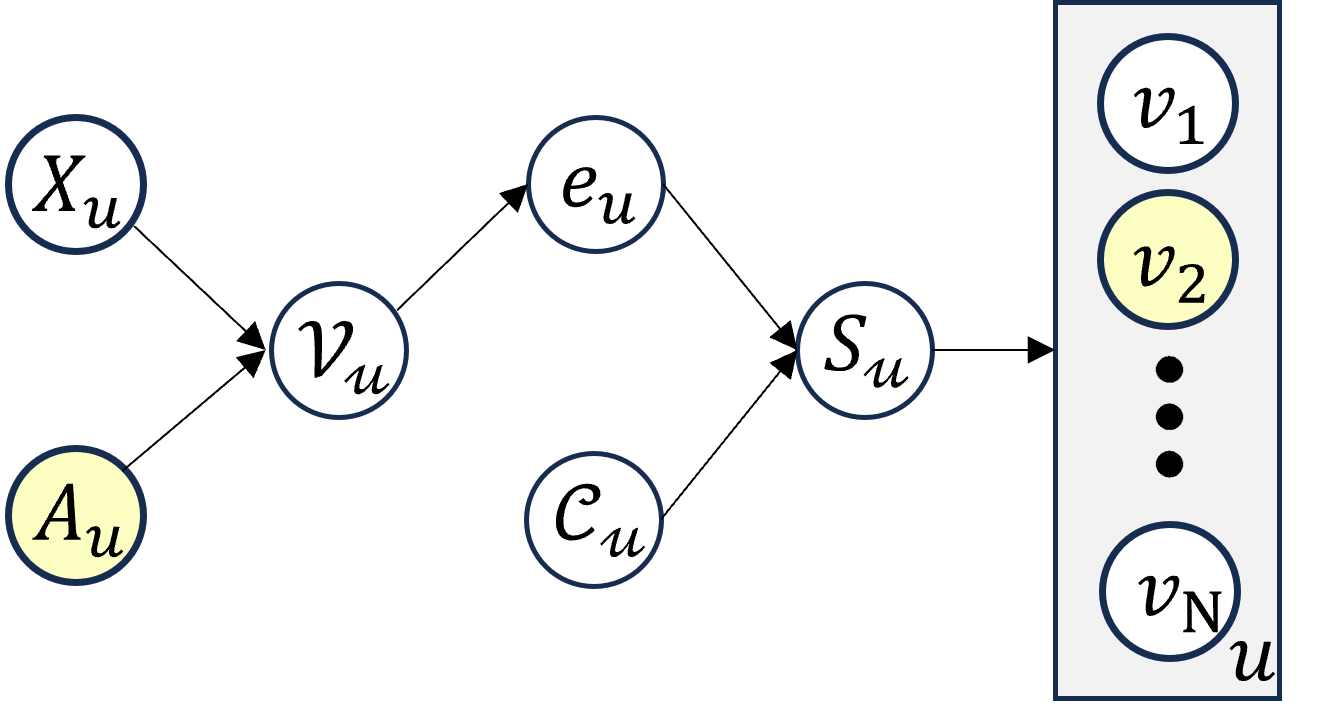}
    \caption{Causal relationships of general RS}
    \label{fig:general_causal_graph}
  \end{subfigure}%
  \hspace{\fill}
  \begin{subfigure}[b]{0.42\textwidth}
    \includegraphics[width=1\textwidth]{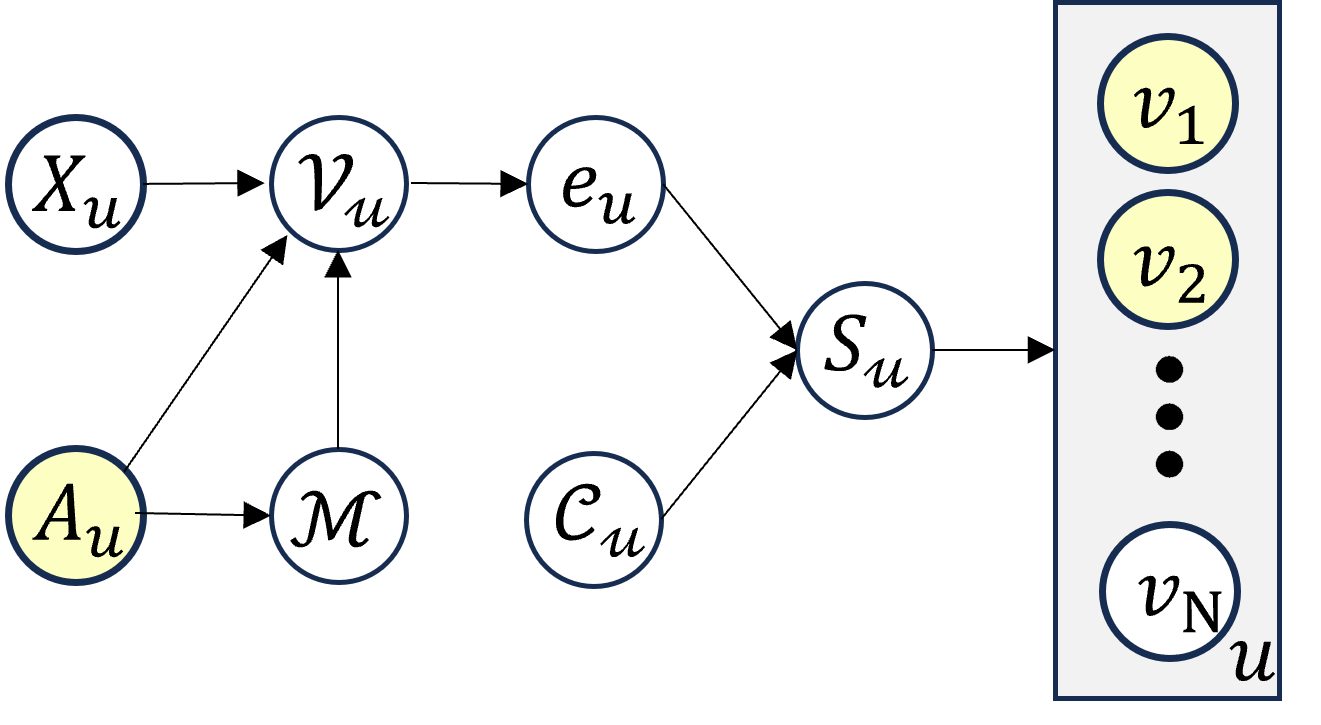}
    \caption{Causal relationships of multimodal RS}
    \label{fig:subfig2}
  \end{subfigure}%
  \hspace{\fill}
  \caption{The causal relationships of general recommender systems (RS) (Figure~\textit{a})~\cite{PCFR} and multimodal RS (Figure~\textit{b}).
    $A_u$ and $X_u$ are sensitive attributes and non-sensitive attributes of user $u$, respectively. $\mathcal{V}_u$ is the set of historical items interacted (or preferred) by user $u$, and $\mathcal{M}$ denotes the multimodal content of items.
    $\boldsymbol{e}_u$ and $\mathcal{C}_u$ are the user representation and the candidate item set for user $u$, respectively. 
    $S_u$ contains the predicted scores of all candidate items. 
    The yellow color highlights the causal impact of sensitive attributes $A_u$ on the recommended items.
    }
  \label{fig:causal_graph}
\end{figure}

However, integrating multimodal information introduces complexities in understanding the causal effects of item modalities on user preferences~\cite{scholkopf2021toward, pearl2009causality}, and may lead to more severe leakage of users' sensitive information.
Empirically, our studies show that an increase in the \textit{quantity} or \textit{variety} of modalities (such as movie posters and plots) can lead to a higher degree of users' sensitive information leakage (\textit{i.e.}, more accurate prediction of users' sensitive attributes like gender, age, and occupation as shown in Figure~\ref{fig:sens_leak_exp}). 
Such sensitive information leakage from multimodal content poses a significant risk to fair representation learning from a causal perspective.
\textbf{Fair representation learning} aims to eliminate sensitive information in user preference modeling, thereby ensuring that recommendations can be independent of users' sensitive attribute(s)~\cite{FairGo, PCFR, zhao2023fair, zhu2024adaptive}.
As shown in the causal relationships in Figure~\ref{fig:causal_graph}, multimodal content can act as confounding variables that introduce spurious correlations between users' sensitive attributes and their preferences~\cite{pearl2009causality, DBLP:conf/nips/KilbertusRPHJS17}.
To achieve counterfactual fairness in recommendations, it is essential to control for the causal effects of sensitive attributes on the recommendation outcomes~\cite{DBLP:conf/nips/KusnerLRS17, PCFR}.



Recently, plenty of fair representation learning methods for RS~\cite{shao2022faircf, DBLP:conf/sigir/WuXZZ0ZL022, hua2023up5, zhu2024adaptive, zhao2023fair} have been proposed to filter out sensitive information in user representations.
For instance, AL~\cite{wadsworth2018achieving} first introduces adversarial learning to filter out sensitive information in user representations via a min-max game. 
CAL~\cite{bose2019compositional} and PCFR~\cite{PCFR} leverage compositional filters to enable a personalized selection of sensitive attributes.
FairGo~\cite{FairGo} examines and eliminates users' sensitive information in different levels of user-centric graph representations.
However, they often assume that sensitive information is solely leaked from users' interaction data and overlook the sensitive information leakage from multimodal content, which limits their applicability in multimodal scenarios.
Particularly, these methods do not explicitly model the causal relationships introduced by multimodal data, which can impact the fairness of recommendations.

Intuitively, one promising approach for fairness-aware multimodal recommendations is to incorporate multimodal representations as additional knowledge in learning fair representations.
However, without proper causal intervention, simply integrating multimodal data may not effectively eliminate the undesired causal effects of sensitive attributes on recommendations~\cite{chiappa2019path, DBLP:conf/nips/KusnerLRS17}.
This approach faces two key challenges:

\begin{itemize}[leftmargin=*]
    \item \textit{C1}: The \textit{entanglement} present in multimodal content poses difficulty in eliminating sensitive information while utilizing non-sensitive information to ensure accuracy.
    From a causal inference standpoint, it is important to disentangle the causal effects of sensitive attributes from the multimodal features to address fairness concerns without significant accuracy loss.
    \item \textit{C2}: The \textit{heterogeneity} between items' multimodal representations and user representations hinders leveraging the modality-based knowledge to promote fair user representation learning. 
    Specifically, the multimodal representations of items and the user representations are in highly distinct semantic spaces, posing a considerable obstacle to their interactions.
\end{itemize}

To tackle these two challenges, we propose a \textbf{\underline{f}}air \textbf{\underline{m}}ulti\textbf{\underline{m}}odal \textbf{\underline{rec}}ommendation approach (referred to \textbf{FMMRec}) through causality-inspired \textbf{fairness-oriented modal disentanglement} and \textbf{relation-aware fairness learning}.
For \textit{C1} (entanglement), we disentangle modal embeddings to maximize the potential sensitive information of the biased embeddings and minimize that of filtered embeddings, while maintaining sufficient non-sensitive information of filtered embeddings for preserving personalized information.
For \textit{C2} (heterogeneity), instead of forcing interactions between multimodal item representations and user representations, FMMRec mines dual user-user relations given the disentangled modal embeddings to learn fair and informative user representations.
Specifically, unfair relations (\textit{w.r.t.} biased embeddings) are identified to promote the fairness of user representations, while fair relations (\textit{w.r.t.} filtered embeddings) are for expressiveness.
By addressing the causal effects of sensitive attributes on user preferences, our approach aims to achieve counterfactual fairness (see Definition~\ref{def:counterfactual_fairness}) in multimodal recommendations.
Extensive experiments, conducted on two public datasets MovieLens and MicroLens, show the superiority of our proposed FMMRec in learning fair representations while maintaining comparable accuracy.
Our key contributions are four-fold:
\begin{itemize}[]
    \item We empirically demonstrate the sensitive information leakage in multimodal scenarios from a causal perspective, revealing that an increase in quantity or variety of modalities yields more leakage of users’ sensitive information due to entangled causal relationships.
    \item We propose fairness-oriented modal disentanglement and relation-aware fairness learning based on adversarial learning for fair multimodal recommendations, inspired by causal inference techniques to eliminate sensitive causal effects while preserving sufficient personalized information.
    \item We have conducted comprehensive experiments in practical multimodal scenarios, which demonstrate the effectiveness of our method in terms of accuracy-fairness trade-off, in comparison with several state-of-the-art baselines. 
    \item 
    To the best of our knowledge, this is the first work that aims to improve fairness in multimodal recommendations by developing a causality-inspired relation-aware fairness learning framework with disentangled modal embeddings.
\end{itemize}


%% file: 2.related_work.tex
\subsection{Multimodal Recommendation}
Conventional collaborative filtering methods~\cite{BPR, he2020lightgcn} are widely used for recommendations, which capture users' preferences on items by learning their historical interactions.
However, their effectiveness often relies on the density of user-item interactions~\cite{zhou2023comprehensive}. 

By effectively leveraging the multimodal content of items, multimodal recommendation models could alleviate the data sparsity problem~\cite{FREEDOM, DRAGON, MGCN, liu2024multimodal}.
As a pioneer of multimodal recommendation (MMR), VBPR~\cite{VBPR} incorporates the items' visual features extracted from Convolutional Neural Network (CNN) into BPR paradigm~\cite{BPR} along with ID-based features.
After that, various MMR models have exploited different mechanisms to model user preferences.
For example, VECF~\cite{DBLP:conf/sigir/ChenCXZ0QZ19} captures users' attention on different regions of content like images or reviews based on a multimodal attention network.
To effectively and efficiently recommend items under the multimodal and sequential context, MMMLP~\cite{MMMLP} was proposed based on Multilayer Perceptron (MLP) to capture users’ fine-grained preferences with only linear complexity.
Notably, Graph Neural Networks (GNNs) have gained increasing attention in MMR as its ability to capture high-order semantic information into user/item representations~\cite{FREEDOM}.
For instance, MMGCN~\cite{MMGCN} utilizes the message-passing concept of graph neural networks to construct the modality-specific representations of users and micro-videos.
To explicitly learn the semantic item relationship, LATTICE~\cite{LATTICE} mines the modality-based item-item structure via a modality-aware graph learning layer and graph convolutional layers.
Built upon LATTICE, FREECOM~\cite{FREEDOM} freezes the latent graph structure learning for efficiency and competitive accuracy.
SLMRec~\cite{SLMRec} generates and differentiates multiple views of items based on GNN-based self-supervised learning. 
To address the high computational complexity of negative sampling and the exorbitant memory cost of constructing/learning on large-scale auxiliary graphs, BM3~\cite{BM3} proposes a multimodal contrastive loss (MMCL) with a dropout layer without negative sampling and auxiliary graphs.
{MCDRec~\cite{ma2024multimodal} leverages a diffusion model to explicitly model and fuse multimodal features while denoising the user-item graph, alleviating ill-posed embeddings and improving high-order multimodal representation learning.
MENTOR~\cite{xu2025mentor} employs multi-level self-supervised alignment guided by ID embeddings to reduce modality gaps while preserving interaction semantics for robust multimodal recommendation.
LGMRec~\cite{guo2024lgmrec} jointly models local and global user interests by decoupling collaborative and multimodal signals through local graph embeddings and global hypergraph learning, enhancing recommendation robustness.
IISAN~\cite{fu2024iisan} introduces a decoupled parameter-efficient fine-tuning framework that adapts multimodal representations efficiently, significantly reducing GPU memory usage and training time in sequential recommendations.
}

However, though being demonstrated effective in accuracy, current MMR methods may suffer from the unfair representation learning issues and thus harm user experience.

\subsection{Fairness-aware Recommendation}
Generally, fairness definitions in RS could be categorized into \textit{group} fairness and \textit{individual} fairness~\cite{lifairness, wang2022survey, deldjoo2024fairness, chen2025investigating}.
Group fairness mainly focuses on equity among user groups with varying sensitive attribute(s), in terms of recommendation distribution or performance.
For example, statistical parity~\cite{calders2009building} encourages that different groups \textit{w.r.t.} sensitive attributes are treated similarly in terms of recommendation outcome.
Equal opportunity~\cite{DBLP:conf/nips/HardtPNS16} takes into account the true preference of each user group, with corresponding measurable rating-based recommendation metrics~\cite{FOCF}.

Different from group fairness, individual fairness requires similar users to be treated similarly at the individual level~\cite{biega2018equity}.
Actually, most group fairness definitions could be transferred to individual fairness by making each individual user belong to a unique group.
However, this would result in too many fairness constraints, which makes fairness learning difficult.
As a result,
some fairness definitions specific to the individual level are proposed. 
For example, envy-free fairness~\cite{EnvyFree} requires that individual users should be free of envy when knowing others' recommendations.
Notably, based on causal notions, counterfactual fairness~\cite{DBLP:conf/nips/KusnerLRS17, PCFR, TFR, hua2023up5} requires the same recommendation distribution in both the actual and counterfactual worlds, where the values of users' sensitive attributes are randomly intervened to other attainable values.
To achieve counterfactual fairness in recommender systems, the learned representations should contain no sensitive information~\cite{wang2022survey, PCFR}.

Adversarial learning~\cite{goodfellow2014generative} is the dominant technique for learning fair representations~\cite{wang2022survey, lifairness, zhu2024adaptive, FairRec, DBLP:conf/sigir/WuXZZ0ZL022} via a min-max game between a filter and a discriminator (\textit{a.k.a.} the sensitive attribute predictor).
For example, AL~\cite{wadsworth2018achieving} introduces the adversary network to eliminate sensitive information with basic filter-discriminator architecture.
CAL~\cite{bose2019compositional} proposes compositional filters for graph embeddings in RS.
Based on CAL, FairGo~\cite{FairGo} finds that the user's sensitive information is also exposed in their graph-based representation, thus applying adversarial learning for both explicit user representation and graph-based user representation.
PCFR~\cite{PCFR} introduces both integrated and combined modes of adversarial learning for achieving personalized counterfactual fairness in recommendations. 
{UP5~\cite{hua2023up5} proposes a counterfactually-fair prompting method for foundation models with adversarial training, mitigating biases in recommendations by ensuring fairness across sensitive user attributes.}
Zhu et al.~\cite{zhu2024adaptive} explored adaptive fair representation learning for personalized fairness in recommendations via information alignment.
{FairCoRe~\cite{bin2025faircore} constructs counterfactual scenarios and disentangles sensitive information through bias-aware learning and mutual information optimization to ensure sensitive attribute-invariant representations.}
{DistFair~\cite{yang2024distributional} achieves fairness by aligning performance distributions across groups via a generative adversarial framework and dual curriculum learning strategy.}
{DALFRec~\cite{liu2024dual} integrates user-side and item-side adversarial learning to mitigate biases in recommendations, enhancing fairness across both user and item representations.}
{FairDgcl~\cite{chen2024fairdgcl} introduces a dynamic graph contrastive learning framework that employs adversarial view generation to enhance fairness in recommendations, ensuring equitable treatment across diverse user groups.}
{FACTER~\cite{fayyazi2025facter} integrates conformal prediction with adversarial prompt generator in large language models to enforce fairness constraints, reducing demographic biases in recommendation outputs.}

However, existing fairness approaches assume users' sensitive attributes information is solely leaked in users' historical interactions, limiting their effectiveness in multimodal scenarios.
Moreover, many of these methods do not explicitly model the causal relationships introduced by multimodal data, which can impact the fairness of recommendations.

\subsection{Causal Infernece in Recommender Systems}

Causal inference has recently gained attention in the field of recommender systems to address issues such as bias, fairness, and explainability~\cite{DBLP:conf/kdd/WangF0WC21, xu2023causal, tan2021counterfactual}. 
Counterfactual reasoning has been applied to achieve fairness in recommendations. 
Kusner et al.~\cite{DBLP:conf/nips/KusnerLRS17} introduced the concept of counterfactual fairness, ensuring that the model outcome remains the same in a counterfactual world where a user's sensitive attributes are different.
Li et al.~\cite{PCFR} introduced this concept to recommender systems and proposed a personalized adversarial framework to allow users or developers to choose sensitive attributes to be protected after training.
Pearl's do-calculus~\cite{pearl2009causality} provides a framework for causal inference, which has been used to control for confounding factors in recommender systems~\cite{he2023addressing}.

In the context of recommender systems, Schnabel et al.~\cite{50} proposed methods to debias learning and evaluation by treating recommendations as treatments in causal inference. 
Wang et al.~\cite{DBLP:conf/kdd/WangF0WC21} introduced deconfounded recommender systems that use causal inference to remove biases from the data. 
Bonner and Vasile~\cite{bonner2018causal} proposed causal embeddings to integrate causal inference into embedding learning for recommendations. 
{Zhao et al.~\cite{zhao2023disentangled} introduce DCCL, a framework that disentangles user interest and conformity in recommender systems using contrastive learning, addressing the challenge of confounding factors in user interactions.}
{PreRec~\cite{pretrained2024causal} adopts a causal debiasing perspective by pre-training on multi-domain user-item interactions to capture universal patterns, facilitating rapid adaptation to new domains with limited data and mitigating selection bias in recommendations.}
{CausalD~\cite{zhang2023causal} employs front-door adjustment via multi-teacher distillation to estimate causal effects in recommendations, addressing performance heterogeneity by mitigating biases from unobserved confounders.}

Our work differs from previous approaches by focusing on the causal effects introduced by multimodal content in recommender systems. 
We propose a causality-inspired framework that explicitly models and intervenes on causal relationships to improve fairness in multimodal recommendations. 
Specifically, we address the entanglement of multimodal content and the heterogeneity between item and user representations by disentangling modal embeddings and utilizing relation-aware fairness learning. 
To the best of our knowledge, this is the first work that addresses fairness in multimodal recommendations from a causal perspective by controlling the causal effects of sensitive attributes through modal disentanglement and relation-aware learning.

%% file: 3.preliminary.tex

\subsection{Recommendation Problem}

Conventionally, we denote user set by $\mathcal{U}=\left\{u_1, u_2, \cdots, u_N\right\}$  and item set by $\mathcal{V}=\left\{v_1, v_2, \cdots, v_M\right\}$.
Let $\mathcal{R} \in \mathbb{R}^{N \times M}$ {denote} the historical binary interaction matrix,  each unit $r_{ij} \in \mathcal{R}$ would be filled by 1 if user $u_i$ has interacted with item $v_j$, otherwise by 0.
For brevity, we denote $u$ for $u_i$, $v$ for $v_j$, and $r_{uv}$ for $r_{ij}$, respectively. 
For each user $u$, her/his preference would be learned and contained in user embedding $\boldsymbol{e}_u \in \mathbb{R}^d$, where $d$ is the latent vector size.
Similarly, for each item $v$, item embedding $\boldsymbol{e}_v \in \mathbb{R}^d$ is learned.
Based on the learned representations, the user's preferences for items can be predicted, thereby generating a top-N recommendation list for each user $u$. 
The loss of the recommendation task is denoted as $\mathcal{L}_{\text{Rec}}$.

Specifically for multimodal recommendations, we denote the modality embedding of item $v$ extracted from the modality encoder (\textit{e.g.}, ResNet as the visual encoder) by $\boldsymbol{e}^m_v \in \mathbb{R}^{d_m}$, where $m$ denotes a specific modality (\textit{e.g.}, visual modality) and $d_m$ is the embedding dimension for the $m$ modality.
In this paper, we consider visual, textual and audio modalities, \textit{i.e.}, $m \in \mathcal{M}=\{\mathrm{v}, \mathrm{t}, \mathrm{a}\}$.

\subsection{Counterfactual Fairness}
From a causal perspective, fairness in recommender systems can be rigorously defined using the concept of counterfactuals.
Counterfactual fairness aims to ensure that recommendations are not influenced by users' sensitive attributes in a causal sense. 
That is, changing the sensitive attributes while keeping everything else (\textit{e.g.,} non-sensitive attributes) constant should not affect the recommendation outcome.
The formal definition of counterfactual fairness in recommender systems is as follows:
\begin{definition} [Counterfactually fair recommendation~\cite{PCFR}]
\label{def:counterfactual_fairness}
\textit{A recommender system is counterfactually fair if it generates a recommendation list $L$ to any users with non-sensitive attribute $\boldsymbol{X} = \boldsymbol{x}$ and sensitive attribute $\boldsymbol{A} = \boldsymbol{a}$ as below:}
\begin{equation}
\operatorname{Pr}\left(L_{\boldsymbol{A} \leftarrow \boldsymbol{a}} \mid \boldsymbol{X} = \boldsymbol{x}, \boldsymbol{A} = \boldsymbol{a}\right)=\operatorname{Pr}\left(L_{\boldsymbol{A} \leftarrow\boldsymbol{a}^\prime} \mid \boldsymbol{X} = \boldsymbol{x}, \boldsymbol{A} = \boldsymbol{a}\right) 
\end{equation}
\textit{for any $L$ and for any value $\boldsymbol{a}^\prime$ attainable by $\boldsymbol{A}$.} 
\end{definition}

In other words, the causal effect of the sensitive attributes $\boldsymbol{A}$ on the recommendation outcome $L$ should be nullified.
To achieve counterfactual fairness in recommendations, we need to guarantee the independence between users' sensitive attribute $\boldsymbol{A}_u$ and the recommendation list $L_u$ through representations~\cite{PCFR}.
Figure~\ref{fig:causal_graph} illustrates the causal relationships in general RS and multimodal RS.
In multimodal RS, the multimodal content $\mathcal{M}$ can introduce additional paths through which the sensitive attributes $\boldsymbol{A}_u$ causally affect the predicted scores $\mathcal{S}_u$ and the subsequent recommendation list $L$, making it more challenging to achieve counterfactual fairness.

As the recommendation list $L_u$ is generated based on the predicted preference $\hat{r}_{uv} = \boldsymbol{e}_u \cdot \boldsymbol{e}_v$ typically, we need to guarantee the independence between the sensitive attribute $\boldsymbol{A}_u$ and the user and item representations, \textit{i.e.}, $\boldsymbol{e}_u \perp \boldsymbol{A}_u$ and $ \boldsymbol{e}_v \perp \boldsymbol{A}_u$ for each $u \in \mathcal{U}$ and all $v \in \mathcal{V}_u$, where $\mathcal{V}_u$ is the item set interacted by user $u$.

\subsection{Empirical Analysis of Sensitive Information Leakage in Multimodal Content} 
\label{sec:evidence}
\begin{figure*}[htbp]
  \centering

   \captionsetup[subfigure]{justification=raggedleft, singlelinecheck=false}
   \hfill
  \begin{subfigure}[b]{0.42\textwidth}
    \includegraphics[width=1\textwidth]{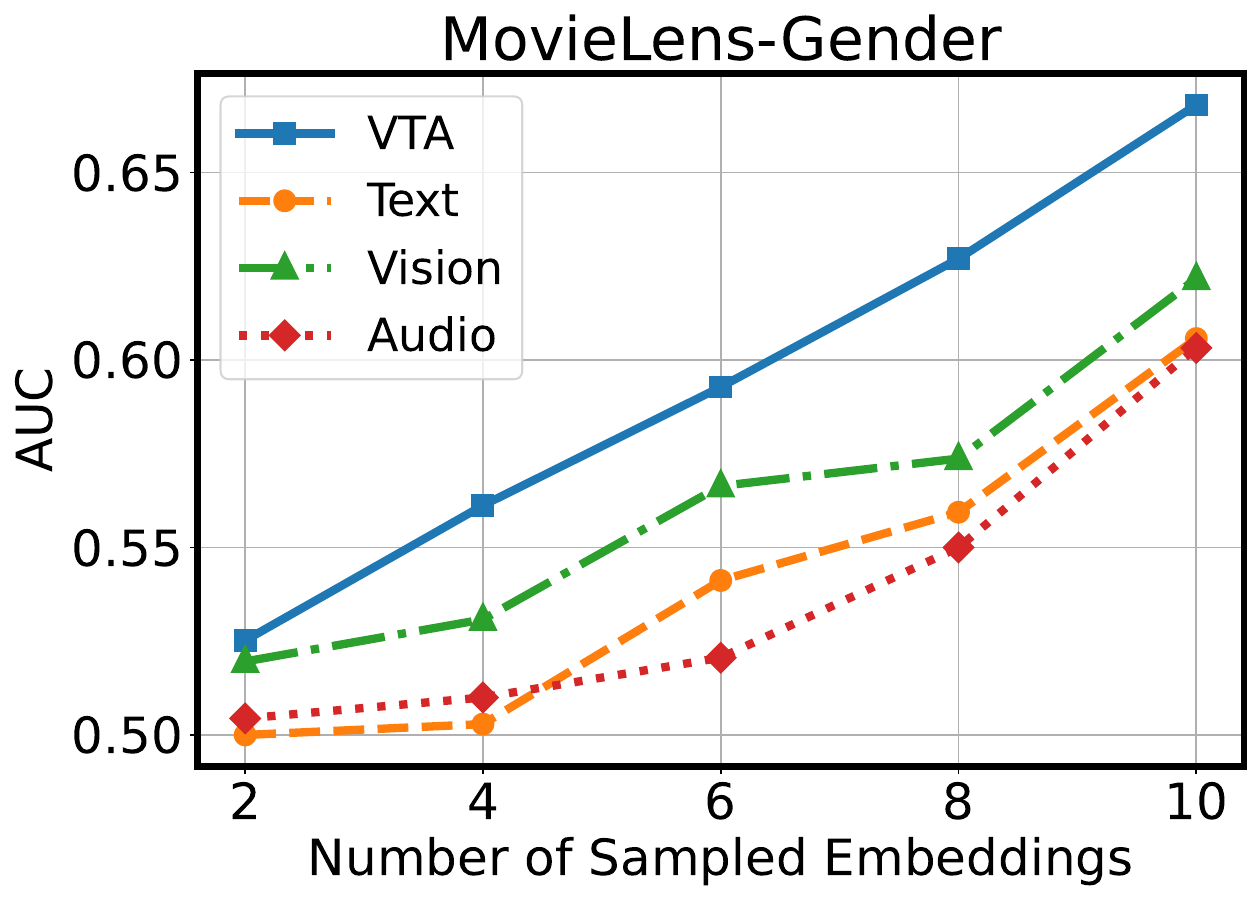}
    \label{fig:subfig1}
  \end{subfigure}%
   \hfill
  \begin{subfigure}[b]{0.42\textwidth}
    \includegraphics[width=1\textwidth]{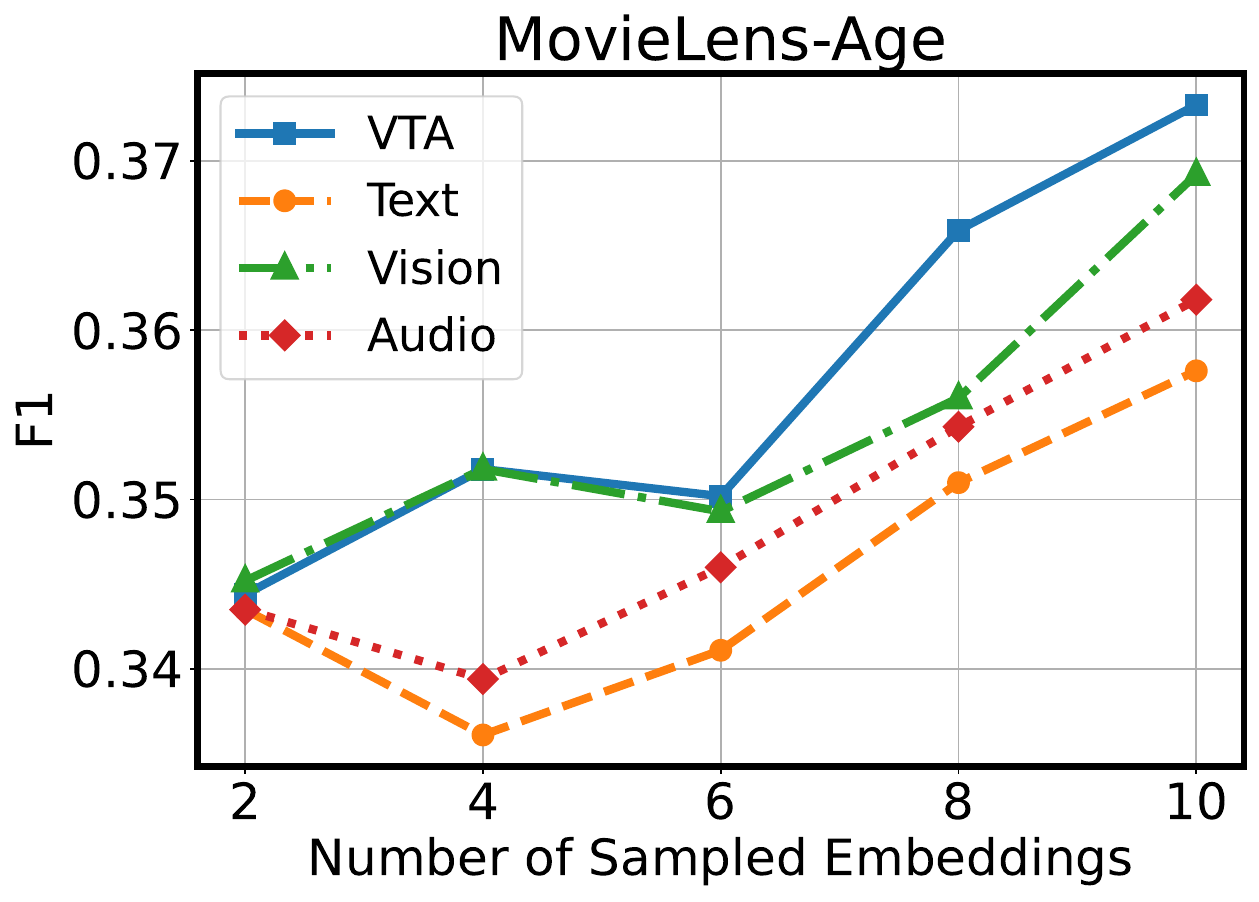}
    \label{fig:subfig2}
  \end{subfigure}%
    \hspace{\fill}

    \hfill
  \begin{subfigure}[b]{0.42\textwidth}
    \includegraphics[width=1\textwidth]{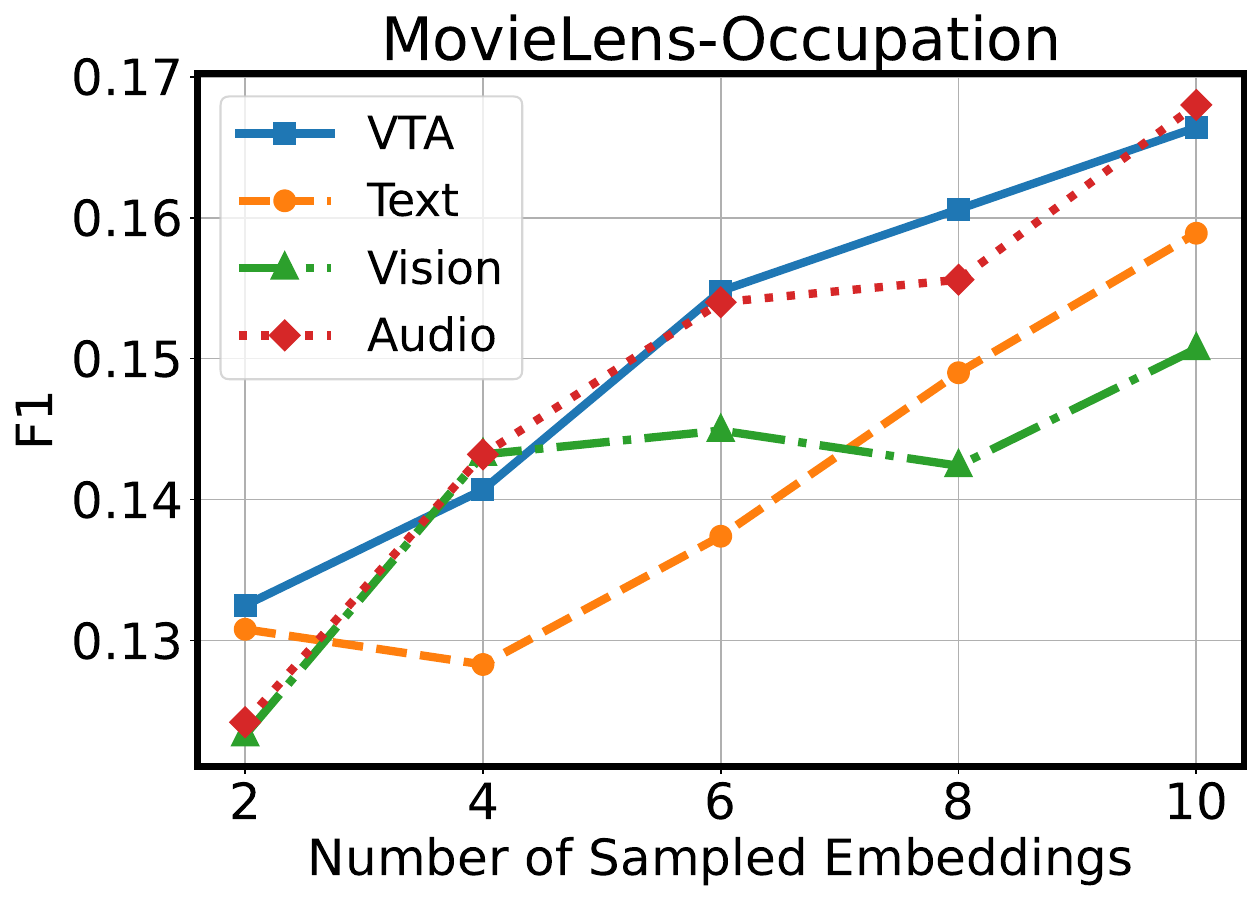}
    \label{fig:subfig3}
  \end{subfigure}%
   \hfill
  \begin{subfigure}[b]{0.42\textwidth}
    \includegraphics[width=1\textwidth]{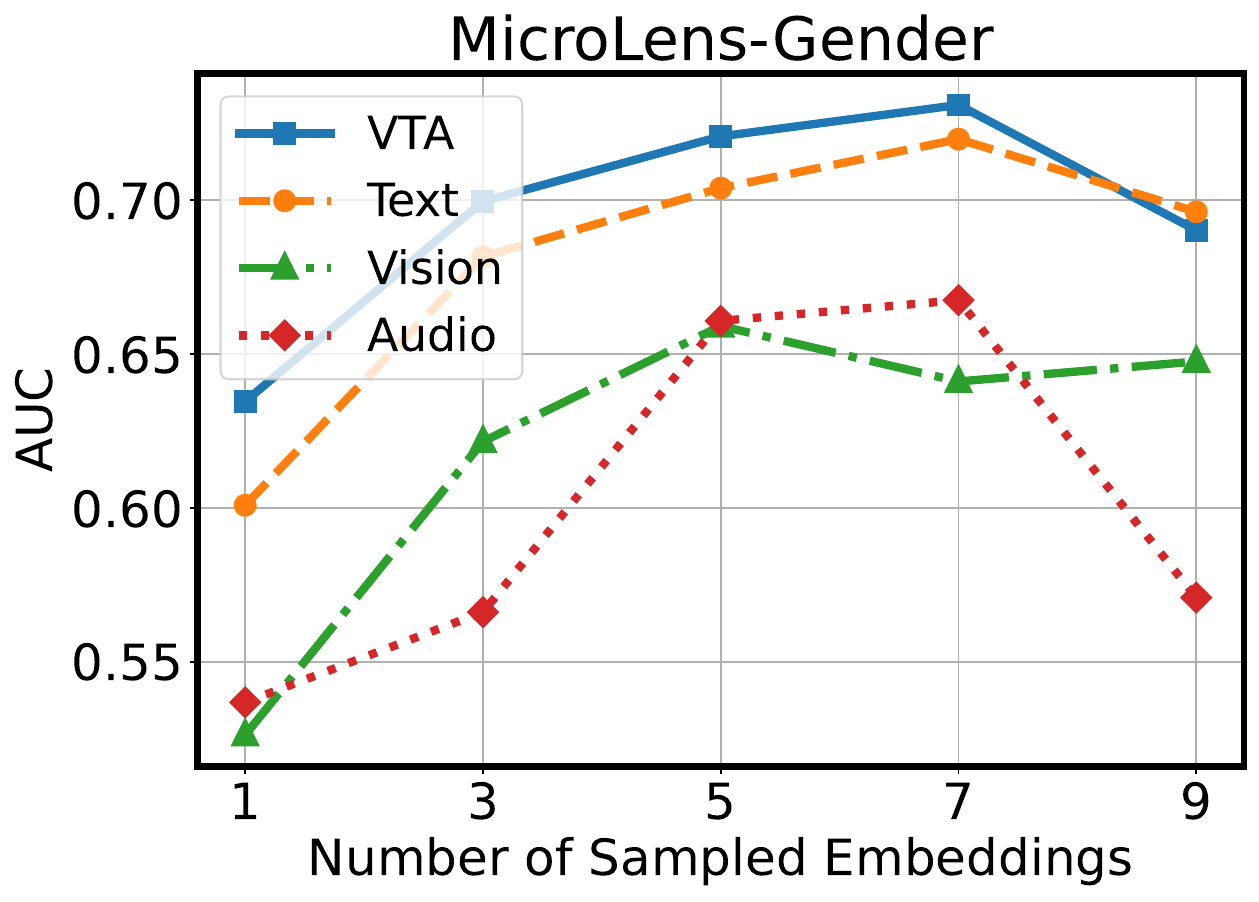}
    \label{fig:subfig4}
  \end{subfigure}%
    \hspace{\fill}
  
  \caption{Sensitive attribute prediction performance given different numbers (\textit{quantity}) and types (\textit{variety}) of modalities on two datasets MovieLens and MicroLens. 
  More details of how we train and evaluate attackers can be found in Section~\ref{sec:evaluation}.
  }
  \label{fig:sens_leak_exp}
\end{figure*}

In this preliminary study, we probe users' sensitive attribute(s) using varying numbers and types of modalities on two datasets MovieLens and MicroLens (see Section~\ref{sec:datasets}). 
The results are illustrated in Figure~\ref{fig:sens_leak_exp}.
The unimodal representation (\textit{e.g.}, text, vision, and audio) of a user is generated by aggregating the representations of items sampled from the user's interactions.
The multimodal representation \textit{VTA} is denoted by the concatenation of unimodal representations.

From a causal perspective, these modalities can be viewed as mediators or proxies that potentially carry information about sensitive attributes.
The trend observed across our experiments indicates that both the quantity and variety of modalities increase the leakage of users’ sensitive information. 
This suggests that the multimodal content introduces additional causal pathways from the sensitive attributes to the user representations and ultimately to the recommendations.

There are two notable observations:
\begin{itemize}[leftmargin=*]
    \item 
    \textit{O1}: As the \textit{quantity} of accessible modal content increases, while the type of modality remains constant, there is a more profound leakage of users' sensitive information \textit{w.r.t.} demographic attributes (\textit{e.g.}, gender, age, and occupation).
    This implies that the causal influence of sensitive attributes is amplified through more abundant modal data.
    \item \textit{O2}: As a wider \textit{variety} of modalities (\textit{e.g.}, \textit{VTA}) is introduced, while the quantity of available modalities is fixed, there is a more severe leakage of users' sensitive information.
    This indicates that different modalities may capture different aspects of the causal relationships between sensitive attributes and user preferences.
\end{itemize}

Hence, this study highlights that the risk of sensitive information leakage can be introduced in the process of learning user preferences in multimodal recommendations due to complex causal dependencies. 
Consequently, ensuring the independence between multimodal recommendations and users' sensitive attributes becomes more challenging than in traditional recommendations without leveraging multimodal content.

%% file: 4.methodology.tex
\begin{figure*}[htbp]
\setlength{\abovecaptionskip}{3pt plus 2pt minus 2pt}
  \centering
  \includegraphics[width=1.02\textwidth]{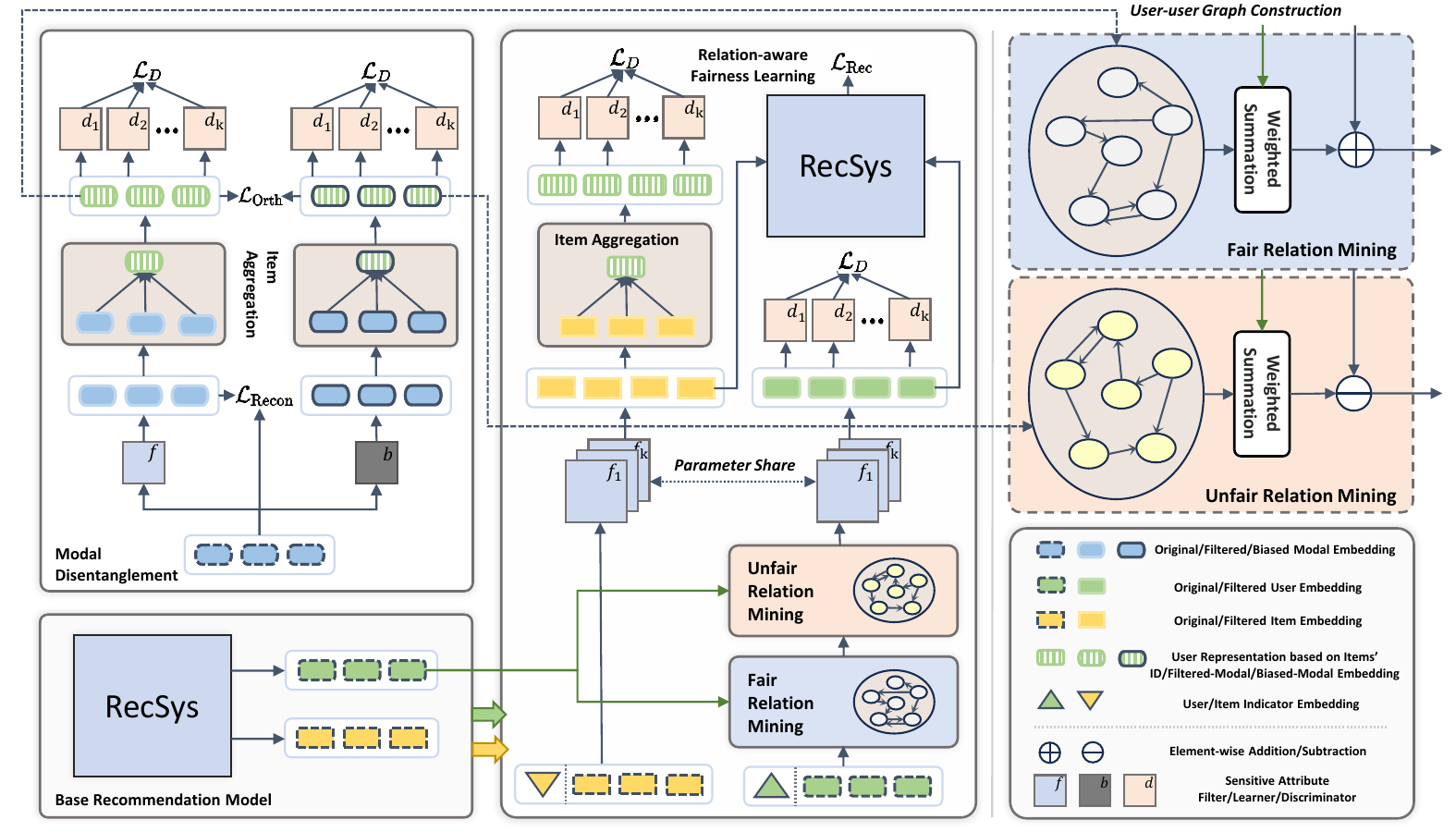}
  \caption{The overall flowchart of our FMMRec.}
  \label{fig:model}
\end{figure*}

The overall illustration of our proposed FMMRec is shown in Figure~\ref{fig:model}.
Specifically, \textbf{fairness-oriented modal disentanglement} is to separate fair and unfair modal embeddings, and \textbf{relation-aware fairness learning} is to mine dual user-user relations given the disentangled modal embeddings for eliminating sensitive information in user representations while preserving non-sensitive information.

\subsection{Fairness-oriented Modal Disentanglement}

As discussed, separating sensitive and non-sensitive information from multimodal content is crucial to address fairness issues without significant accuracy loss.
From a causal inference standpoint, the multimodal content can be seen as confounders or mediators that introduce spurious associations between sensitive attributes and user-item interactions~\cite{pearl2009causality, scholkopf2021toward}.
Thus, to address \textit{C1} (entanglement), our objective is to disentangle the original modal embedding $\boldsymbol{e}^m_v \in \mathbb{R}^{d_m}$ into two separate embeddings~\cite{Denoising, Aug, lai2024matryoshka}: 
\begin{itemize}
    \item $\boldsymbol{\bar{e}}^m_v \in \mathbb{R}^{d_m}$, which contains minimal sensitive information (representing the non-causal part with respect to sensitive attributes),
    \item $\boldsymbol{\tilde{e}}^m_v \in \mathbb{R}^{d_m}$, which captures the maximal sensitive information (representing the causal effects of sensitive attributes).
\end{itemize} 

To generate the mentioned two views of modal embedding, we employ a filter network $f_m: \mathbb{R}^{d_m} \mapsto \mathbb{R}^{d_m}$ and a biased learner network $b_m: \mathbb{R}^{d_m} \mapsto \mathbb{R}^{d_m}$ to original modal embedding $\boldsymbol{e}^m_v$:

\begin{equation}
\label{eqn:L_1}
\begin{split}
\boldsymbol{\bar{e}}^m_v = f_m(\boldsymbol{e}^m_v), \\
\boldsymbol{\tilde{e}}^m_v = b_m(\boldsymbol{e}^m_v).
\end{split}
\end{equation}

This process can be viewed as performing a causal intervention~\cite{pearl2009causality} to separate the representations into non-sensitive (causal effect removed) and sensitive components.
To detect how much sensitive information is learned in the filtered and biased modal embeddings $\boldsymbol{\bar{e}}^m_v$ and $\boldsymbol{\tilde{e}}^m_v$, we train two sets of $K$ discriminators $D^f_m = \{d^f_{m, 1}, d^f_{m, 2}, \ldots, d^f_{m, K}\}$ and $D^b_m = \{d^b_{m, 1}, d^b_{m, 2},\ldots, d^b_{m, K}\}$ to infer the ground-truth values of $K$ sensitive attributes of users.
By minimizing the sensitive information in $\boldsymbol{\bar{e}}^m_v$ and maximizing it in $\boldsymbol{\tilde{e}}^m_v$, we aim to control the causal influence of sensitive attributes on the recommendations.
Concretely, each sub discriminator $d^f_{m, k}$ (or $d^b_{m, k}$) works as a classifier to predict the $k$-th sensitive attribute, given user $u$'s aggregated modal embedding $\boldsymbol{\bar{e}}_u^m$ (or $\boldsymbol{\tilde{e}}_u^m$):

\begin{equation}
\label{eqn:e_4}
\begin{split}
\boldsymbol{\bar{a}}^m_{uk} = d^f_{m, k} (\boldsymbol{\bar{e}}^m_u),\; \boldsymbol{\bar{e}}_u^m = \frac{\sum _{v \in \mathcal{V}_u} \boldsymbol{\bar{e}_v^m}}{|\mathcal{V}_u|},  \\
\boldsymbol{\tilde{a}}^m_{uk} = d^b_{m, k} (\boldsymbol{\tilde{e}}^m_u),\; \boldsymbol{\tilde{e}}_u^m = \frac{\sum _{v \in \mathcal{V}_u} \boldsymbol{\tilde{e}_v^m}}{|\mathcal{V}_u|},
\end{split}
\end{equation}

where $\boldsymbol{\bar{a}}^m_{uk}$ and $\boldsymbol{\tilde{a}}^m_{uk}$ are the predicted values of $k$-th sensitive attribute of user $u$ given the filtered and biased modal embeddings $\boldsymbol{\bar{e}}^m_u$ and $\boldsymbol{\tilde{e}}^m_u$, respectively.
With the output of these two sets of discriminators, we employ two sensitive attribute prediction losses $\mathcal{L}_{D^b_m}$ and $\mathcal{L}_{D^f_m}$ for biased and filtered embeddings, adopting binary cross entropy loss (BCE loss) for binary attributes (\textit{e.g.}, \textit{gender}) and negative log likelihood loss (NLL loss) for multi-class attributes (\textit{e.g.}, \textit{occupation}).
Taking the binary-class case as an example:

\begin{equation}
\label{eq:l_d_f_m}
\mathcal{L}_{D^f_m} = \sum_{k=1}^K \boldsymbol{a}_{uk} \cdot \log \left(\boldsymbol{\bar{a}}^m_{uk}\right)+\left(1-\boldsymbol{a}_{uk}\right) \cdot \log \left(1-\boldsymbol{\bar{a}}^m_{uk}\right),
\end{equation}
\begin{equation}
\label{eq:l_d_b_m}
\mathcal{L}_{D^b_m} = \sum_{k=1}^K \boldsymbol{a}_{uk} \cdot \log \left(\boldsymbol{\tilde{a}}^m_{uk}\right)+\left(1-\boldsymbol{a}_{uk}\right) \cdot \log \left(1-\boldsymbol{\tilde{a}}^m_{uk}\right),
\end{equation}

where $\boldsymbol{a}_{uk}$ is the user $u$'s ground-truth value of the $k$-th sensitive attribute (\textit{e.g.}, 0 for \textit{male} and 1 for \textit{female}).
Moreover, we aim to preserve the non-sensitive representative information of the filtered modal embedding $\boldsymbol{\bar{e}}^m_v$, making it to be similar to the original modal embedding $\boldsymbol{e}^m_v$ for retaining recommendation performance.
Therefore, we consider a reconstruction loss for filtered modal embedding $\boldsymbol{\bar{e}}^m_v$ to retain non-sensitive information as much as in the original embedding $\boldsymbol{e}^m_v$ of items interacted by user $u$:

\begin{equation}
\mathcal{L}_{\text{Recon}} = \frac{1}{|\mathcal{V}_u|} \sum_{v \in \mathcal{V}_u} 1 - \frac{\boldsymbol{\bar{e}}^m_v \cdot \boldsymbol{e}^m_v}{ \lVert \boldsymbol{\bar{e}}^m_v \rVert  \lVert \boldsymbol{e}^m_v \rVert }.
\end{equation}

Though we directly optimize the Equations~\eqref{eq:l_d_f_m} and \eqref{eq:l_d_b_m}, there would still be some sensitive information leakage in the filtered modal embedding $\boldsymbol{\bar{e}}^m_v$ in practice since the filter network $f^m$ could cheat the corresponding discriminators instead of fully removing the sensitive information~\cite{FairRec}.
Analogously, the learner $b^m$ may not perfectly capture the sensitive information into the biased modal embedding $\boldsymbol{\tilde{e}}^m_v$.
To further filter out the sensitive information in the filtered embedding and elicit sensitive information into the biased embedding, we push the filtered and biased modal embeddings away from each other in the latent space by an orthogonality loss:

\begin{equation}
\label{eq:l_orth}
\mathcal{L}_{\text{Orth}} = \frac{1}{|\mathcal{V}_u|} \sum_{v \in \mathcal{V}_u} \max (0, \frac{\boldsymbol{\bar{e}}^m_v \cdot \boldsymbol{\tilde{e}}^m_v}{\lVert \boldsymbol{\bar{e}}^m_v \rVert 
 \lVert \boldsymbol{\tilde{e}}^m_v \rVert}).
\end{equation}

The adversarial training setup here aligns with causal representation learning frameworks that aim to isolate and control causal factors~\cite{scholkopf2021toward}.
Similar to adversarial training, we adopt a min-max game for the optimization of the filter network and the corresponding discriminator, and optimize the disentanglement learning by jointly optimizing these four losses with different importance.
The overall optimization objectives for the filter and learner networks are:

\begin{equation}
\label{eqn:optimization_dis_2}
\operatorname*{arg\,min} _{f^m, b^m} \mathcal{L}_{\text{Recon}} + \lambda_{m_0}( \mathcal{L}_{D^b_m} - \mathcal{L}_{D^f_m}) + \lambda_{m_1} \mathcal{L}_{\text{Orth}},
\end{equation}
\begin{equation}
\label{eqn:optimization_dis_1}
\operatorname*{arg\,min} _{D^f_m, D^b_m} \mathcal{L}_{D^f_m} + \mathcal{L}_{D^b_m},
\end{equation}

where $\lambda_{m_0}$ and $\lambda_{m_1}$ are hyperparameters controlling the trade-off between fairness and reconstruction.
By disentangling the embeddings in this way, we attempt to control the causal effects from the sensitive attributes to the recommendation outcomes through the multimodal content.

\subsection{Relation-aware Fairness Learning}
For \textit{C2} (heterogeneity), we exploit the biased and filtered modal embeddings to mine unfair and fair user-user relations for promoting fairness and expressiveness of user representations, respectively.
This step corresponds to modeling the causal relationships among users, aiming to mitigate the influence of sensitive attributes~\cite{DBLP:conf/nips/KilbertusRPHJS17}.
\subsubsection{Fair/Unfair Relation Mining}
In fact, a significant obstacle in enhancing fairness by utilizing multimodal information stems from the substantial disparity in latent space between multimodal embedding and user embedding.
What's more, there is no direct correspondence between them.
Hence, directly using multimodal information for fairness learning remains challenging.
To overcome this challenge, we propose an innovative approach called fair/unfair relation mining to utilize biased and filtered modal embeddings to mine dual user-user relations for learning fair and informative user representations with the awareness of the mined relations.
Specifically, unfair relations (\textit{w.r.t.} biased embeddings) are identified to eliminate the sensitive information in user representations, while fair relations (\textit{w.r.t.} filtered embeddings) are to preserve personalized non-sensitive information.

Inspired by~\cite{LATTICE, yang2021enhanced, ye2023towards,HACD}, we aim to construct modality-based user-user relations to find "fair" and "unfair" neighbors of each user.
We adopt the simple and parameter-free cosine similarity to compute the modality-based user-user "fair" and "unfair" similarity {matrices}{, denoted as $\bar{S}^m \in \mathbb{R}^{N \times N}$ and $\tilde{S}^m \in \mathbb{R}^{N \times N}$, respectively. These matrices are constructed for each user-pair entry $u, u^{\prime} \in \mathcal{U}$} based on users' filtered/biased aggregated modality-based embedding $\boldsymbol{\bar{e}}_u^m$/$\boldsymbol{\tilde{e}}_u^m$, as defined in Equation~\eqref{eqn:e_4}:
\begin{equation}
\label{eqn:s_uu}
\begin{split}
{
\bar{S}_{u u^{\prime}}^m=\frac{\boldsymbol{\bar{e}}_u^m \cdot \boldsymbol{\bar{e}}_{u^{\prime}}^m}{\left\lVert\boldsymbol{\bar{e}}_u^m\right\rVert\left\lVert\boldsymbol{\bar{e}}_{u^{\prime}}^m\right\rVert},}\\
{\tilde{S}_{u u^{\prime}}^m=\frac{\boldsymbol{\tilde{e}}_u^m \cdot \boldsymbol{\tilde{e}}_{u^{\prime}}^m}{\left\lVert\boldsymbol{\tilde{e}}_u^m\right\rVert\left\lVert\boldsymbol{\tilde{e}}_{u^{\prime}}^m\right\rVert}.}
\end{split} 
\end{equation}

The graph adjacency matrix is conventionally expected to have non-negative values. 
However, in the case of $\bar{S}_{u u^{\prime}}^m$ (or $\tilde{S}_{u u^{\prime}}^m$), which ranges between [-1, 1], we set any negative entries to zeros following convention~\cite{LATTICE} as users' similarities are supposed to be non-negative.
What's more, following~\cite{LATTICE}, we perform $\mathbb{k}$-nearest-neighbor ($\mathbb{k}$NN) sparsification~\cite{chen2009fast} to convert the dense graphs to sparse graphs for computational efficiency and normalization to address the exploding/vanishing gradient problem~\cite{kipf2016semi}.

These similarity measures reflect the causal influence of sensitive attributes on user similarities. 
By separating fair and unfair relations, we aim to adjust the learning process to mitigate undesired causal effects~\cite{DBLP:conf/nips/KusnerLRS17, Wang2019blessings}.
With the mined adjacency matrices $\bar{S}^m$ and $\tilde{S}^m$ (\textit{i.e.}, fair and unfair relations) for each modality $m \in \mathcal{M}$, we propose to fuse multiple unimodal matrices to an integrated multimodal matrix\footnote{We opt for disentanglement prior to integrating adjacency matrices due to the potential information loss during modality aggregation and the increased difficulty in disentangling integrated modal representations.}.
Considering that the different modalities may exhibit different levels of sensitive information leakage (see Figure~\ref{fig:sens_leak_exp}), we assign different importance to different modalities when integrating multiple filtered/biased unimodal adjacency matrices into a filtered/biased multimodal adjacency matrix:
\begin{equation}
\label{eq:agg_graphs}
\begin{split}
\bar{S}=\sum_{m \in \mathcal{M}} \alpha_m \bar{S}^m,\\
\tilde{S}=\sum_{m \in \mathcal{M}} \alpha_m \tilde{S}^m,
\end{split}
\end{equation}

where $\bar{S} \in \mathbb{R}^{N \times N}$ and $\tilde{S} \in \mathbb{R}^{N \times N}$ are the filtered and biased multimodal adjacency matrices, respectively, and $\alpha_m$ denotes the weight of modality $m$.
We restrict $\sum_{m \in \mathcal{M}} \alpha_m=1$ to keep the integrated graphs $\bar{S}$ and $\tilde{S}$ normalized.
We then aggregate the filtered/biased neighbors' representations of each user $u$ with the constructed modality-based similarities as weights:
\begin{equation}
\label{eq:nb_rep}
\begin{split}
\boldsymbol{\bar{h}}_u=\sum_{u^{\prime} \in \mathcal{N}(u; \bar{S})} \bar{S}_{u u^{\prime}} \boldsymbol{e_{u^{\prime}}},\\
\boldsymbol{\tilde{h}}_u=\sum_{u^{\prime} \in \mathcal{N}(u; \tilde{S})} \tilde{S}_{u u^{\prime}} \boldsymbol{e_{u^{\prime}}},
\end{split}
\end{equation}

where $\mathcal{N}(u; \bar{S})$ and $\mathcal{N}(u; \tilde{S})$ are the filtered and biased neighbor sets of user $u$ respectively, and $\boldsymbol{\bar{h}}_u \in \mathbb{R}^d$ and $\boldsymbol{\tilde{h}}_u \in \mathbb{R}^d$ are the filtered and biased neighbor representations respectively.
With the filtered and biased neighbor representations, we incorporate the modality-based fair and unfair relations into the user representation $\boldsymbol{e_u}$\footnote{We did not enhance the item representation due to the training instability in experiments. One of the possible reasons is that the learned multimodal information of item representation may be essential in multimodal recommendations, and enhancing the item representation by enforcing the margin between it and its relation-based neighbors may hinder the learning of multimodal information.}~\cite{chen2022global} for enhancing its fairness and expressiveness:
\begin{equation}
\label{eq:enhance}
\boldsymbol{\hat{e}_u} = \boldsymbol{e_u} + \lambda_h (\boldsymbol{\bar{h}}_u - \boldsymbol{\tilde{h}}_u),
\end{equation}

where $\boldsymbol{\hat{e}_u} \in \mathbb{R}^d$ is the relation-aware user representation, and $\lambda_h$ is a hyperparameter to control the weight of enhancement.
This adjustment can be seen as a form of causal intervention, where we enhance the user representation by incorporating information from fair neighbors and reducing the influence of unfair ones~\cite{pearl2009causality}.

\subsubsection{Adversarial Learning for Causal Fairness}
In this section, we introduce the adversarial learning framework to both the relation-aware user representation and item representation.
This process aligns with the approach of using adversarial networks to achieve fair representations by removing the influence of sensitive attributes~\cite{DBLP:conf/icml/MadrasCPZ18}.
Basically, the filter network for eliminating the information of sensitive attributes in representations and the discriminator network for predicting the values of sensitive attributes are applied together and adversarially optimized via a min-max game.

One of the related works is FairGo~\cite{FairGo} which applied a composition of filters based on~\cite{bose2019compositional} to both the user and item representations for learning fair representations.
However, it cannot distinguish whether the current input is a user or an item, which may lead to confusing learning and be more ineffective.
To address this issue, we first introduce two simple yet effective \textit{role indicator embeddings} $\boldsymbol{r}_u \in \mathbb{R}^{d_r}$ for indicating user and $\boldsymbol{r}_v \in \mathbb{R}^{d_r}$ for indicating item.
By concatenating the role indicator embedding $\boldsymbol{r}_u$ (or $\boldsymbol{r}_v$) to the enhanced user representation $\boldsymbol{\hat{e}}_u$ (or $\boldsymbol{e}_v$) as the input of the parameter-shared compositional filters $F = \{f_1, f_2, \ldots, f_K\}$, thus the filters could not only share the filtering knowledge among the user and item sides but also be able to distinguish the input role each time.
Formally, we can obtain the filtered user and item representations by:
\begin{equation}
\label{eq:filter_u_i}
\begin{split}
\boldsymbol{\bar{e}}_u=\frac{\sum_{k=1}^K {f}_k ([\boldsymbol{r}_u ; \boldsymbol{\hat{e}}_u])}{K}, \\
\boldsymbol{\bar{e}}_v=\frac{\sum_{k=1}^K {f}_k ([\boldsymbol{r}_v ;\boldsymbol{e}_v])}{K},
\end{split}
\end{equation}

where each sub filter $f_k: \mathbb{R}^{d_r + d} \mapsto \mathbb{R}^{d_r + d}$ has the same structure and capacity. 
Correspondingly, to learn the filter network $F$, we train two sets of discriminators $D_u = \{d^u_1, d^u_2, \ldots, d^u_K\}$ and $D_v = \{d^v_1, d^v_2, \ldots, d^v_K\}$ to predict the value of user $u$'s  sensitive attributes from the filtered user representation $\boldsymbol{\bar{e}}_u$ and the implicit representation of user by aggregating her/his interacted items' filtered representations $\boldsymbol{\bar{e}_v}, \forall v \in \mathcal{V}_u$: 
\begin{equation}
\label{eqn:e_u_e_v_u}
\begin{split}
\boldsymbol{\bar{a}}_{uk} = d^u_{k} (\boldsymbol{\bar{e}}_u), \;\;\;\;\;\;\;\;\;\;\;\;\;\;\;\;\;\;  \\
\boldsymbol{\bar{a}}_{\mathcal{V}_u k} = d^v_{k} (\boldsymbol{\bar{e}}_{\mathcal{V}_u}),\; \boldsymbol{\bar{e}}_{\mathcal{V}_u} = \frac{\sum _{v \in \mathcal{V}_u} \boldsymbol{\bar{e}_v}}{|\mathcal{V}_u|},
\end{split}
\end{equation}

where $\boldsymbol{\bar{a}}_{uk}$ and $\boldsymbol{\bar{a}}_{\mathcal{V}_u k}$ are the predicted values of user $u$'s $k$-th sensitive attribute given her/his explicit representation $\boldsymbol{\bar{e}}_u$ and implicit representation $\boldsymbol{\bar{e}}_{\mathcal{V}_u}$, respectively. 
Similar to Equations~\eqref{eq:l_d_f_m}-\eqref{eq:l_d_b_m}, we can then calculate prediction losses given a binary attribute as an example:
\begin{equation}
\label{eq:l_d_u}
\mathcal{L}_{D_u} = \sum_{k=1}^K \boldsymbol{a}_{uk} \cdot \log \left(\boldsymbol{\bar{a}}_{uk}\right)+\left(1-\boldsymbol{a}_{uk}\right) \cdot \log \left(1-\boldsymbol{\bar{a}}_{uk}\right),
\end{equation}
\begin{equation}
\label{eq:l_d_v}
\mathcal{L}_{D_v} = \sum_{k=1}^K \boldsymbol{a}_{uk} \cdot \log \left(\boldsymbol{\bar{a}}_{\mathcal{V}_u k}\right)+\left(1-\boldsymbol{a}_{uk}\right) \cdot \log \left(1-\boldsymbol{\bar{a}}_{\mathcal{V}_u k}\right).
\end{equation}

To balance accuracy and fairness, we adversarially optimize the above losses $\mathcal{L}_{D_u}$ and $\mathcal{L}_{D_v}$, jointly with the recommendation loss $\mathcal{L}_{\text{Rec}}$ optimization:
\begin{equation}
\label{eqn:L_r}
\operatorname*{arg\,min} _{\boldsymbol{\Theta \backslash \{D_u, D_v\}}} \mathcal{L}_{\text{Rec}} - \lambda_{D_u} \mathcal{L}_{D_u} - \lambda_{D_v} \mathcal{L}_{D_v},
\end{equation}
\begin{equation}
\label{eqn:L_d}
\operatorname*{arg\,min} _{D_u, D_v} \mathcal{L}_{D_u} + \mathcal{L}_{D_v},
\end{equation}

where $\Theta$ are the all parameters for relation-aware fairness learning, and $\lambda_{D_u}$ and $\lambda_{D_v}$ are the hyperparameters to control the accuracy-fairness trade-off.
By adversarially training the filters and discriminators, we further aim to remove the causal effects of sensitive attributes from the representations, achieving counterfactual fairness~\cite{DBLP:conf/nips/KusnerLRS17, PCFR}.
The pseudo-code of our fairness-aware multimodal recommendation process is shown in Algorithm \ref{alg:cap}.

\subsection{{Time Complexity Analysis}}\label{sec:complexity}

{In this section, we analyze the time complexity of the main components of our proposed FMMRec algorithm.}
\begin{itemize}
    \item {\textbf{Complexity of Modal Disentanglement.} The modal disentanglement process involves passing the original modal embeddings $\mathbf{e}_v^m$ through the filter and learner networks, each of which consists of simple neural network layers. For each item $v$ and modality $m$, the computational cost is $O(d_m^2)$, where $d_m$ is the dimensionality of the modal embeddings. Since this operation is performed for all items and modalities, the time complexity for modal disentanglement is $O(M \cdot |\mathcal{V}| \cdot d_m^2)$, where $M$ is the number of modalities and $|\mathcal{V}|$ is the number of items.}
    \item {\textbf{Complexity of Relation-Aware Fairness Learning.} In this step, each user's representation is updated by aggregating information from their neighbors. Without considering sparsification, each user potentially connects to \( N \) neighbors on average, leading to a complexity of \( O(|\mathcal{U}| \cdot N \cdot d) \), where \( |\mathcal{U}| \) is the number of users, \( N \) is the average number of user neighbors, and \( d \) is the embedding dimension.}
    \item {\textbf{Complexity of Adversarial Training.} The adversarial components involve updating user and item embeddings through discriminators to eliminate sensitive information. The complexity for this process is \( O(|\mathcal{U}| \cdot d^2 + |\mathcal{V}| \cdot d^2) \), due to operations involving embedding matrices of size \( d \).}
\end{itemize}


{Combining these components, the overall training time complexity is:}

\begin{equation}
{
O\left( M \cdot |\mathcal{V}| \cdot d_m^2 + |\mathcal{U}| \cdot N \cdot d + |\mathcal{U}| \cdot d^2 + |\mathcal{V}| \cdot d^2 \right).}
\end{equation}

{Since \( M \) is a small constant (\textit{e.g.,} 3), we can omit it in the big \( O \) notation. Also, in practice, the dimensions \( d_m \) and \( d \) are of similar sizes, so we can approximate \( d_m^2 \) with \( d^2 \). Therefore, the overall training complexity simplifies to:}

\begin{equation}
{
O\left(|\mathcal{U}| \cdot N \cdot d + |\mathcal{V}| \cdot d^2 + |\mathcal{U}| \cdot d^2 \right)}
\end{equation}

{To further reduce the training complexity, we apply $\mathbb{k}$-nearest neighbors ($\mathbb{k}$NN) to limit the number of neighbors per user to \( \mathbb{k} \), where \( \mathbb{k} \ll N \). This reduces the neighbor aggregation complexity from \( O(|\mathcal{U}| \cdot N \cdot d) \) to \( O(|\mathcal{U}| \cdot \mathbb{k} \cdot d) \). Since $\mathbb{k}$ is a small constant, the updated overall training complexity becomes:}

\begin{equation}
{
O\left(|\mathcal{U}| \cdot d^2 + |\mathcal{V}| \cdot d^2 \right)}
\end{equation}

{This final expression indicates that the training time complexity of FMMRec scales linearly with the number of users \( |\mathcal{U}| \) and items \( |\mathcal{V}| \), and quadratically with the embedding dimension \( d \). Notably, this is at the same level as basic adversarial-based fairness approaches, effectively demonstrating that our method maintains practical computational efficiency suitable for real-world recommender systems.}

\begin{algorithm}
\caption{Fair Multimodal Recommendation}
\label{alg:cap}
\begin{algorithmic}
\renewcommand{\algorithmicrequire}{\textbf{Input:}}
\renewcommand{\algorithmicensure}{\textbf{Output:}}

\renewcommand{\algorithmicrequire}{\textbf{Phase 1:}}
\Require Fairness-oriented Modal
Disentanglement

\renewcommand{\algorithmicrequire}{\textbf{Input:}}
\Require Training user set $\mathcal{U}$; Item set $\mathcal{V}_u$ of user $u$; User embedding $\boldsymbol{e}_u, \forall u \in \mathcal{U}$; Modal embedding $ \boldsymbol{{e}}_v^m \;, \forall v \in \mathcal{V}, \forall m \in \mathcal{M}$

\Ensure $\boldsymbol{\bar{e}}_u^m$ and  $\boldsymbol{\tilde{e}}_u^m, \; \forall u \in \mathcal{U}, \forall m \in \mathcal{M}$;

\renewcommand{\algorithmicensure}{\textbf{Initialize:}}

\Ensure $f_m$, $b_m$, $D^f_m$, $D^b_m$


\For{$m \in \mathcal{M}$}

\Repeat 
    \For{$u \in \mathcal{U}$}
        \State \Comment{Get filtered and biased modal embeddings and the corresponding predicted attribute values\;\;\;\;\;\;\;\;\;\;\;\;\;\;\;\;}
        \State  $\boldsymbol{\bar{e}}_u^m, \boldsymbol{\tilde{e}}_u^m,  \boldsymbol{\bar{a}}^m_{uk}, \boldsymbol{\tilde{a}}^m_{uk} \gets$ Eq.~\eqref{eqn:e_4}
    
        \State \Comment{Calculate filtered and biased attribute prediction losses, reconstruction loss, and orthogonality loss\;\;\;\;\;}
        \State $\mathcal{L}_{D^f_m}, \mathcal{L}_{D^b_m}, \mathcal{L}_{\text{Recon}}, \mathcal{L}_{\text{Orth}} \gets$ Eq.~\eqref{eq:l_d_f_m}-\eqref{eq:l_orth}

        \State Optimize $f_m$, $b_m$, $D^f_m$, $D^b_m$ according to Eq.~\eqref{eqn:optimization_dis_1} and \eqref{eqn:optimization_dis_2}
    \EndFor
\Until{\textit{stopping criterion is met};}

\EndFor



\State \hrulefill
\renewcommand{\algorithmicrequire}{\textbf{Phase 2:}}
\Require Relation-aware Fairness Learning

\renewcommand{\algorithmicrequire}{\textbf{Input:}}
\renewcommand{\algorithmicensure}{\textbf{Output:}}

\Require Training user set $\mathcal{U}$; Item set $\mathcal{V}$; Item set $\mathcal{V}_u$ of user $u$; User embedding $\boldsymbol{e}_u, \forall u \in \mathcal{U}$; Item embedding $\boldsymbol{e}_v, \forall v \in \mathcal{V}$; Filtered and biased modal embedding $\boldsymbol{\bar{e}}_u^m$, $\boldsymbol{\tilde{e}}_u^m, \; \forall u \in \mathcal{U}, \forall m \in \mathcal{M}$; Role indicator embeddings $\boldsymbol{r}_u$ and $\boldsymbol{r}_v$;

\Ensure The all parameters $\Theta$ for relation-aware fairness learning

\renewcommand{\algorithmicensure}{\textbf{Initialize:}}

\Ensure $\Theta$

\State \Comment{Construct fair and unfair modality-based user-user graph structures for each modality\;\;\;\;\;\;\;\;\;\;\;\;\;\;\;\;\;\;\;\;\;\;\;\;\;\;\;\;\;\;\;\;\;\;\;\;\;\;\;\;}

\State $\bar{S}^m, \tilde{S}^m$ for $m \in \mathcal{M}$ $\gets$  Eq.~\eqref{eqn:s_uu}

\State \Comment{Aggregate fair/unfair multimodal structures \;\;\;\;\;\;\;\;\;\;\;\;\;\;\;\;\;\;\;\;\;\;\;\;\;\;\;\;\;\;\;\;\;\;\;\;\;\;\;\;\;\;\;\;\;\;\;\;\;\;\;\;\;\;\;\;\;\;\;\;\;\;\;\;\;\;\;\;\;\;\;\;\;\;\;\;\;\;\;\;\;\;\;\;\;\;\;\;\;\;\;\;\;\;\;\;\;\;\;\;\;\;}

\State $\bar{S}, \tilde{S} \gets$ Eq.~\eqref{eq:agg_graphs}

\Repeat

\For{$u \in \mathcal{U}, v \in \mathcal{V}$}

\State \Comment{Generate filtered and biased neighbor representations \;\;\;\;\;\;\;\;\;\;\;\;\;\;\;\;\;\;\;\;\;\;\;\;\;\;\;\;\;\;\;\;\;\;\;\;\;\;\;\;\;\;\;\;\;\;\;\;\;\;\;\;\;\;\;\;\;\;\;\;\;\;\;\;\;\;\;\;\;\;\;\;\;\;\;}

\State $\boldsymbol{\bar{h}}_u, \boldsymbol{\tilde{h}}_u \gets$ Eq.~\eqref{eq:nb_rep}

\State \Comment{Enhance user representation by the filtered and biased neighbor representations\;\;\;\;\;\;\;\;\;\;\;\;\;\;\;\;\;\;\;\;\;\;\;\;\;\;\;\;\;\;\;\;\;\;\;\;\;}

\State $\boldsymbol{\hat{e}_u} \gets \boldsymbol{e_u} + \lambda_h (\boldsymbol{\bar{h}}_u - \boldsymbol{\tilde{h}}_u)$

\State \Comment{Get filtered user and item representations \;\;\;\;\;\;\;\;\;\;\;\;\;\;\;\;\;\;\;\;\;\;\;\;\;\;\;\;\;\;\;\;\;\;\;\;\;\;\;\;\;\;\;\;\;\;\;\;\;\;\;\;\;\;\;\;\;\;\;\;\;\;\;\;\;\;\;\;\;\;\;\;\;\;\;\;\;\;\;\;\;\;\;\;\;\;\;\;\;\;\;\;\;}

\State $\boldsymbol{\bar{e}}_u \gets \frac{\sum_{k=1}^K {f}_k ([\boldsymbol{r}_u ; \boldsymbol{\hat{e}}_u])}{K}, \boldsymbol{\bar{e}}_v \gets \frac{\sum_{k=1}^K {f}_k ([\boldsymbol{r}_v ;\boldsymbol{e}_v])}{K}$

\State \Comment{Get predicted attribute values given explicit and implicit user representations respectively\;\;\;\;\;\;\;\;\;\;\;\;\;\;\;\;\;\;\;\;\;\;\;}

\State $\boldsymbol{\bar{a}}_{uk} \gets d^u_{k} (\boldsymbol{\bar{e}}_u)$
\State $\boldsymbol{\bar{a}}_{\mathcal{V}_u k} \gets d^v_{k} (\boldsymbol{\bar{e}}_{\mathcal{V}_u}),\; \boldsymbol{\bar{e}}_{\mathcal{V}_u} \gets \frac{\sum _{v \in \mathcal{V}_u} \boldsymbol{\bar{e}_v}}{|\mathcal{V}_u|}$

\State \Comment{Calculate attribute prediction losses given explicit and implicit representations, and recommendation loss}

\State $\mathcal{L}_{D_u}, \mathcal{L}_{D_v} \gets$ Eq.~\eqref{eq:l_d_u} and \eqref{eq:l_d_v}

\State $\mathcal{L}_{\text{Rec}} \gets \operatorname{RS}(\boldsymbol{\bar{e}}_u, \boldsymbol{\bar{e}}_v)$

\State Optimize $\Theta$ according to Eq.~\eqref{eqn:L_r} and \eqref{eqn:L_d}

\EndFor

\Until{\textit{stopping criterion is met};}

\end{algorithmic}
\end{algorithm}

%% file: 5.experiment.tex
In our study, we conducted experiments to evaluate the effectiveness of our proposed method from both accuracy and causal fairness perspectives.
Particularly, we aim to address the following research questions:

\begin{itemize}[]
    \item \textbf{RQ1}: How does FMMRec perform as compared to state-of-the-art (SOTA) multimodal recommendation baselines and fairness baselines from a causal fairness perspective?
    \item \textbf{RQ2}: How does FMMRec perform in modal disentanglement in terms of controlling causal effects?
    \item \textbf{RQ3}: How do fairness-oriented modal disentanglement (FMD), fair/unfair relation mining (FRM/UFRM), and role indicator embedding (RI) contribute to the performance of FMMRec?
    \item {\textbf{RQ4}: How do the hyper-parameters affect the performance of FMMRec?}
    \item \textbf{RQ5}: How does FMMRec perform in group fairness assessment?
    \item \textbf{RQ6}: How does FMMRec perfrom in unimodal settings compared to multimodal settings?
\end{itemize}

\subsection{Experimental Settings}

\begin{table}[htbp]
\centering
\caption{Statistics of the two datasets used in our experiments, wherein V, T, and A denote the dimensions of visual, textual, and audio modalities, respectively.}
\begin{tabular}{@{}cccccccc@{}}
\toprule Dataset&\#Interactions&\#Users&\#Items&Sparsity&V&T&A\\
\midrule
MovieLens&$1,000,209$&6,040&3,706&$95.53\%$&1,000&384&128\\
MicroLens&$123,368$&5,936&12,414&$99.83\%$&1,000&768&128\\
\bottomrule
\end{tabular}
\label{tab:statistics}
\end{table}

\subsubsection{Datasets}
\label{sec:datasets}
We evaluate FMMRec and baselines on two public datasets and regard available demographics of users as sensitive attributes following~\cite{PCFR, FairGo}:
\begin{itemize}[]
    \item \textbf{MovieLens\footnote{https://grouplens.org/datasets/movielens/1m}.}
This is a widely used benchmark dataset with multiple modalities and several user attributes for movie recommendation. 
We consider three modalities including movie posters as the visual modality, movie plots as the textual modality, and the extracted audio tracks of movie trailers from YouTube\footnote{https://www.youtube.com} as the audio modality.
We regard users' \textit{gender} (binary classes), \textit{age} (seven classes), and \textit{occupation} (21 classes) as sensitive attributes.

\item \textbf{MicroLens~\cite{ni2023contentdriven}\footnote{https://recsys.westlake.edu.cn/MicroLens-Fairness-Dataset}.}
This is a multimodal dataset with gender information for micro-video recommendations.
Specifically, we leverage items' textual titles, audio tracks, and five frames extracted from video as multimodal features, and treat \textit{gender} (binary classes) as the sensitive attribute of users.
\end{itemize}

For both datasets, we split the historical interactions into training, validation, and test sets in a ratio of 8:1:1.
We use the pretrained modality encoder to extract the modal representation for each modality.
Specifically, we utilize ResNet50~\cite{he2016deep} for encoding the visual modality, sentence-transformers~\cite{reimers2019sentence} for the textual modality, and VGGish~\cite{hershey2017cnn} for the audio modality.
The statistics of the preprocessed datasets are shown in Table~\ref{tab:statistics}.

\subsubsection{Baselines}


Three state-of-the-art (SOTA) fair representation learning methods are compared, as fairness baselines:
\begin{itemize}[]
    \item \textbf{AL} \cite{wadsworth2018achieving} applies adversarial learning to eliminate sensitive information in user representations via a min-max game.
    \item \textbf{CAL} \cite{bose2019compositional} introduces compositional filters for fair representation learning in multi-attribute scenarios.
    \item \textbf{FairGo} \cite{FairGo} applies compositional filters to both user and item representations, and applies discriminators to explicit user representation and graph-based high-order user representation.
\end{itemize}

In addition, we consider five SOTA multimodal recommendation models as recommendation baselines:

\begin{itemize}[]
    \item \textbf{VBPR} \cite{VBPR} is the first model that incorporates the visual features into recommender systems, treating visual features as another view of item representations;
    \item \textbf{MMGCN} \cite{MMGCN} learns modality-specific representations of users and items based on the message-passing mechanism of graph neural network (GNN) for each modality, enhanced by a user-item bipartite graph;
    \item \textbf{LATTICE} \cite{LATTICE} mines the latent structure for multimodal recommendation to explicitly learn the semantic item-item relationships for each modality, and learn high-order item affinities based on graph convolutional network with the mined modality-based graphs;
    \item \textbf{FREEDOM} \cite{FREEDOM} leverages the modality-based item-item graphs following the same approach as LATTICE~\cite{LATTICE}, but with two notable differences that it freezes the mined graphs during training and incorporates the degree-sensitive edge pruning techniques to effectively reduce noise in the user-item graph;
    \item \textbf{DRAGON} \cite{DRAGON} improves dyadic relations in multimodal recommendations by constructing homogeneous graphs and learning dual representations for both users and items.
\end{itemize}

As our FMMRec and fairness baselines are model-agnostic, we selected the model that demonstrated the highest accuracy on each dataset as the foundational recommender.
Specifically, we utilized the LATTICE model for the MovieLens dataset and the DRAGON model for the MicroLens dataset. 
For a fair comparison, we employed the same model as the foundational recommender, and the same structure and capacity for each filter or discriminator, for our method and the fairness-aware baselines mentioned above.
We implemented our fairness method FMMRec based on the publicly available multimodal recommendation framework\footnote{https://github.com/enoche/MMRec}~\cite{zhou2023comprehensive} for high-quality reproducibility.

\subsubsection{Evaluation Protocols}
\label{sec:evaluation}
For measuring recommendation accuracy, we adopt two widely recognized metrics \textit{Recall} and \textit{NDCG (Normalized Discounted Cumulative Gain)} on Top-20 recommendations.
We choose Top-20 metrics because they offer a balanced evaluation, providing sufficient differentiation in performance without being overly coarse or overly fine-grained.

For fairness evaluation, we focused on measuring the residual causal effects of sensitive attributes.
Following the common setting of fair representation learning~\cite{PCFR, TFR, FairGo, FairRec}, we train a surrogate classifier to classify each attribute using the learned representations of the user on attacker train set (80\%).
Subsequently, on attacker test set (20\%), we examine users' (filtered) explicit representation $\boldsymbol{\bar{e}}_u$ and implicit representation $\boldsymbol{\bar{e}}_{\mathcal{V}_u}$ from Equation~\eqref{eqn:e_u_e_v_u} respectively, and report \textit{AUC} for binary attribute(s) and micro-averaged \textit{F1} for multi-class attribute(s) as fairness metrics~\cite{FairGo}.
Lower AUC and F1 scores indicate that the representations contain less information about sensitive attributes, aligning with the goal of reducing causal influence~\cite{PCFR}.

\subsection{Implementation Details}
\label{appd:implementation}

\subsubsection{Common Settings}
We set the hidden size $d$ and $d_r$ to 64, the batch size to 2048, and the learning rate to 0.001 for all methods.
The embedding size $d_m$ of each modality is shown in Table~\ref{tab:statistics}.
We initialize model parameters with the Xavier method~\cite{glorot2010understanding} and adopt Adam~\cite{DBLP:journals/corr/KingmaB14} as the optimizer.

Early stopping is adopted to choose the best models based on the validation performance.
Each filter is implemented by a two-layer neural network with LeakyReLU activation.
Each classifier (discriminator or attacker) is implemented as a multi-layer perceptron with two layers. 
The activation function used is LeakyReLU, and a dropout rate of 0.2 is applied.

\subsubsection{Specific Settings}
\begin{itemize}
    \item 
    \textbf{For FMMRec.}
    The hyperparameters $\lambda_{m_0}$ and $\lambda_{m_1}$ in disentanglement learning are both set to 0.1.
    The weights of modalities $\alpha_m$ for $m \in \mathcal{M} = \{v, t, a\}$ are set to 0.6, 0.2, and 0.2 on the MovieLens dataset respectively, and set to 0.2, 0.6, and 0.2 on the MicroLens dataset respectively, according to the disentanglement performance of each modality.
    The number of neighbors $\mathbb{k}$ is searched within the range from 1 to 10.
    The hyperparameters $\lambda_h$ and $\lambda_{D_u}$ in relation-aware fairness learning are both set to 0.1, and $\lambda_{D_v}$ is 0.05.
    We update Equation~\eqref{eqn:L_d} for ten steps after each update for Equation~\eqref{eqn:L_r} to allow discriminators to reach their optimal values, following~\cite{PCFR, TFR}.
    \item \textbf{For the baselines.}
    For the hyperparameters specific to baselines, we use grid search over the ranges provided in the original paper or tune them on the validation data of each dataset, to find the optimal combination on our datasets.
    For example, for FairGo, we considered the first-order and second-order neighborhood of users for the ego-centric network, referring to the original paper~\cite{FairGo}.
    We ensure that each method has the same capacity in terms of neural network layers and hidden sizes for a fair comparison.
\end{itemize}

%% file: 6.result.tex

\begin{table*}[htbp]
\centering
\caption{Experimental results of baselines for accuracy and fairness (\textit{w.r.t.} gender, age, and occupation) performance on the MovieLens dataset. \textbf{Bold} and \underline{underline} for best and second-best results of fairness methods, respectively.}
\begin{tabular}{@{}lcccccccc@{}}
\toprule
\multirow{2}{*}{\textbf{Methods}} & \multicolumn{2}{c}{\textbf{Accuracy}} & \multicolumn{2}{c}{\textbf{Fairness-Gen.}} & \multicolumn{2}{c}{\textbf{Fairness-Age}} & \multicolumn{2}{c}{\textbf{Fairness-Occ.}} \\ \cmidrule(l){2-9} 
 & Recall $\uparrow$ & NDCG $\uparrow$ & AUC-E $\downarrow$ & AUC-I $\downarrow$ & F1-E $\downarrow$ & F1-I $\downarrow$ & F1-E $\downarrow$ & F1-I $\downarrow$ \\ \midrule
VBPR & 0.2136 & 0.2033 & 0.7338 & 0.6472 & 0.4983 & 0.3882 & 0.2144 & 0.1722 \\
MMGCN & 0.2180 & 0.2110 & 0.7314 & 0.6279 & 0.4925 & 0.3858 & 0.2243 & 0.1689 \\
LATTICE & 0.2476 & 0.2378 & 0.7397 & 0.5428 & 0.5025 & 0.3725 & 0.2202 & 0.1747 \\
FREEDOM & 0.2423 & 0.2357 & 0.7158 & 0.6306 & 0.4826 & 0.3990 & 0.2012 & 0.1755 \\
DRAGON & 0.2387 & 0.2331 & 0.7117 & 0.5803 & 0.4884 & 0.3725 & 0.2169 & 0.1614 \\ \cmidrule(l){1-9}
AL & \underline{0.2163} & \underline{0.2066} & \textbf{0.5172} & 0.5750 & \textbf{0.3560} & 0.3891 & 0.1656 & 0.1780 \\
CAL & 0.2143 & 0.2035 & 0.5340 & 0.5594 & 0.3609 & 0.3717 & 0.1672 & 0.1780 \\ 
FairGo & 0.2133 & 0.2003 & 0.5431 & \textbf{0.5000} & 0.3659 & \underline{0.3535} & \underline{0.1639} & \underline{0.1647} \\ \cmidrule(l){1-9}
\textbf{FMMRec} & \textbf{0.2214} & \textbf{0.2079} & \underline{0.5224} & \textbf{0.5000} & \underline{0.3576} & \textbf{0.3526} & \textbf{0.1573} & \textbf{0.1507} \\ \bottomrule
\end{tabular}
\label{tab:main_movielens}
\end{table*}

\begin{table*}[htbp]
\centering
\caption{Experimental results of baselines for accuracy and fairness (\textit{w.r.t.} gender) performance on the MicroLens dataset. \textbf{Bold} and \underline{underline} for best and second-best results of fairness methods, respectively.}
\begin{tabular}{@{}lcccc@{}}
\toprule
\multirow{2}{*}{\textbf{Methods}} & \multicolumn{2}{c}{\textbf{Accuracy}} & \multicolumn{2}{c}{\textbf{Fairness-Gen.}} \\ \cmidrule(l){2-5} 
 & Recall $\uparrow$ & NDCG $\uparrow$ & AUC-E $\downarrow$ & AUC-I $\downarrow$ \\ \midrule
VBPR & 0.0652 & 0.0331 & 0.6888 & 0.7533 \\
MMGCN & 0.0465 & 0.0231 & 0.7692 & 0.7813 \\
LATTICE & 0.0745 & 0.0382 & 0.7773 & 0.7781 \\
FREEDOM & 0.0648 & 0.0339 & 0.7644 & 0.7691 \\
DRAGON & 0.0860 & 0.0432 & 0.7582 & 0.7728 \\ \cmidrule(l){1-5}
AL & \underline{0.0734} & 0.0368 & 0.6387 & 0.7748 \\
CAL & 0.0733 & \underline{0.0375} & 0.6266 & 0.7786 \\ 
FairGo & 0.0720 & 0.0356 & \underline{0.5932} & \underline{0.5329} \\ \cmidrule(l){1-5}
\textbf{FMMRec} & \textbf{0.0746} & \textbf{0.0379} & \textbf{0.5599} & \textbf{0.5050} \\ \bottomrule
\end{tabular}
\label{tab:main_microlens}
\end{table*}

\subsection{Overall Performance (RQ1)}
The results in Tables~\ref{tab:main_movielens}-\ref{tab:main_microlens} yield the following observations:
\begin{itemize}[]
    \item Our FMMRec outperforms all baselines in terms of fairness performance (\textit{i.e.}, lowest AUC/F1 values overall), indicating that it is vital to consider the sensitive information in multimodal representations for improving fairness in multimodal recommendations.
    In particular, our FMMRec eliminates the most sensitive information on implicit representations, \textit{i.e.}, achieving the lowest values of unfairness metrics ending with `-I'.
    This demonstrates that leveraging the unfair and fair user-user relations from disentangled modal representations to fair user representation learning is effective in eliminating sensitive information learned in multimodal recommender systems.
    These results suggest that FMMRec effectively reduces the causal influence of sensitive attributes on recommendations, achieving better performance on counterfactual fairness~\cite{DBLP:conf/nips/KusnerLRS17, PCFR}.

    \item Compared with AL and CAL which do not contribute to fairness improvement on implicit representations, FMMRec and FairGo largely improve fairness performance on implicit representations.
    The main reason is that, among all fairness methods, only FairGo and our FMMRec apply filters to implicit user representations.
    Considering the fairness requirement that no sensitive information is learned by the recommender system, it is nontrivial to eliminate sensitive information on implicit user representations.
    However, FMMRec outperforms FairGo significantly. 
    While FairGo does not distinguish the role of filters’ input and neglects the modal impact on sensitive information learning, our FMMRec incorporates causal disentanglement and relation-aware fairness learning, leading to better elimination of sensitive information.
    \item Compared with the base multimodal recommender baselines (\textit{i.e.}, LATTICE on the MovieLens dataset and DRAGON on the MicroLens dataset), applying all fairness-aware methods leads to an accuracy drop.
    As mentioned, achieving counterfactual fairness requires the independence between sensitive information and recommendation outcomes, and accuracy normally drops depending on how much information is lost.
    Thus, it is generally acceptable that accuracy slightly decreases when improving fairness performance significantly in recommendations.
    Notably, FMMRec not only delivers superior fairness performance but also outperforms other fairness-aware methods in terms of accuracy, effectively balancing accuracy and fairness.
\item The multimodal recommendation baselines always show a significant level of sensitive information leakage (\textit{i.e.}, high AUC/F1 values) on both explicit and implicit representations of users.
Compared with conventional recommendation models that learn sensitive information merely from user-item historical interactions~\cite{PCFR}, the multimodal recommendation models may inherit additional sensitive information from the multimodal representations.
Therefore, it is nontrivial to eliminate the potentially sensitive information in multimodal recommendations.
Our method addresses this by explicitly controlling the causal effects of sensitive attributes, leading to fairer recommendations.
\end{itemize}

{We conducted paired t-tests comparing FMMRec with two baselines to evaluate the statistical significance of our experimental results.
Specifically, we repeated the experiments four times to ensure reliability, and performed paired t-tests between FMMRec and the state-of-the-art multimodal recommender model LATTICE, as well as between FMMRec and the fairness-aware method FairGo.
The resulting \textit{p}-values are 6.28e-10 between FMMRec and LATTICE and 3.71e-3 between FMMRec and FairGo, both below the significance threshold of 0.05, indicating that FMMRec achieves statistically significant performance improvements over these baselines.
}

\subsection{Disentanglement Performance (RQ2)}
In fact, the effectiveness of our FMMRec highly relies on the disentanglement performance as we aim to leverage the biased and filtered neighbors based on each modality for users to enhance user representation's fairness.
By disentangling sensitive and non-sensitive information, we control the causal pathways from sensitive attributes to recommendations.

To this end, compared with the original embedding $\boldsymbol{{e}}^m_v$, we expect more sensitive information to be learned in the biased modal embedding $\boldsymbol{\tilde{e}}^m_v$ while less sensitive information to be leaked in the filtered modal embedding $\boldsymbol{\bar{e}}^m_v$.
This aligns with the goal of isolating the causal effects of sensitive attributes.

\begin{figure}[htbp]
\setlength{\abovecaptionskip}{3pt plus 2pt minus 2pt}
  \centering
  \includegraphics[width=0.76\textwidth]{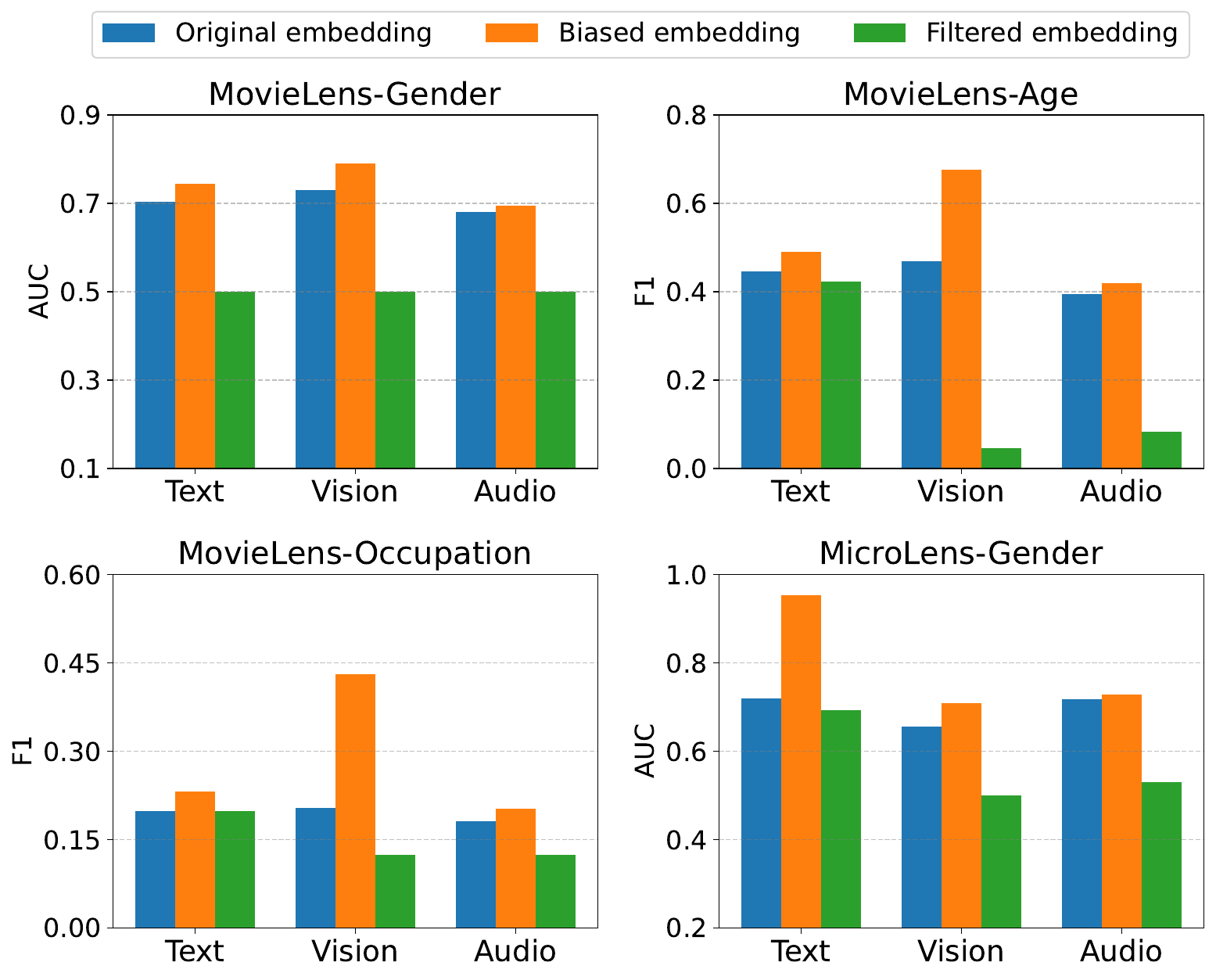}
  \caption{Disentanglement performance (\textit{a.k.a.} sensitive attribute prediction accuracy) of each modality on all the four dataset-attribute pairs (\textit{e.g.}, MovieLens-Gender denotes the protected attribute gender of the MovieLens dataset).
  }
  \label{fig:disentanglement}
\end{figure}

We report the sensitive attribute classification accuracy AUC/F1 given the input of the original embedding $\boldsymbol{{e}}^m_v$, the biased modal embedding $\boldsymbol{\tilde{e}}^m_v$ and the filtered modal embedding $\boldsymbol{\bar{e}}^m_v$ in Figure~\ref{fig:disentanglement}, to verify the disentanglement performance.

The trend observed across all four dataset-attribute pairs indicates that the degree of sensitive information leakage in biased embeddings is higher than that in the original embeddings, which, in turn, exceeds the leakage in filtered embeddings. 
This finding aligns with our expectations and validates the effectiveness of our modal disentanglement learning in controlling causal influences.

\begin{table*}[htbp]
\centering
\caption{Performance of FMMRec with different variants on MovieLens dataset.}
\begin{tabular}{@{}lcccccccc@{}}
\toprule
\multirow{3}{*}{\textbf{Variants}} & \multicolumn{8}{c}{\textbf{MovieLens}} \\ \cmidrule(l){2-9} 
 & \multicolumn{2}{c}{\textbf{Accuracy}} & \multicolumn{2}{c}{\textbf{Fairness-Gen.}} & \multicolumn{2}{c}{\textbf{Fairness-Age}} & \multicolumn{2}{c}{\textbf{Fairness-Occ.}} \\ \cmidrule(l){2-9} 
 & Recall $\uparrow$ & NDCG $\uparrow$ & AUC-E $\downarrow$ & AUC-I $\downarrow$ & F1-E $\downarrow$ & F1-I $\downarrow$ & F1-E $\downarrow$ & F1-I $\downarrow$ \\ \midrule

{w/o FMD}     & {0.2048} & {0.1952} & {0.5902} & {0.5037} & {0.3709} & {\textbf{0.3526}} & {0.1755} & {\underline{0.1531}} \\
{w/o FRM}     & \underline{0.2194} & \textbf{0.2082} & 0.5587 & \textbf{0.5000} & 0.3584 & \textbf{0.3526} & 0.1623 & 0.1623 \\
{w/o UFRM}    & 0.2162 & 0.2050 & 0.5321 & \textbf{0.5000} & 0.3609 & \textbf{0.3526} & \underline{0.1598} & 0.1689 \\
{w/o RI}      & 0.2048 & 0.1923 & \underline{0.5225} & 0.5003 & \textbf{0.3568} & 0.3535 & 0.1623 & 0.1623 \\
\textbf{FMMRec} & \textbf{0.2214} & \underline{0.2079} & \textbf{0.5224} & \textbf{0.5000} & \underline{0.3576} & \textbf{0.3526} & \textbf{0.1573} & \textbf{0.1507} \\

\bottomrule
\end{tabular}%
\label{tab:ablation_movielens}
\end{table*}

\begin{table}[htbp]
\centering
\caption{Performance of FMMRec with different variants on MicroLens dataset.}
\begin{tabular}{@{}lcccc@{}}
\toprule
\multirow{2}{*}{\textbf{Variants}} & \multicolumn{2}{c}{\textbf{Accuracy}} & \multicolumn{2}{c}{\textbf{Fairness-Gen.}} \\ \cmidrule(l){2-5} 
 & Recall $\uparrow$ & NDCG $\uparrow$ & AUC-E $\downarrow$ & AUC-I $\downarrow$ \\ \midrule

{w/o FMD}     & {0.0660} & {0.0331} & {0.7588} & {0.7543} \\
{w/o FRM}     & \textbf{0.0765} & \textbf{0.0393} & \underline{0.5678} & \underline{0.5553} \\
{w/o UFRM}    & \underline{0.0755} & \underline{0.0381} & 0.7044 & 0.6995 \\
{w/o RI}      & 0.0697 & 0.0347 & 0.6882 & 0.6601 \\
\textbf{FMMRec} & 0.0746 & 0.0379 & \textbf{0.5599} & \textbf{0.5050} \\

\bottomrule
\end{tabular}%
\label{tab:ablation_microlens}
\end{table}

\subsection{Ablation Study (RQ3)}
\label{appd:ablation}

To figure out the contributions of different components of our FMMRec, we consider three variants of FMMRec for ablation study:
\begin{itemize}[]
    \item {\textbf{w/o FMD}: We remove fairness-oriented modal disentanglement of FMMRec.}
    \item \textbf{w/o FRM}: We remove fair relation mining of FMMRec.
    \item \textbf{w/o UFRM}: We remove unfair relation mining of FMMRec.
    \item \textbf{w/o RI}: We remove role indicator embedding of FMMRec.
\end{itemize}
We report the accuracy and fairness results in Tables~\ref{tab:ablation_movielens}-\ref{tab:ablation_microlens}.
The following observations are made:
\begin{itemize}
    \item All components positively contribute to fairness performance. As sensitive attribute prediction accuracy AUC/F1 given user or item representation is increased for each variant, it indicates that removing any component reduces the method's ability to eliminate sensitive information effectively.
    \item {Comparing with the variant \textbf{`w/o FMD'}, FMMRec achieves both higher accuracy and significantly better fairness performance on both datasets. This highlights the critical role of fairness-oriented modal disentanglement in controlling sensitive information leakage from multimodal content. Without the fairness-oriented modal disentanglement (\textit{i.e.}, in the w/o FMD variant), the model lacks the mechanism to separate modal embeddings into biased and filtered components. This leads to inadequate control over sensitive information embedded in the multimodal representations, allowing sensitive attributes to influence the user representations. Consequently, the sensitive attribute prediction accuracy increases (indicating worse fairness), and the recommendation accuracy decreases due to the entanglement of sensitive and non-sensitive information. This demonstrates that modal disentanglement effectively isolates sensitive information, enhancing both fairness and accuracy in our method.}

    {On the MovieLens dataset, for instance, FMMRec improves the Recall from 0.2048 (in w/o FMD) to 0.2214, and reduces the AUC-E from 0.5902 to 0.5224. Similarly, on the MicroLens dataset, FMMRec raises the Recall from 0.0660 to 0.0746, and decreases the AUC-E from 0.7588 to 0.5599. These significant improvements in both accuracy and fairness metrics confirm the effectiveness of the fairness-oriented modal disentanglement component in our framework.}
    
    \item Compared with the variant \textbf{`w/o FRM'}, FMMRec achieves better fairness performance, emphasizing the importance of fair relation mining in enhancing user representation expressiveness without introducing sensitive information.
    
    {Regarding the accuracy performance, FMMRec also achieves higher Recall and comparable NDCG on the MovieLens dataset but exhibits suboptimal results on the MicroLens dataset. 
    The decreased performance on the MicroLens dataset may be attributed to a stronger dependence on users' sensitive attributes. For example, the sensitive attribute prediction accuracy on the MicroLens dataset is higher than that on the MovieLens dataset, as shown in Figure~\ref{fig:sens_leak_exp}. 
    Consequently, the application of fair relation mining may lead to a greater reduction in accuracy on the MicroLens dataset compared to the MovieLens dataset due to the stronger reliance of recommendation accuracy on sensitive attributes.
    }
    \item Similarly, the variant \textbf{‘w/o UFRM’} shows worse fairness performance compared to FMMRec, highlighting the role of unfair relation mining in identifying and mitigating the influence of sensitive attributes.
    {We also observe similar trends in accuracy with the variant ‘w/o FRM’, which can be attributed to the stronger reliance of the recommendation accuracy on sensitive attribute information.}
    \item The variant \textbf{‘w/o RI’} exhibits both lower accuracy and fairness, confirming that role indicator embeddings are crucial for the filters to distinguish between user and item inputs, thus avoiding learning confusion and maintaining performance.
\end{itemize}

These results validate the effectiveness of each component in our proposed method, aligning with our aim to control causal effects for achieving counterfactual fairness.


\subsection{Sensitivity Analyses (RQ4)}
{We conducted sensitive analyses on four hyperparamters, $\mathbb{k}$  (the number of neighbors), $\lambda_h$ (the strength of fair and unfair relation mining), $\lambda_{D_u}$ (the strength of the discrimination loss on explicit user representation), and $\lambda_{D_v}$ (the strength of the discrimination loss on implicit user representation)}.

\begin{itemize}
    \item \textbf{Sensitivity analysis on $\mathbb{k}$. }
We adjust the number of neighbors $\mathbb{k} \in \{1, 2, \ldots, 10\}$ and report the {accuracy and} fairness performance on the MovieLens dataset in Figure~\ref{fig:sensitivity_k}.
This analysis examines how the neighborhood size in relation-aware fairness learning affects the ability to control causal influences.

From the results, the sensitive attribute prediction accuracy AUC/F1 is stably low \textit{w.r.t.} both explicit and implicit user representations, indicating that our FMMRec is relatively insensitive to the chosen number $\mathbb{k}$ for eliminating sensitive information.
This robustness suggests that the causal adjustment performed through relation-aware fairness learning is effective across different neighborhood sizes.

While we can observe slight fluctuations in the AUC/F1 values \textit{w.r.t.} explicit representation on three attributes over different values of $\mathbb{k}$, they are still in a relatively low range. 
One of the possible reasons for the slight fluctuations may be the varying influence of neighboring data points on the training of filters, as the neighbor number $\mathbb{k}$ changes.
Overall, our approach demonstrates consistency in achieving counterfactual fairness for multimodal recommendations.



    \begin{figure}[htbp]
    \setlength{\abovecaptionskip}{3pt plus 2pt minus 2pt}
      \centering
      \includegraphics[width=0.95\textwidth]{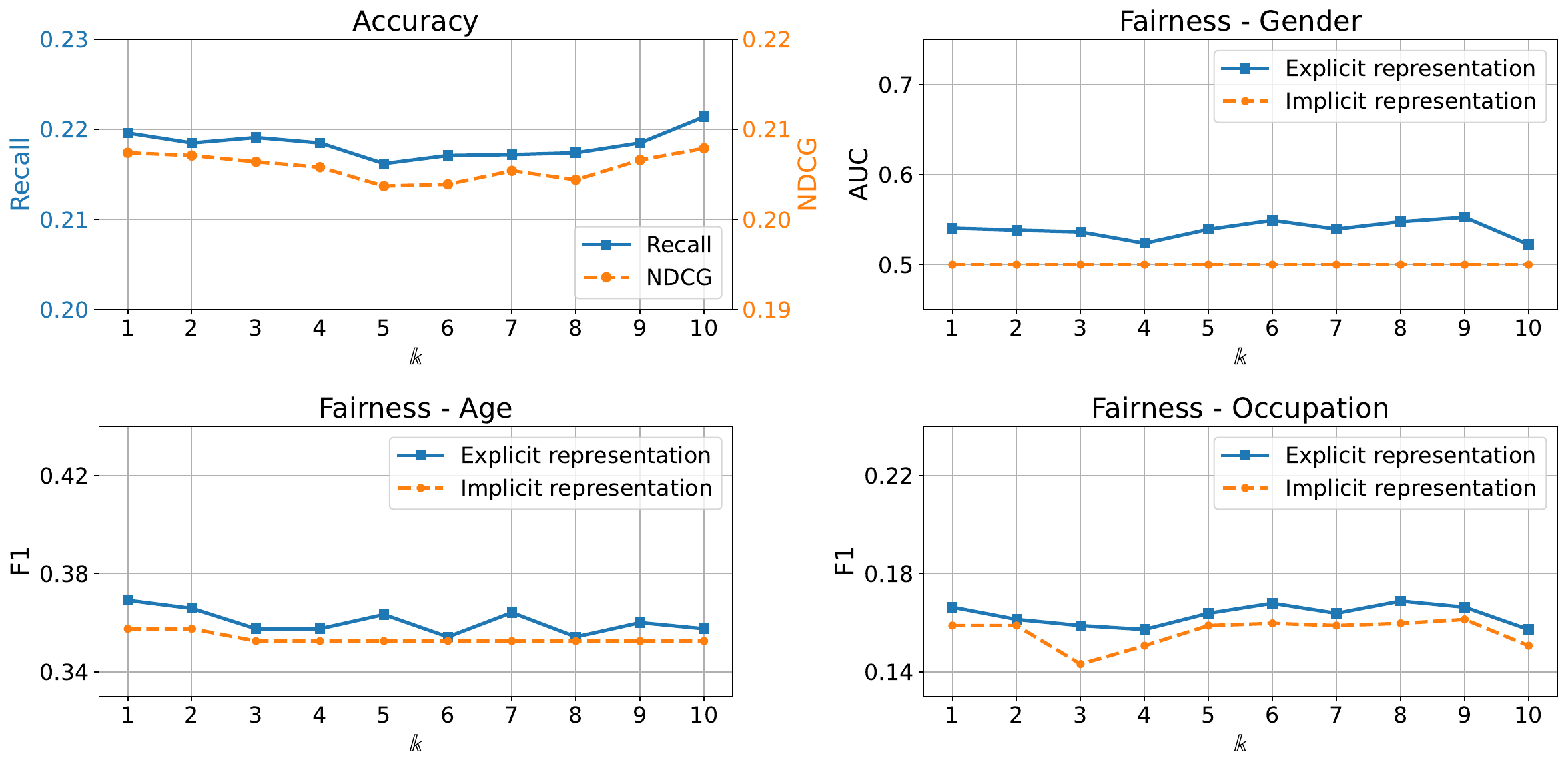}
      \caption{{Recommendation accuracy and sensitive attribute prediction accuracy (\textit{i.e.}, causal fairness) on explicit and implicit user representation under different values of $\mathbb{k}$ (the number of modality-based neighbors) on the MovieLens dataset.}
      }
      \label{fig:sensitivity_k}
    \end{figure}

    \item \textbf{{Sensitivity analysis on $\lambda_h$.}} 
    {We vary \(\lambda_h\), which controls the strength of fair and unfair relation mining, across values \(\{0.01, 0.05, 0.1, 0.2, 0.3, 0.5, 1.0\}\) and report the accuracy and fairness performance on the MovieLens dataset in Figure~\ref{fig:sensitivity_lambda_h}. From the results, we observe that as \(\lambda_h\) increases from 0.01 to 0.1, the recommendation accuracy metrics (Recall@20 and NDCG@20) slightly improve, reaching peak performance at \(\lambda_h = 0.1\). This suggests that incorporating fair and unfair relation mining with an appropriate strength enhances the expressiveness of user representations, improving accuracy. However, as \(\lambda_h\) increases beyond 0.1, the accuracy metrics start to decline, indicating that overly emphasizing relation mining may introduce noise or overfitting, adversely affecting recommendation performance.}
    
    {Regarding fairness metrics, the AUC scores for sensitive attribute prediction on explicit user representations remain relatively stable across different \(\lambda_h\) values, with slight fluctuations. The implicit representations maintain low AUC scores close to 0.5, demonstrating effective elimination of sensitive information regardless of the \(\lambda_h\) value. These results imply that our method is robust to the choice of \(\lambda_h\) within a reasonable range, and \(\lambda_h = 0.1\) offers a good trade-off between accuracy and fairness in our experiments.}
    \begin{figure}[htbp]
    \setlength{\abovecaptionskip}{3pt plus 2pt minus 2pt}
      \centering
      \includegraphics[width=0.95\textwidth]{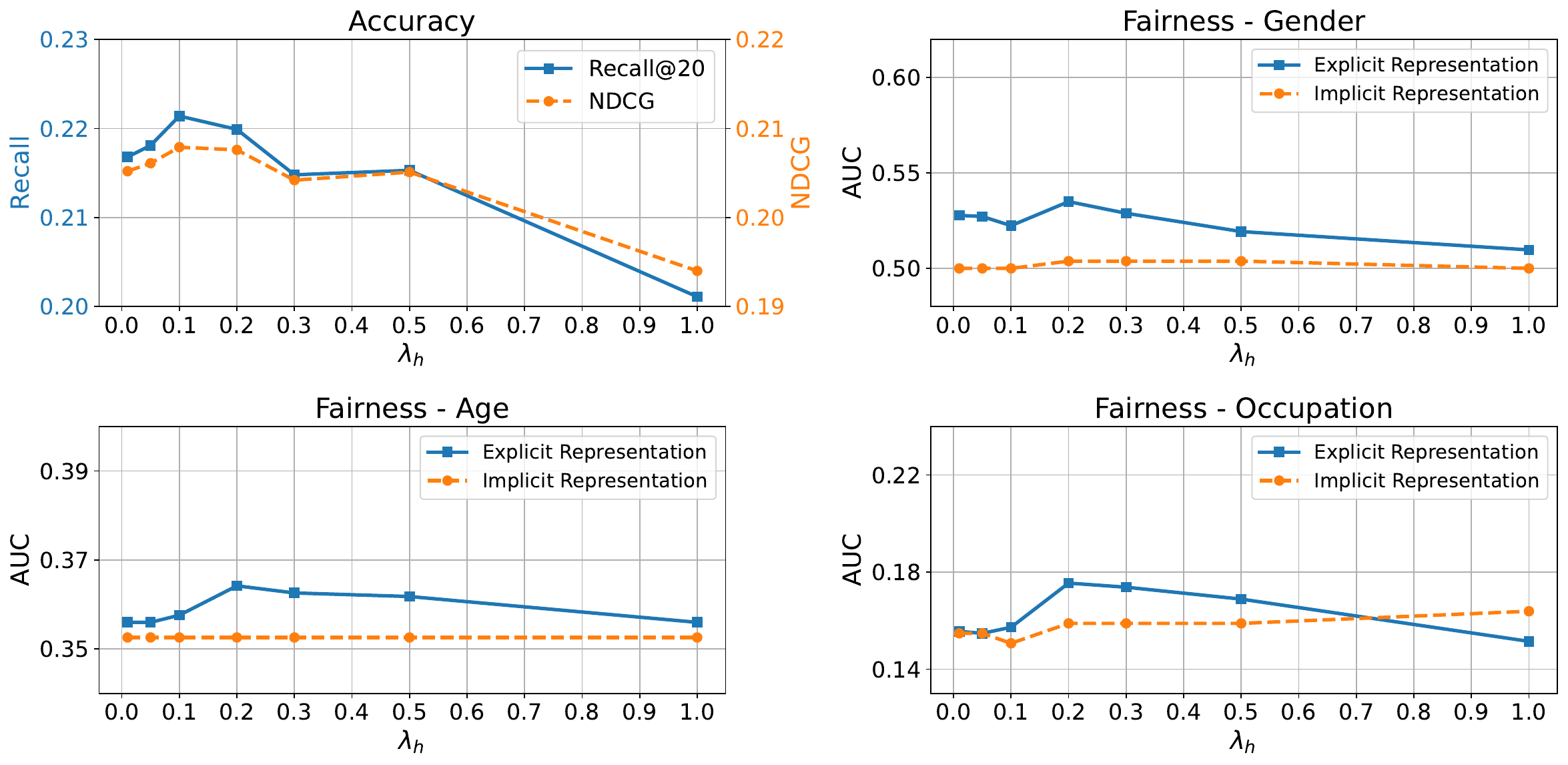}
      \caption{{Recommendation accuracy and sensitive attribute prediction accuracy (\textit{i.e.}, causal fairness) on explicit and implicit user representation under different values of $\lambda_h$ (the strength of fair and unfair relation mining) on the MovieLens dataset.}
      }
      \label{fig:sensitivity_lambda_h}
    \end{figure}

    \item \textbf{{Sensitivity analysis on $\lambda_{D_u}$.}}
     {We adjust \(\lambda_{D_u}\), the weight of the discrimination loss on explicit user representations, over the set \(\{0.01, 0.05, 0.1, 0.2, 0.3, 0.5, 1.0\}\), and present the results in Figure~\ref{fig:sensitivity_lambda_du}. As \(\lambda_{D_u}\) increases, we observe a decline in recommendation accuracy metrics. Specifically, Recall@20 decreases from 0.2379 to 0.1897, and NDCG@20 decreases from 0.2252 to 0.1782 when \(\lambda_{D_u}\) increases from 0.01 to 1.0. This trend indicates that stronger adversarial training on explicit user representations leads to greater removal of information, which can negatively impact the model's ability to capture user preferences accurately. On the other hand, the fairness metrics improve with increasing \(\lambda_{D_u}\). The AUC scores for sensitive attribute prediction on explicit user representations decrease significantly, indicating enhanced fairness. For example, the AUC for gender decreases from 0.7314 to 0.5011 as \(\lambda_{D_u}\) increases.} 
     
     {This trade-off suggests that while higher \(\lambda_{D_u}\) values improve fairness by effectively eliminating sensitive information from explicit user representations, they can also reduce recommendation accuracy due to loss of useful information. Therefore, selecting an optimal \(\lambda_{D_u}\) requires balancing fairness and accuracy based on specific application requirements.}

    \begin{figure}[htbp]
    \setlength{\abovecaptionskip}{3pt plus 2pt minus 2pt}
      \centering
      \includegraphics[width=0.95\textwidth]{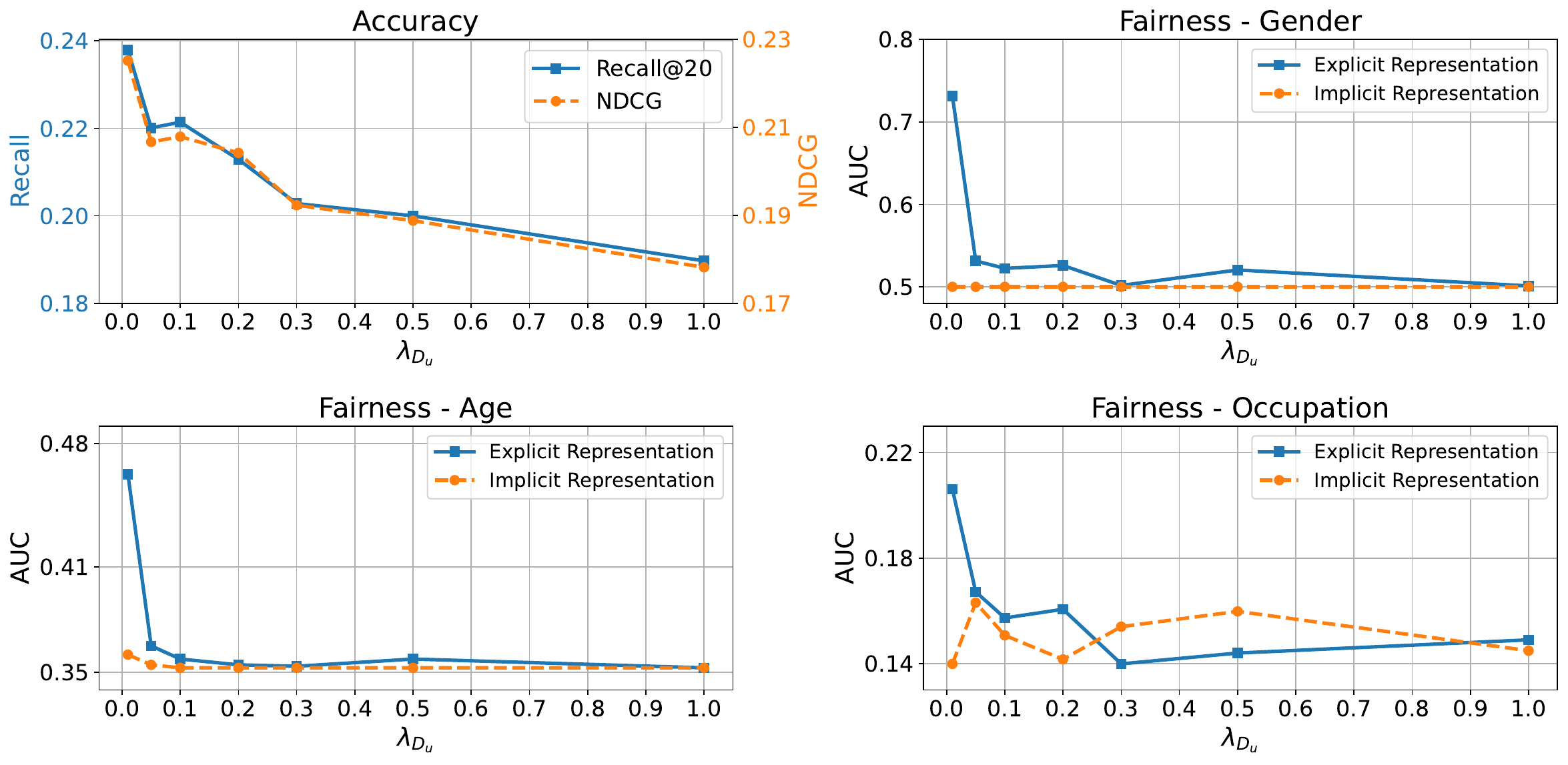}
      \caption{{Recommendation accuracy and sensitive attribute prediction accuracy (\textit{i.e.}, causal fairness) on explicit and implicit user representation under different values of $\lambda_{D_u}$ (the strength of fair and unfair relation mining) on the MovieLens dataset.}
      }
      \label{fig:sensitivity_lambda_du}
    \end{figure}

    \item \textbf{{Sensitivity analysis on $\lambda_{D_v}$.}}
    {We vary \(\lambda_{D_v}\), the weight of the discrimination loss on implicit user representations, over the set \(\{0.01, 0.05, 0.1, 0.2, 0.3, 0.5, 1.0\}\), and display the results in Figure~\ref{fig:sensitivity_lambda_dv}. The recommendation accuracy metrics (Recall@20 and NDCG@20) remain relatively stable across different values of \(\lambda_{D_v}\), with only minor fluctuations. This indicates that the adversarial training on implicit user representations has minimal impact on the model's ability to capture user preferences. In terms of fairness metrics, as \(\lambda_{D_v}\) increases, the AUC scores for sensitive attribute prediction on implicit user representations decrease notably, especially for attributes like age and occupation. For instance, the AUC for age prediction on implicit representations drops from 0.3543 to 0.0455 when \(\lambda_{D_v}\) increases from 0.05 to 1.0. This significant reduction demonstrates the effectiveness of increasing \(\lambda_{D_v}\) in eliminating sensitive information from implicit representations.}
    
    {These findings suggest that higher values of \(\lambda_{D_v}\) enhance fairness without compromising recommendation accuracy, making it beneficial to set \(\lambda_{D_v}\) to relatively large values to achieve better fairness in practice.}

    \begin{figure}[htbp]
    \setlength{\abovecaptionskip}{3pt plus 2pt minus 2pt}
      \centering
      \includegraphics[width=0.95\textwidth]{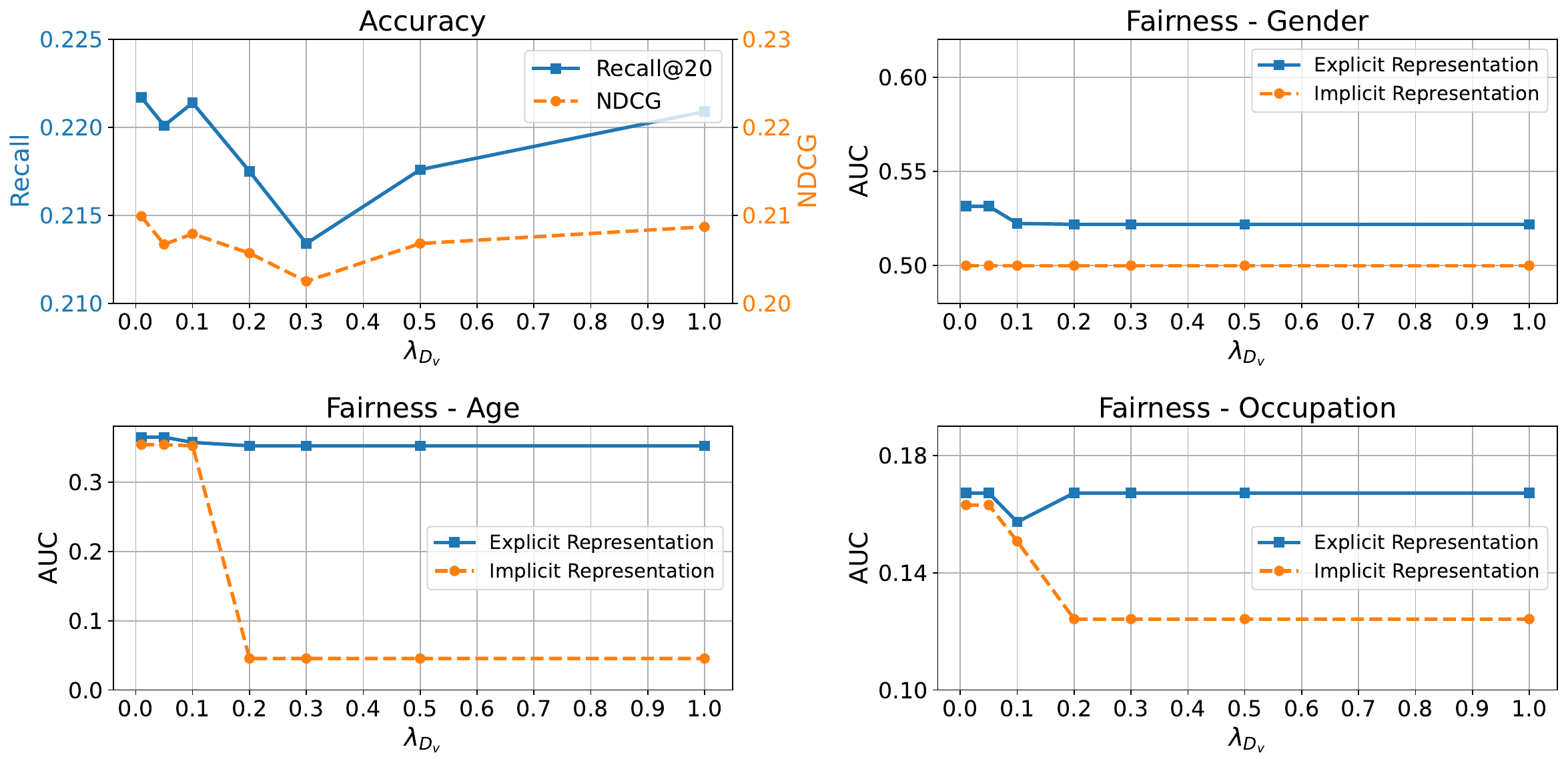}
      \caption{{Recommendation accuracy and sensitive attribute prediction accuracy (\textit{i.e.}, causal fairness) on explicit and implicit user representation under different values of $\lambda_{D_v}$ (the strength of fair and unfair relation mining) on the MovieLens dataset.}
      }
      \label{fig:sensitivity_lambda_dv}
    \end{figure}
\end{itemize}

\subsection{Group Fairness Performance (RQ5)}
While our primary focus is on achieving counterfactual fairness at the individual level, we also assess the group fairness performance of our proposed FMMRec in comparison to other state-of-the-art fairness approaches.
Group fairness ensures that different demographic groups receive equitable treatment in the distributions or quality of recommendations.

To measure group fairness, we adopt user-oriented group fairness (UGF)~\cite{UGF}, which reflects equal opportunity in recommender systems from a user's perspective.
Lower UGF values indicate better group fairness performance, and the specific formulation of UGF can be found in~\cite{UGF}.

\begin{figure}[htbp]
  \centering

   \captionsetup[subfigure]{justification=raggedleft}

    \hspace{\fill}
  \begin{subfigure}[b]{0.40\textwidth}
    \includegraphics[width=1\textwidth]{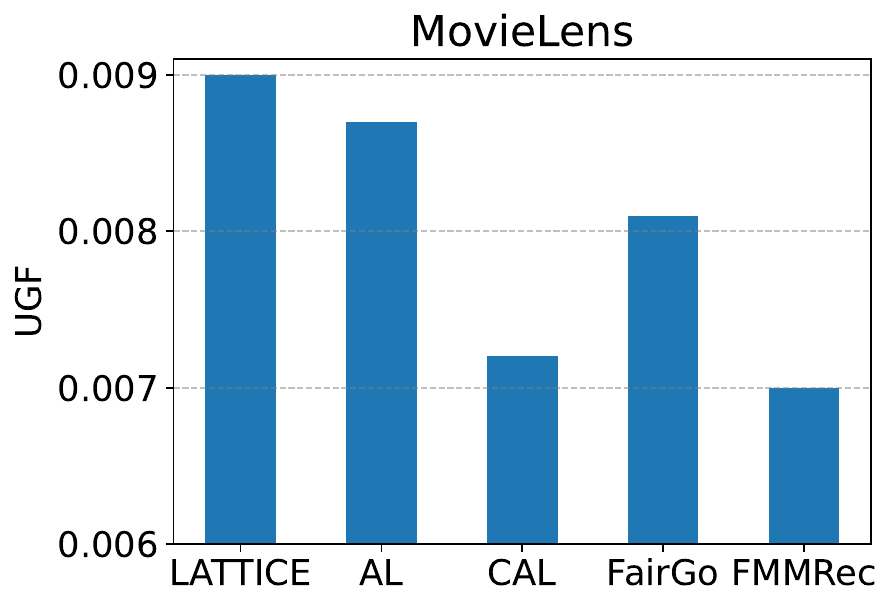}
    \label{fig:subfig1}
  \end{subfigure}%
  \hspace{\fill}
  \begin{subfigure}[b]{0.40\textwidth}
    \includegraphics[width=1\textwidth]{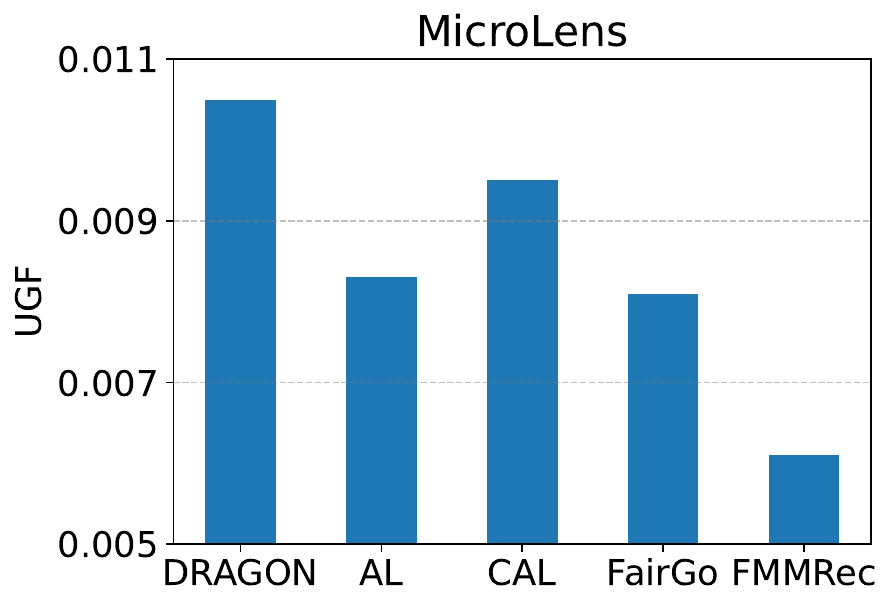}
    \label{fig:subfig2}
  \end{subfigure}%
  \hspace{\fill}
  \caption{Group unfairness performance on users' gender.}
  \label{fig:group_unfairness}
\end{figure}

The results of our group fairness assessment are presented in Figure~\ref{fig:group_unfairness}.
Notably, our FMMRec method outperforms others in group fairness performance on both the MovieLens and MicroLens datasets.
This indicates that our method not only controls individual causal effects but also contributes to reducing disparities among demographic groups in recommendation outcomes.

\subsection{Performance Comparison with Unimodality (RQ6)}
\begin{figure}[htbp]
  \centering

   \captionsetup[subfigure]{justification=raggedleft}
    \hfill
  \begin{subfigure}[b]{0.4518\textwidth}
    \centering
    \includegraphics[width=1\textwidth]{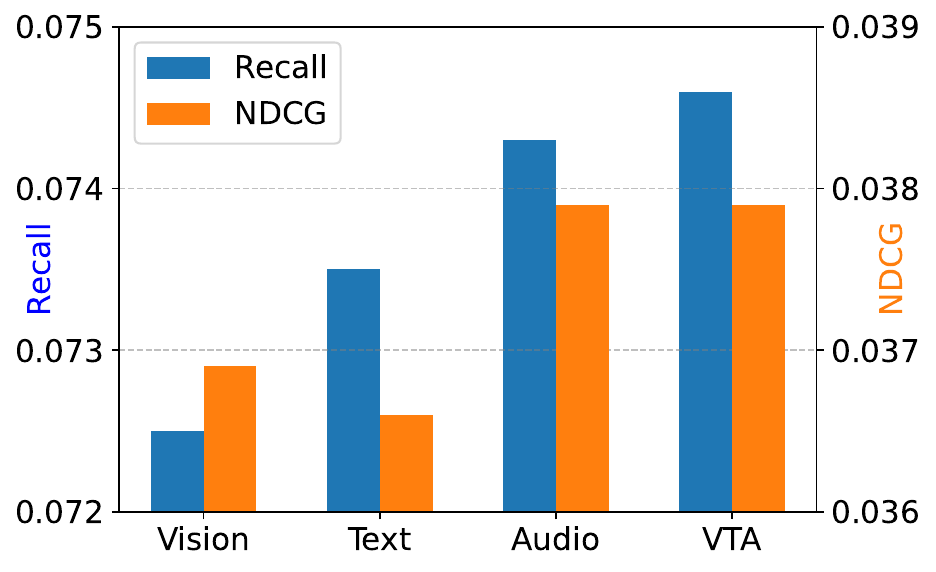}
    \caption{Accuracy}
    \label{fig:subfig1}
  \end{subfigure}%
  \hfill
  \begin{subfigure}[b]{0.4122\textwidth}
    \centering
    \includegraphics[width=1\textwidth]{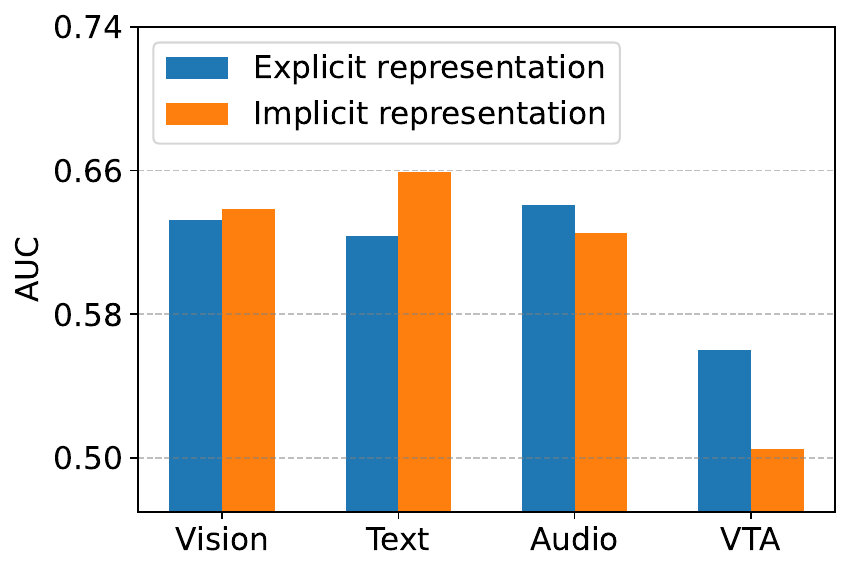}
    \caption{\raggedleft Unfairness\;\;\;\;\;\;\;\;\;\;\;\;\;\;\;\;\;\;\;\;\;\;}
    \label{fig:subfig2}
  \end{subfigure}%
  \hspace{\fill}
  \caption{Performance comparison of FMMRec in unimodal and multimodal settings.}
  \label{fig:unimodal_fairness}
\end{figure}

Our study showcases a comparative evaluation of FMMRec's performance in multimodal and unimodal settings on the MicroLens dataset, as shown in Figure~\ref{fig:unimodal_fairness}.
This analysis assesses whether integrating multimodal content enhances the effectiveness of causal fairness interventions compared to using a single modality.

FMMRec notably excels in multimodal settings, exhibiting the best performance in both accuracy and fairness.
Corresponding to the finding shown in Figure~\ref{fig:sens_leak_exp} that a wider range of modalities encompasses more user-sensitive information, integrating multimodal relations surpasses unimodal relations in sensitive information elimination.
This suggests that FMMRec was able to effectively leverage inter-modality correlations, further eliminating sensitive information and thus improving fairness beyond what was achieved on a unimodality level.

Moreover, the results indicate that multimodal content provides richer information that can be used to adjust for causal effects more effectively, helping achieve better counterfactual fairness in recommendations.

%% file: 7.conclusion.tex
In this paper, we addressed the problem of unfair representation learning in multimodal recommender systems from a causal perspective.
We proposed FMMRec, a novel approach that incorporates causal inference principles to disentangle sensitive and non-sensitive information and to adjust for undesired causal effects through relation-aware fairness learning.
Through causality-inspired modal disentanglement and relation-aware fairness learning, FMMRec effectively addresses the \textit{entanglement} of multimodal content and the \textit{heterogeneity} between item and user representations, which are the two key challenges in incorporating multimodal knowledge into fair representation learning.
Technically, for disentanglement learning, we first maximize and minimize the sensitive attribute prediction accuracy for the biased and filtered modal representations respectively, so that the causal effects of sensitive and non-sensitive attributes can be controlled in the learning process.
Utilizing the disentangled modal representations, we mine the modality-based unfair and fair (corresponding to biased and filtered respectively) user-user relations to adjust undesired causal effects and learn fair and informative user representations.
Extensive empirical studies on two public datasets demonstrate that our FMMRec achieves superior causal fairness performance compared with several state-of-the-art multimodal recommendation models and fairness-aware methods while maintaining highly comparable accuracy performance.

To the best of our knowledge, this is the first work that explicitly models and intervenes on causal relationships to improve fairness in multimodal recommendations.
Future work includes exploring more sophisticated causal models and interventions to further enhance fairness and robustness in recommender systems.
Exploring the causal impact of the introduced multimodal content on item-side unfairness (\textit{e.g.}, popularity bias~\cite{chen2023bias, resample, liu2023mitigating}) and to devise strategies to balance two-sided fairness in multimodal scenarios, are also promising directions in future.

%% file: acks.tex
\begin{acks}
This work is supported by Hong Kong Baptist University IG-FNRA Project (RC-FNRA-IG/21-22/SCI/01), Key Research Partnership Scheme (KRPS/23-24/02), and NSFC/RGC Joint Research Scheme (N\_HKBU214/24).
\end{acks}

%% file: main.bbl

\begin{thebibliography}{88}


\ifx \showCODEN    \undefined \def \showCODEN     #1{\unskip}     \fi
\ifx \showDOI      \undefined \def \showDOI       #1{#1}\fi
\ifx \showISBNx    \undefined \def \showISBNx     #1{\unskip}     \fi
\ifx \showISBNxiii \undefined \def \showISBNxiii  #1{\unskip}     \fi
\ifx \showISSN     \undefined \def \showISSN      #1{\unskip}     \fi
\ifx \showLCCN     \undefined \def \showLCCN      #1{\unskip}     \fi
\ifx \shownote     \undefined \def \shownote      #1{#1}          \fi
\ifx \showarticletitle \undefined \def \showarticletitle #1{#1}   \fi
\ifx \showURL      \undefined \def \showURL       {\relax}        \fi
\providecommand\bibfield[2]{#2}
\providecommand\bibinfo[2]{#2}
\providecommand\natexlab[1]{#1}
\providecommand\showeprint[2][]{arXiv:#2}

\bibitem[Biega et~al\mbox{.}(2018)]%
        {biega2018equity}
\bibfield{author}{\bibinfo{person}{Asia~J Biega}, \bibinfo{person}{Krishna~P Gummadi}, {and} \bibinfo{person}{Gerhard Weikum}.} \bibinfo{year}{2018}\natexlab{}.
\newblock \showarticletitle{Equity of attention: Amortizing individual fairness in rankings}. In \bibinfo{booktitle}{\emph{Proceedings of the 41st International ACM SIGIR Conference on Research and Development in Information Retrieval (SIGIR'18)}}. \bibinfo{pages}{405--414}.
\newblock


\bibitem[Bin et~al\mbox{.}(2025)]%
        {bin2025faircore}
\bibfield{author}{\bibinfo{person}{Chenzhong Bin}, \bibinfo{person}{Wenqiang Liu}, \bibinfo{person}{Feng Zhang}, \bibinfo{person}{Liang Chang}, {and} \bibinfo{person}{Tianlong Gu}.} \bibinfo{year}{2025}\natexlab{}.
\newblock \showarticletitle{FairCoRe: Fairness-aware recommendation through counterfactual representation learning}.
\newblock \bibinfo{journal}{\emph{IEEE Transactions on Knowledge and Data Engineering}} (\bibinfo{year}{2025}).
\newblock


\bibitem[Bonner and Vasile(2018)]%
        {bonner2018causal}
\bibfield{author}{\bibinfo{person}{Stephen Bonner} {and} \bibinfo{person}{Flavian Vasile}.} \bibinfo{year}{2018}\natexlab{}.
\newblock \showarticletitle{Causal embeddings for recommendation}. In \bibinfo{booktitle}{\emph{Proceedings of the 12th ACM conference on recommender systems (RecSys'18)}}. \bibinfo{pages}{104--112}.
\newblock


\bibitem[Bose and Hamilton(2019)]%
        {bose2019compositional}
\bibfield{author}{\bibinfo{person}{Avishek Bose} {and} \bibinfo{person}{William Hamilton}.} \bibinfo{year}{2019}\natexlab{}.
\newblock \showarticletitle{Compositional fairness constraints for graph embeddings}. In \bibinfo{booktitle}{\emph{Proceedings of the 36th International Conference on Machine Learning (ICML'19)}}. \bibinfo{pages}{715--724}.
\newblock


\bibitem[Calders et~al\mbox{.}(2009)]%
        {calders2009building}
\bibfield{author}{\bibinfo{person}{Toon Calders}, \bibinfo{person}{Faisal Kamiran}, {and} \bibinfo{person}{Mykola Pechenizkiy}.} \bibinfo{year}{2009}\natexlab{}.
\newblock \showarticletitle{Building classifiers with independency constraints}. In \bibinfo{booktitle}{\emph{Proceedings of the {IEEE} International Conference on Data Mining Workshops (ICDM'09)}}. \bibinfo{pages}{13--18}.
\newblock


\bibitem[Chen et~al\mbox{.}(2023)]%
        {chen2023bias}
\bibfield{author}{\bibinfo{person}{Jiawei Chen}, \bibinfo{person}{Hande Dong}, \bibinfo{person}{Xiang Wang}, \bibinfo{person}{Fuli Feng}, \bibinfo{person}{Meng Wang}, {and} \bibinfo{person}{Xiangnan He}.} \bibinfo{year}{2023}\natexlab{}.
\newblock \showarticletitle{Bias and debias in recommender system: A survey and future directions}.
\newblock \bibinfo{journal}{\emph{ACM Transactions on Information Systems}} (\bibinfo{year}{2023}), \bibinfo{pages}{67:1--67:39}.
\newblock


\bibitem[Chen et~al\mbox{.}(2009)]%
        {chen2009fast}
\bibfield{author}{\bibinfo{person}{Jie Chen}, \bibinfo{person}{Haw-ren Fang}, {and} \bibinfo{person}{Yousef Saad}.} \bibinfo{year}{2009}\natexlab{}.
\newblock \showarticletitle{Fast approximate kNN graph construction for high dimensional data via recursive Lanczos bisection}.
\newblock \bibinfo{journal}{\emph{Journal of Machine Learning Research}}  \bibinfo{volume}{10} (\bibinfo{year}{2009}), \bibinfo{pages}{1989--2012}.
\newblock


\bibitem[Chen et~al\mbox{.}(2025)]%
        {chen2025investigating}
\bibfield{author}{\bibinfo{person}{Weixin Chen}, \bibinfo{person}{Li Chen}, {and} \bibinfo{person}{Yuhan Zhao}.} \bibinfo{year}{2025}\natexlab{}.
\newblock \showarticletitle{Investigating User-side fairness in outcome and process for multi-type sensitive attributes in recommendations}.
\newblock \bibinfo{journal}{\emph{ACM Transactions on Recommender Systems}} (\bibinfo{year}{2025}).
\newblock


\bibitem[Chen et~al\mbox{.}(2022)]%
        {chen2022global}
\bibfield{author}{\bibinfo{person}{Weixin Chen}, \bibinfo{person}{Mingkai He}, \bibinfo{person}{Yongxin Ni}, \bibinfo{person}{Weike Pan}, \bibinfo{person}{Li Chen}, {and} \bibinfo{person}{Zhong Ming}.} \bibinfo{year}{2022}\natexlab{}.
\newblock \showarticletitle{Global and personalized graphs for heterogeneous sequential recommendation by learning behavior transitions and user intentions}. In \bibinfo{booktitle}{\emph{Proceedings of the 16th {ACM} Conference on Recommender Systems (RecSys'22)}}. \bibinfo{pages}{268--277}.
\newblock


\bibitem[Chen et~al\mbox{.}(2024)]%
        {chen2024fairdgcl}
\bibfield{author}{\bibinfo{person}{Wei Chen}, \bibinfo{person}{Meng Yuan}, \bibinfo{person}{Zhao Zhang}, \bibinfo{person}{Ruobing Xie}, \bibinfo{person}{Fuzhen Zhuang}, \bibinfo{person}{Deqing Wang}, {and} \bibinfo{person}{Rui Liu}.} \bibinfo{year}{2024}\natexlab{}.
\newblock \showarticletitle{FairDgcl: Fairness-aware recommendation with dynamic graph contrastive learning}.
\newblock \bibinfo{journal}{\emph{arXiv preprint arXiv:2410.17555}} (\bibinfo{year}{2024}).
\newblock


\bibitem[Chen et~al\mbox{.}(2019)]%
        {DBLP:conf/sigir/ChenCXZ0QZ19}
\bibfield{author}{\bibinfo{person}{Xu Chen}, \bibinfo{person}{Hanxiong Chen}, \bibinfo{person}{Hongteng Xu}, \bibinfo{person}{Yongfeng Zhang}, \bibinfo{person}{Yixin Cao}, \bibinfo{person}{Zheng Qin}, {and} \bibinfo{person}{Hongyuan Zha}.} \bibinfo{year}{2019}\natexlab{}.
\newblock \showarticletitle{Personalized fashion recommendation with visual explanations based on multimodal attention network: Towards visually explainable recommendation}. In \bibinfo{booktitle}{\emph{Proceedings of the 42nd International {ACM} {SIGIR} Conference on Research and Development in Information Retrieval (SIGIR'19)}}. \bibinfo{pages}{765--774}.
\newblock


\bibitem[Chiappa(2019)]%
        {chiappa2019path}
\bibfield{author}{\bibinfo{person}{Silvia Chiappa}.} \bibinfo{year}{2019}\natexlab{}.
\newblock \showarticletitle{Path-specific counterfactual fairness}. In \bibinfo{booktitle}{\emph{Proceedings of the 33rd AAAI Conference on Artificial Intelligence (AAAI'19)}}. \bibinfo{pages}{7801--7808}.
\newblock


\bibitem[Deldjoo et~al\mbox{.}(2024)]%
        {deldjoo2024fairness}
\bibfield{author}{\bibinfo{person}{Yashar Deldjoo}, \bibinfo{person}{Dietmar Jannach}, \bibinfo{person}{Alejandro Bellogin}, \bibinfo{person}{Alessandro Difonzo}, {and} \bibinfo{person}{Dario Zanzonelli}.} \bibinfo{year}{2024}\natexlab{}.
\newblock \showarticletitle{Fairness in recommender systems: research landscape and future directions}.
\newblock \bibinfo{journal}{\emph{User Modeling and User-Adapted Interaction}} \bibinfo{volume}{34}, \bibinfo{number}{1} (\bibinfo{year}{2024}), \bibinfo{pages}{59--108}.
\newblock


\bibitem[Ekstrand et~al\mbox{.}(2018)]%
        {resample}
\bibfield{author}{\bibinfo{person}{Michael~D. Ekstrand}, \bibinfo{person}{Mucun Tian}, \bibinfo{person}{Ion~Madrazo Azpiazu}, \bibinfo{person}{Jennifer~D. Ekstrand}, \bibinfo{person}{Oghenemaro Anuyah}, \bibinfo{person}{David McNeill}, {and} \bibinfo{person}{Maria~Soledad Pera}.} \bibinfo{year}{2018}\natexlab{}.
\newblock \showarticletitle{All the cool kids, how do they fit in?: Popularity and demographic biases in recommender evaluation and effectiveness}. In \bibinfo{booktitle}{\emph{Proceedings of the 1st Conference on Fairness, Accountability and Transparency (FAccT'18)}}. \bibinfo{pages}{172–186}.
\newblock


\bibitem[Fayyazi et~al\mbox{.}(2025)]%
        {fayyazi2025facter}
\bibfield{author}{\bibinfo{person}{Arya Fayyazi}, \bibinfo{person}{Mehdi Kamal}, {and} \bibinfo{person}{Massoud Pedram}.} \bibinfo{year}{2025}\natexlab{}.
\newblock \showarticletitle{FACTER: Fairness-aware conformal thresholding and prompt engineering for enabling fair LLM-based recommender systems}.
\newblock \bibinfo{journal}{\emph{arXiv preprint arXiv:2502.02966}} (\bibinfo{year}{2025}).
\newblock


\bibitem[Fu et~al\mbox{.}(2024)]%
        {fu2024iisan}
\bibfield{author}{\bibinfo{person}{Junchen Fu}, \bibinfo{person}{Xuri Ge}, \bibinfo{person}{Xin Xin}, \bibinfo{person}{Alexandros Karatzoglou}, \bibinfo{person}{Ioannis Arapakis}, \bibinfo{person}{Jie Wang}, {and} \bibinfo{person}{Joemon~M Jose}.} \bibinfo{year}{2024}\natexlab{}.
\newblock \showarticletitle{IISAN: Efficiently adapting multimodal representation for sequential recommendation with decoupled PEFT}. In \bibinfo{booktitle}{\emph{Proceedings of the 47th International ACM SIGIR Conference on Research and Development in Information Retrieval (SIGIR'24)}}. \bibinfo{pages}{687--697}.
\newblock


\bibitem[Ghodsi et~al\mbox{.}(2018)]%
        {EnvyFree}
\bibfield{author}{\bibinfo{person}{Mohammad Ghodsi}, \bibinfo{person}{MohammadTaghi HajiAghayi}, \bibinfo{person}{Masoud Seddighin}, \bibinfo{person}{Saeed Seddighin}, {and} \bibinfo{person}{Hadi Yami}.} \bibinfo{year}{2018}\natexlab{}.
\newblock \showarticletitle{Fair allocation of indivisible goods: Improvements and generalizations}. In \bibinfo{booktitle}{\emph{Proceedings of the ACM Conference on Economics and Computation (EC'18)}}. \bibinfo{pages}{539--556}.
\newblock


\bibitem[Glorot and Bengio(2010)]%
        {glorot2010understanding}
\bibfield{author}{\bibinfo{person}{Xavier Glorot} {and} \bibinfo{person}{Yoshua Bengio}.} \bibinfo{year}{2010}\natexlab{}.
\newblock \showarticletitle{Understanding the difficulty of training deep feedforward neural networks}. In \bibinfo{booktitle}{\emph{Proceedings of the 13th International Conference on Artificial Intelligence and Statistics (AISTATS'10)}}. \bibinfo{pages}{249--256}.
\newblock


\bibitem[Goodfellow et~al\mbox{.}(2014)]%
        {goodfellow2014generative}
\bibfield{author}{\bibinfo{person}{Ian Goodfellow}, \bibinfo{person}{Jean Pouget-Abadie}, \bibinfo{person}{Mehdi Mirza}, \bibinfo{person}{Bing Xu}, \bibinfo{person}{David Warde-Farley}, \bibinfo{person}{Sherjil Ozair}, \bibinfo{person}{Aaron Courville}, {and} \bibinfo{person}{Yoshua Bengio}.} \bibinfo{year}{2014}\natexlab{}.
\newblock \showarticletitle{Generative adversarial nets}. In \bibinfo{booktitle}{\emph{Proceedings of the 27th International Conference on Neural Information Processing Systems (NeurIPS'14)}}. \bibinfo{pages}{2672--2680}.
\newblock


\bibitem[Guo et~al\mbox{.}(2024)]%
        {guo2024lgmrec}
\bibfield{author}{\bibinfo{person}{Zhiqiang Guo}, \bibinfo{person}{Jianjun Li}, \bibinfo{person}{Guohui Li}, \bibinfo{person}{Chaoyang Wang}, \bibinfo{person}{Si Shi}, {and} \bibinfo{person}{Bin Ruan}.} \bibinfo{year}{2024}\natexlab{}.
\newblock \showarticletitle{LGMRec: Local and global graph learning for multimodal recommendation}. In \bibinfo{booktitle}{\emph{Proceedings of the AAAI Conference on Artificial Intelligence (AAAI'24)}}, Vol.~\bibinfo{volume}{38}. \bibinfo{pages}{8454--8462}.
\newblock


\bibitem[Hardt et~al\mbox{.}(2016)]%
        {DBLP:conf/nips/HardtPNS16}
\bibfield{author}{\bibinfo{person}{Moritz Hardt}, \bibinfo{person}{Eric Price}, {and} \bibinfo{person}{Nati Srebro}.} \bibinfo{year}{2016}\natexlab{}.
\newblock \showarticletitle{Equality of opportunity in supervised learning}. In \bibinfo{booktitle}{\emph{Proceedings of the 29th International Conference on Neural Information Processing Systems (NeurIPS'16)}}. \bibinfo{pages}{3315--3323}.
\newblock


\bibitem[He et~al\mbox{.}(2016)]%
        {he2016deep}
\bibfield{author}{\bibinfo{person}{Kaiming He}, \bibinfo{person}{Xiangyu Zhang}, \bibinfo{person}{Shaoqing Ren}, {and} \bibinfo{person}{Jian Sun}.} \bibinfo{year}{2016}\natexlab{}.
\newblock \showarticletitle{Deep residual learning for image recognition}. In \bibinfo{booktitle}{\emph{Proceedings of the 2016 {IEEE} Conference on Computer Vision and Pattern Recognition (CVPR'16)}}. \bibinfo{pages}{770--778}.
\newblock


\bibitem[He and McAuley(2016)]%
        {VBPR}
\bibfield{author}{\bibinfo{person}{Ruining He} {and} \bibinfo{person}{Julian~J. McAuley}.} \bibinfo{year}{2016}\natexlab{}.
\newblock \showarticletitle{{VBPR:} Visual bayesian personalized ranking from implicit feedback}. In \bibinfo{booktitle}{\emph{Proceedings of the 30th {AAAI} Conference on Artificial Intelligence (AAAI'16)}}. \bibinfo{pages}{144--150}.
\newblock


\bibitem[He et~al\mbox{.}(2020)]%
        {he2020lightgcn}
\bibfield{author}{\bibinfo{person}{Xiangnan He}, \bibinfo{person}{Kuan Deng}, \bibinfo{person}{Xiang Wang}, \bibinfo{person}{Yan Li}, \bibinfo{person}{Yongdong Zhang}, {and} \bibinfo{person}{Meng Wang}.} \bibinfo{year}{2020}\natexlab{}.
\newblock \showarticletitle{LightGCN: Simplifying and powering graph convolution network for recommendation}. In \bibinfo{booktitle}{\emph{Proceedings of the 43rd International {ACM} {SIGIR} Conference on Research and Development in Information Retrieval (SIGIR'20)}}. \bibinfo{pages}{639--648}.
\newblock


\bibitem[He et~al\mbox{.}(2023)]%
        {he2023addressing}
\bibfield{author}{\bibinfo{person}{Xiangnan He}, \bibinfo{person}{Yang Zhang}, \bibinfo{person}{Fuli Feng}, \bibinfo{person}{Chonggang Song}, \bibinfo{person}{Lingling Yi}, \bibinfo{person}{Guohui Ling}, {and} \bibinfo{person}{Yongdong Zhang}.} \bibinfo{year}{2023}\natexlab{}.
\newblock \showarticletitle{Addressing confounding feature issue for causal recommendation}.
\newblock \bibinfo{journal}{\emph{ACM Transactions on Information Systems}} \bibinfo{volume}{41}, \bibinfo{number}{3} (\bibinfo{year}{2023}), \bibinfo{pages}{1--23}.
\newblock


\bibitem[Hershey et~al\mbox{.}(2017)]%
        {hershey2017cnn}
\bibfield{author}{\bibinfo{person}{Shawn Hershey}, \bibinfo{person}{Sourish Chaudhuri}, \bibinfo{person}{Daniel~PW Ellis}, \bibinfo{person}{Jort~F Gemmeke}, \bibinfo{person}{Aren Jansen}, \bibinfo{person}{R~Channing Moore}, \bibinfo{person}{Manoj Plakal}, \bibinfo{person}{Devin Platt}, \bibinfo{person}{Rif~A Saurous}, \bibinfo{person}{Bryan Seybold}, {et~al\mbox{.}}} \bibinfo{year}{2017}\natexlab{}.
\newblock \showarticletitle{CNN architectures for large-scale audio classification}. In \bibinfo{booktitle}{\emph{Proceedings of the 2017 {IEEE} International Conference on Acoustics, Speech and Signal Processing (ICASSP'17)}}. \bibinfo{pages}{131--135}.
\newblock


\bibitem[Hua et~al\mbox{.}(2024)]%
        {hua2023up5}
\bibfield{author}{\bibinfo{person}{Wenyue Hua}, \bibinfo{person}{Yingqiang Ge}, \bibinfo{person}{Shuyuan Xu}, \bibinfo{person}{Jianchao Ji}, {and} \bibinfo{person}{Yongfeng Zhang}.} \bibinfo{year}{2024}\natexlab{}.
\newblock \showarticletitle{UP5: Unbiased foundation model for fairness-aware recommendation}. In \bibinfo{booktitle}{\emph{Proceedings of the 18th Conference of the European Chapter of the Association for Computational Linguistics (EACL'24)}}. \bibinfo{pages}{1899--1912}.
\newblock


\bibitem[Kilbertus et~al\mbox{.}(2017)]%
        {DBLP:conf/nips/KilbertusRPHJS17}
\bibfield{author}{\bibinfo{person}{Niki Kilbertus}, \bibinfo{person}{Mateo Rojas{-}Carulla}, \bibinfo{person}{Giambattista Parascandolo}, \bibinfo{person}{Moritz Hardt}, \bibinfo{person}{Dominik Janzing}, {and} \bibinfo{person}{Bernhard Sch{\"{o}}lkopf}.} \bibinfo{year}{2017}\natexlab{}.
\newblock \showarticletitle{Avoiding discrimination through causal reasoning}. In \bibinfo{booktitle}{\emph{Proceedings of the 30th International Conference on Neural Information Processing Systems (NeurIPS'17)}}. \bibinfo{pages}{656--666}.
\newblock


\bibitem[Kingma and Ba(2015)]%
        {DBLP:journals/corr/KingmaB14}
\bibfield{author}{\bibinfo{person}{Diederik~P. Kingma} {and} \bibinfo{person}{Jimmy Ba}.} \bibinfo{year}{2015}\natexlab{}.
\newblock \showarticletitle{Adam: {A} method for stochastic optimization}. In \bibinfo{booktitle}{\emph{Proceedings of the 3rd International Conference on Learning Representations (ICLR'15)}}.
\newblock


\bibitem[Kipf and Welling(2016)]%
        {kipf2016semi}
\bibfield{author}{\bibinfo{person}{Thomas~N Kipf} {and} \bibinfo{person}{Max Welling}.} \bibinfo{year}{2016}\natexlab{}.
\newblock \showarticletitle{Semi-supervised classification with graph convolutional networks}.
\newblock \bibinfo{journal}{\emph{arXiv preprint arXiv:1609.02907}} (\bibinfo{year}{2016}).
\newblock


\bibitem[Kusner et~al\mbox{.}(2017)]%
        {DBLP:conf/nips/KusnerLRS17}
\bibfield{author}{\bibinfo{person}{Matt~J. Kusner}, \bibinfo{person}{Joshua~R. Loftus}, \bibinfo{person}{Chris Russell}, {and} \bibinfo{person}{Ricardo Silva}.} \bibinfo{year}{2017}\natexlab{}.
\newblock \showarticletitle{Counterfactual fairness}. In \bibinfo{booktitle}{\emph{Proceedings of the 30th International Conference on Neural Information Processing Systems (NeurIPS'17)}}. \bibinfo{pages}{4066--4076}.
\newblock


\bibitem[Lai et~al\mbox{.}(2024)]%
        {lai2024matryoshka}
\bibfield{author}{\bibinfo{person}{Riwei Lai}, \bibinfo{person}{Li Chen}, \bibinfo{person}{Weixin Chen}, {and} \bibinfo{person}{Rui Chen}.} \bibinfo{year}{2024}\natexlab{}.
\newblock \bibinfo{title}{Matryoshka representation learning for recommendation}.
\newblock
\newblock
\showeprint[arxiv]{2406.07432}


\bibitem[Li et~al\mbox{.}(2023b)]%
        {li2023exploring}
\bibfield{author}{\bibinfo{person}{Ruyu Li}, \bibinfo{person}{Wenhao Deng}, \bibinfo{person}{Yu Cheng}, \bibinfo{person}{Zheng Yuan}, \bibinfo{person}{Jiaqi Zhang}, {and} \bibinfo{person}{Fajie Yuan}.} \bibinfo{year}{2023}\natexlab{b}.
\newblock \showarticletitle{Exploring the upper limits of text-based collaborative filtering using large language models: Discoveries and insights}.
\newblock \bibinfo{journal}{\emph{arXiv preprint arXiv:2305.11700}} (\bibinfo{year}{2023}).
\newblock


\bibitem[Li et~al\mbox{.}(2021a)]%
        {UGF}
\bibfield{author}{\bibinfo{person}{Yunqi Li}, \bibinfo{person}{Hanxiong Chen}, \bibinfo{person}{Zuohui Fu}, \bibinfo{person}{Yingqiang Ge}, {and} \bibinfo{person}{Yongfeng Zhang}.} \bibinfo{year}{2021}\natexlab{a}.
\newblock \showarticletitle{User-oriented fairness in recommendation}. In \bibinfo{booktitle}{\emph{Proceedings of the Web Conference (WWW'21)}}. \bibinfo{pages}{624--632}.
\newblock


\bibitem[Li et~al\mbox{.}(2023a)]%
        {lifairness}
\bibfield{author}{\bibinfo{person}{Yunqi Li}, \bibinfo{person}{Hanxiong Chen}, \bibinfo{person}{Shuyuan Xu}, \bibinfo{person}{Yingqiang Ge}, \bibinfo{person}{Juntao Tan}, \bibinfo{person}{Shuchang Liu}, {and} \bibinfo{person}{Yongfeng Zhang}.} \bibinfo{year}{2023}\natexlab{a}.
\newblock \showarticletitle{Fairness in recommendation: Foundations, methods and applications}.
\newblock \bibinfo{journal}{\emph{ACM Transactions on Intelligent Systems and Technology}} \bibinfo{volume}{14}, \bibinfo{number}{5} (\bibinfo{year}{2023}), \bibinfo{pages}{95:1--95:48}.
\newblock


\bibitem[Li et~al\mbox{.}(2021b)]%
        {PCFR}
\bibfield{author}{\bibinfo{person}{Yunqi Li}, \bibinfo{person}{Hanxiong Chen}, \bibinfo{person}{Shuyuan Xu}, \bibinfo{person}{Yingqiang Ge}, {and} \bibinfo{person}{Yongfeng Zhang}.} \bibinfo{year}{2021}\natexlab{b}.
\newblock \showarticletitle{Towards personalized fairness based on causal notion}. In \bibinfo{booktitle}{\emph{Proceedings of the 44th International ACM SIGIR Conference on Research and Development in Information Retrieval (SIGIR'21)}}. \bibinfo{pages}{1054--1063}.
\newblock


\bibitem[Li et~al\mbox{.}(2023c)]%
        {TFR}
\bibfield{author}{\bibinfo{person}{Yunqi Li}, \bibinfo{person}{Dingxian Wang}, \bibinfo{person}{Hanxiong Chen}, {and} \bibinfo{person}{Yongfeng Zhang}.} \bibinfo{year}{2023}\natexlab{c}.
\newblock \showarticletitle{Transferable fairness for cold-start recommendation}.
\newblock \bibinfo{journal}{\emph{arXiv preprint arXiv:2301.10665}} (\bibinfo{year}{2023}).
\newblock


\bibitem[Liang et~al\mbox{.}(2023)]%
        {MMMLP}
\bibfield{author}{\bibinfo{person}{Jiahao Liang}, \bibinfo{person}{Xiangyu Zhao}, \bibinfo{person}{Muyang Li}, \bibinfo{person}{Zijian Zhang}, \bibinfo{person}{Wanyu Wang}, \bibinfo{person}{Haochen Liu}, {and} \bibinfo{person}{Zitao Liu}.} \bibinfo{year}{2023}\natexlab{}.
\newblock \showarticletitle{MMMLP: Multi-modal multilayer perceptron for sequential recommendations}. In \bibinfo{booktitle}{\emph{Proceedings of the {ACM} Web Conference (WWW'23)}}. \bibinfo{pages}{1109--1117}.
\newblock


\bibitem[Lin et~al\mbox{.}(2024)]%
        {pretrained2024causal}
\bibfield{author}{\bibinfo{person}{Ziqian Lin}, \bibinfo{person}{Hao Ding}, \bibinfo{person}{Nghia~Trong Hoang}, \bibinfo{person}{Branislav Kveton}, \bibinfo{person}{Anoop Deoras}, {and} \bibinfo{person}{Hao Wang}.} \bibinfo{year}{2024}\natexlab{}.
\newblock \showarticletitle{Pre-trained Recommender Systems: A Causal Debiasing Perspective}. In \bibinfo{booktitle}{\emph{Proceedings of the 17th ACM International Conference on Web Search and Data Mining (WSDM'24)}}. \bibinfo{pages}{424–433}.
\newblock


\bibitem[Liu et~al\mbox{.}(2024b)]%
        {liu2024multimodal}
\bibfield{author}{\bibinfo{person}{Qijiong Liu}, \bibinfo{person}{Jieming Zhu}, \bibinfo{person}{Yanting Yang}, \bibinfo{person}{Quanyu Dai}, \bibinfo{person}{Zhaocheng Du}, \bibinfo{person}{Xiao-Ming Wu}, \bibinfo{person}{Zhou Zhao}, \bibinfo{person}{Rui Zhang}, {and} \bibinfo{person}{Zhenhua Dong}.} \bibinfo{year}{2024}\natexlab{b}.
\newblock \showarticletitle{Multimodal pretraining, adaptation, and generation for recommendation: A survey}. In \bibinfo{booktitle}{\emph{Proceedings of the 30th ACM SIGKDD Conference on Knowledge Discovery and Data Mining (KDD'24))}}. \bibinfo{pages}{6566--6576}.
\newblock


\bibitem[Liu et~al\mbox{.}(2019)]%
        {DBLP:conf/www/LiuCLH19}
\bibfield{author}{\bibinfo{person}{Shang Liu}, \bibinfo{person}{Zhenzhong Chen}, \bibinfo{person}{Hongyi Liu}, {and} \bibinfo{person}{Xinghai Hu}.} \bibinfo{year}{2019}\natexlab{}.
\newblock \showarticletitle{User-video co-attention network for personalized micro-video recommendation}. In \bibinfo{booktitle}{\emph{The World Wide Web Conference (WWW'19)}}. \bibinfo{pages}{3020--3026}.
\newblock


\bibitem[Liu et~al\mbox{.}(2024a)]%
        {liu2024dual}
\bibfield{author}{\bibinfo{person}{Shenghao Liu}, \bibinfo{person}{Yu Zhang}, \bibinfo{person}{Lingzhi Yi}, \bibinfo{person}{Xianjun Deng}, \bibinfo{person}{Laurence~T Yang}, {and} \bibinfo{person}{Bang Wang}.} \bibinfo{year}{2024}\natexlab{a}.
\newblock \showarticletitle{Dual-side adversarial learning based fair recommendation for sensitive attribute filtering}.
\newblock \bibinfo{journal}{\emph{ACM Transactions on Knowledge Discovery from Data}} \bibinfo{volume}{18}, \bibinfo{number}{7} (\bibinfo{year}{2024}), \bibinfo{pages}{1--20}.
\newblock


\bibitem[Liu et~al\mbox{.}(2023)]%
        {liu2023mitigating}
\bibfield{author}{\bibinfo{person}{Zhongzhou Liu}, \bibinfo{person}{Yuan Fang}, {and} \bibinfo{person}{Min Wu}.} \bibinfo{year}{2023}\natexlab{}.
\newblock \showarticletitle{Mitigating popularity bias for users and items with fairness-centric adaptive recommendation}.
\newblock \bibinfo{journal}{\emph{ACM Transactions on Information Systems}} \bibinfo{volume}{41}, \bibinfo{number}{3} (\bibinfo{year}{2023}), \bibinfo{pages}{55:1--55:27}.
\newblock


\bibitem[Ma et~al\mbox{.}(2024)]%
        {ma2024multimodal}
\bibfield{author}{\bibinfo{person}{Haokai Ma}, \bibinfo{person}{Yimeng Yang}, \bibinfo{person}{Lei Meng}, \bibinfo{person}{Ruobing Xie}, {and} \bibinfo{person}{Xiangxu Meng}.} \bibinfo{year}{2024}\natexlab{}.
\newblock \showarticletitle{Multimodal conditioned diffusion model for recommendation}. In \bibinfo{booktitle}{\emph{Proceedings of the Web Conference (WWW'24)}}. \bibinfo{pages}{1733--1740}.
\newblock


\bibitem[Madras et~al\mbox{.}(2018)]%
        {DBLP:conf/icml/MadrasCPZ18}
\bibfield{author}{\bibinfo{person}{David Madras}, \bibinfo{person}{Elliot Creager}, \bibinfo{person}{Toniann Pitassi}, {and} \bibinfo{person}{Richard~S. Zemel}.} \bibinfo{year}{2018}\natexlab{}.
\newblock \showarticletitle{Learning adversarially fair and transferable representations}. In \bibinfo{booktitle}{\emph{Proceedings of the 35th International Conference on Machine Learning (ICML'18)}}. \bibinfo{pages}{3381--3390}.
\newblock


\bibitem[Ni et~al\mbox{.}(2023)]%
        {ni2023contentdriven}
\bibfield{author}{\bibinfo{person}{Yongxin Ni}, \bibinfo{person}{Yu Cheng}, \bibinfo{person}{Xiangyan Liu}, \bibinfo{person}{Junchen Fu}, \bibinfo{person}{Youhua Li}, \bibinfo{person}{Xiangnan He}, \bibinfo{person}{Yongfeng Zhang}, {and} \bibinfo{person}{Fajie Yuan}.} \bibinfo{year}{2023}\natexlab{}.
\newblock \showarticletitle{A content-driven micro-video recommendation dataset at scale}.
\newblock \bibinfo{journal}{\emph{arXiv preprint arXiv:2309.15379}} (\bibinfo{year}{2023}).
\newblock


\bibitem[Pearl(2009)]%
        {pearl2009causality}
\bibfield{author}{\bibinfo{person}{Judea Pearl}.} \bibinfo{year}{2009}\natexlab{}.
\newblock \bibinfo{booktitle}{\emph{Causality: Models, reasoning and inference}}.
\newblock \bibinfo{publisher}{Cambridge University Press}.
\newblock


\bibitem[Reimers and Gurevych(2019)]%
        {reimers2019sentence}
\bibfield{author}{\bibinfo{person}{Nils Reimers} {and} \bibinfo{person}{Iryna Gurevych}.} \bibinfo{year}{2019}\natexlab{}.
\newblock \showarticletitle{Sentence-BERT: Sentence embeddings using siamese BERT-networks}. In \bibinfo{booktitle}{\emph{Proceedings of the 2019 Conference on Empirical Methods in Natural Language Processing (EMNLP'19)}}. \bibinfo{pages}{3980--3990}.
\newblock


\bibitem[Rendle et~al\mbox{.}(2009)]%
        {BPR}
\bibfield{author}{\bibinfo{person}{Steffen Rendle}, \bibinfo{person}{Christoph Freudenthaler}, \bibinfo{person}{Zeno Gantner}, {and} \bibinfo{person}{Lars Schmidt{-}Thieme}.} \bibinfo{year}{2009}\natexlab{}.
\newblock \showarticletitle{{BPR:} Bayesian personalized ranking from implicit feedback}. In \bibinfo{booktitle}{\emph{Proceedings of the 25th Conference on Uncertainty in Artificial Intelligence (UAI'09)}}. \bibinfo{pages}{452--461}.
\newblock


\bibitem[Schnabel et~al\mbox{.}(2016)]%
        {50}
\bibfield{author}{\bibinfo{person}{Tobias Schnabel}, \bibinfo{person}{Adith Swaminathan}, \bibinfo{person}{Ashudeep Singh}, \bibinfo{person}{Navin Chandak}, {and} \bibinfo{person}{Thorsten Joachims}.} \bibinfo{year}{2016}\natexlab{}.
\newblock \showarticletitle{Recommendations as treatments: Debiasing learning and evaluation}. In \bibinfo{booktitle}{\emph{Proceedings of the 33rd International Conference on Machine Learning (ICML'16)}}. \bibinfo{pages}{1670--1679}.
\newblock


\bibitem[Sch{\"o}lkopf et~al\mbox{.}(2021)]%
        {scholkopf2021toward}
\bibfield{author}{\bibinfo{person}{Bernhard Sch{\"o}lkopf}, \bibinfo{person}{Francesco Locatello}, \bibinfo{person}{Stefan Bauer}, \bibinfo{person}{Nan~Rosemary Ke}, \bibinfo{person}{Nal Kalchbrenner}, \bibinfo{person}{Anirudh Goyal}, {and} \bibinfo{person}{Yoshua Bengio}.} \bibinfo{year}{2021}\natexlab{}.
\newblock \showarticletitle{Toward causal representation learning}.
\newblock \bibinfo{journal}{\emph{Proc. IEEE}} \bibinfo{volume}{109}, \bibinfo{number}{5} (\bibinfo{year}{2021}), \bibinfo{pages}{612--634}.
\newblock


\bibitem[Shao et~al\mbox{.}(2022)]%
        {shao2022faircf}
\bibfield{author}{\bibinfo{person}{Pengyang Shao}, \bibinfo{person}{Le Wu}, \bibinfo{person}{Lei Chen}, \bibinfo{person}{Kun Zhang}, {and} \bibinfo{person}{Meng Wang}.} \bibinfo{year}{2022}\natexlab{}.
\newblock \showarticletitle{FairCF: Fairness-aware collaborative filtering}.
\newblock \bibinfo{journal}{\emph{Science China Information Sciences}} \bibinfo{volume}{65}, \bibinfo{number}{12} (\bibinfo{year}{2022}), \bibinfo{pages}{222102:1–222102:15}.
\newblock


\bibitem[Tan et~al\mbox{.}(2021)]%
        {tan2021counterfactual}
\bibfield{author}{\bibinfo{person}{Juntao Tan}, \bibinfo{person}{Shuyuan Xu}, \bibinfo{person}{Yingqiang Ge}, \bibinfo{person}{Yunqi Li}, \bibinfo{person}{Xu Chen}, {and} \bibinfo{person}{Yongfeng Zhang}.} \bibinfo{year}{2021}\natexlab{}.
\newblock \showarticletitle{Counterfactual explainable recommendation}. In \bibinfo{booktitle}{\emph{Proceedings of the 30th ACM International Conference on Information and Knowledge Management (CIKM'21)}}. \bibinfo{pages}{1784--1793}.
\newblock


\bibitem[Tao et~al\mbox{.}(2022)]%
        {SLMRec}
\bibfield{author}{\bibinfo{person}{Zhulin Tao}, \bibinfo{person}{Xiaohao Liu}, \bibinfo{person}{Yewei Xia}, \bibinfo{person}{Xiang Wang}, \bibinfo{person}{Lifang Yang}, \bibinfo{person}{Xianglin Huang}, {and} \bibinfo{person}{Tat-Seng Chua}.} \bibinfo{year}{2022}\natexlab{}.
\newblock \showarticletitle{Self-supervised learning for multimedia recommendation}.
\newblock \bibinfo{journal}{\emph{IEEE Transactions on Multimedia}}  \bibinfo{volume}{25} (\bibinfo{year}{2022}), \bibinfo{pages}{5107--5116}.
\newblock


\bibitem[Wadsworth et~al\mbox{.}(2018)]%
        {wadsworth2018achieving}
\bibfield{author}{\bibinfo{person}{Christina Wadsworth}, \bibinfo{person}{Francesca Vera}, {and} \bibinfo{person}{Chris Piech}.} \bibinfo{year}{2018}\natexlab{}.
\newblock \showarticletitle{Achieving fairness through adversarial learning: An application to recidivism prediction}. In \bibinfo{booktitle}{\emph{Proceedings of the 5th Workshop on Fairness, Accountability, and Transparency in Machine Learning (FAT/ML'18)}}. \bibinfo{pages}{1:1--1:5}.
\newblock


\bibitem[Wang et~al\mbox{.}(2021)]%
        {DBLP:conf/kdd/WangF0WC21}
\bibfield{author}{\bibinfo{person}{Wenjie Wang}, \bibinfo{person}{Fuli Feng}, \bibinfo{person}{Xiangnan He}, \bibinfo{person}{Xiang Wang}, {and} \bibinfo{person}{Tat{-}Seng Chua}.} \bibinfo{year}{2021}\natexlab{}.
\newblock \showarticletitle{Deconfounded recommendation for alleviating bias amplification}. In \bibinfo{booktitle}{\emph{Proceedings of the 27th {ACM} {SIGKDD} Conference on Knowledge Discovery and Data Mining (KDD'21)}}. \bibinfo{pages}{1717--1725}.
\newblock


\bibitem[Wang and Blei(2019)]%
        {Wang2019blessings}
\bibfield{author}{\bibinfo{person}{Yixin Wang} {and} \bibinfo{person}{David~M. Blei}.} \bibinfo{year}{2019}\natexlab{}.
\newblock \showarticletitle{The blessings of multiple causes}.
\newblock \bibinfo{journal}{\emph{J. Amer. Statist. Assoc.}} \bibinfo{volume}{114}, \bibinfo{number}{528} (\bibinfo{year}{2019}), \bibinfo{pages}{1574--1596}.
\newblock


\bibitem[Wang et~al\mbox{.}(2023)]%
        {wang2022survey}
\bibfield{author}{\bibinfo{person}{Yifan Wang}, \bibinfo{person}{Weizhi Ma}, \bibinfo{person}{Min Zhang}, \bibinfo{person}{Yiqun Liu}, {and} \bibinfo{person}{Shaoping Ma}.} \bibinfo{year}{2023}\natexlab{}.
\newblock \showarticletitle{A survey on the fairness of recommender systems}.
\newblock \bibinfo{journal}{\emph{ACM Transactions on Information Systems}} \bibinfo{volume}{41}, \bibinfo{number}{3} (\bibinfo{year}{2023}), \bibinfo{pages}{52:1--52:43}.
\newblock


\bibitem[Wei et~al\mbox{.}(2023)]%
        {LightGT}
\bibfield{author}{\bibinfo{person}{Yinwei Wei}, \bibinfo{person}{Wenqi Liu}, \bibinfo{person}{Fan Liu}, \bibinfo{person}{Xiang Wang}, \bibinfo{person}{Liqiang Nie}, {and} \bibinfo{person}{Tat{-}Seng Chua}.} \bibinfo{year}{2023}\natexlab{}.
\newblock \showarticletitle{{LightGT}: {A} light graph transformer for multimedia recommendation}. In \bibinfo{booktitle}{\emph{Proceedings of the 46th International {ACM} {SIGIR} Conference on Research and Development in Information Retrieval (SIGIR'23)}}. \bibinfo{pages}{1508--1517}.
\newblock


\bibitem[Wei et~al\mbox{.}(2020)]%
        {GRCN}
\bibfield{author}{\bibinfo{person}{Yinwei Wei}, \bibinfo{person}{Xiang Wang}, \bibinfo{person}{Liqiang Nie}, \bibinfo{person}{Xiangnan He}, {and} \bibinfo{person}{Tat{-}Seng Chua}.} \bibinfo{year}{2020}\natexlab{}.
\newblock \showarticletitle{Graph-refined convolutional network for multimedia recommendation with implicit feedback}. In \bibinfo{booktitle}{\emph{The 28th {ACM} International Conference on Multimedia (MM'20)}}. \bibinfo{pages}{3541--3549}.
\newblock


\bibitem[Wei et~al\mbox{.}(2019)]%
        {MMGCN}
\bibfield{author}{\bibinfo{person}{Yinwei Wei}, \bibinfo{person}{Xiang Wang}, \bibinfo{person}{Liqiang Nie}, \bibinfo{person}{Xiangnan He}, \bibinfo{person}{Richang Hong}, {and} \bibinfo{person}{Tat{-}Seng Chua}.} \bibinfo{year}{2019}\natexlab{}.
\newblock \showarticletitle{{MMGCN:} Multi-modal graph convolution network for personalized recommendation of micro-video}. In \bibinfo{booktitle}{\emph{Proceedings of the 27th {ACM} International Conference on Multimedia (MM'19)}}. \bibinfo{pages}{1437--1445}.
\newblock


\bibitem[Wu et~al\mbox{.}(2021b)]%
        {FairRec}
\bibfield{author}{\bibinfo{person}{Chuhan Wu}, \bibinfo{person}{Fangzhao Wu}, \bibinfo{person}{Xiting Wang}, \bibinfo{person}{Yongfeng Huang}, {and} \bibinfo{person}{Xing Xie}.} \bibinfo{year}{2021}\natexlab{b}.
\newblock \showarticletitle{Fairness-aware news recommendation with decomposed adversarial learning}. In \bibinfo{booktitle}{\emph{Proceedings of the 35th AAAI Conference on Artificial Intelligence (AAAI'21)}}. \bibinfo{pages}{4462--4469}.
\newblock


\bibitem[Wu et~al\mbox{.}(2021a)]%
        {FairGo}
\bibfield{author}{\bibinfo{person}{Le Wu}, \bibinfo{person}{Lei Chen}, \bibinfo{person}{Pengyang Shao}, \bibinfo{person}{Richang Hong}, \bibinfo{person}{Xiting Wang}, {and} \bibinfo{person}{Meng Wang}.} \bibinfo{year}{2021}\natexlab{a}.
\newblock \showarticletitle{Learning fair Representations for recommendation: {A} graph-based perspective}. In \bibinfo{booktitle}{\emph{Proceedings of the Web Conference (WWW'21)}}. \bibinfo{pages}{2198--2208}.
\newblock


\bibitem[Wu et~al\mbox{.}(2022)]%
        {DBLP:conf/sigir/WuXZZ0ZL022}
\bibfield{author}{\bibinfo{person}{Yiqing Wu}, \bibinfo{person}{Ruobing Xie}, \bibinfo{person}{Yongchun Zhu}, \bibinfo{person}{Fuzhen Zhuang}, \bibinfo{person}{Xiang Ao}, \bibinfo{person}{Xu Zhang}, \bibinfo{person}{Leyu Lin}, {and} \bibinfo{person}{Qing He}.} \bibinfo{year}{2022}\natexlab{}.
\newblock \showarticletitle{Selective fairness in recommendation via prompts}. In \bibinfo{booktitle}{\emph{Proceedings of the 45th International ACM SIGIR Conference on Research and Development in Information Retrieval (SIGIR'22)}}. \bibinfo{pages}{2657--2662}.
\newblock


\bibitem[Xu et~al\mbox{.}(2025)]%
        {xu2025mentor}
\bibfield{author}{\bibinfo{person}{Jinfeng Xu}, \bibinfo{person}{Zheyu Chen}, \bibinfo{person}{Shuo Yang}, \bibinfo{person}{Jinze Li}, \bibinfo{person}{Hewei Wang}, {and} \bibinfo{person}{Edith~CH Ngai}.} \bibinfo{year}{2025}\natexlab{}.
\newblock \showarticletitle{Mentor: multi-level self-supervised learning for multimodal recommendation}. In \bibinfo{booktitle}{\emph{Proceedings of the AAAI Conference on Artificial Intelligence (AAAI'25)}}, Vol.~\bibinfo{volume}{39}. \bibinfo{pages}{12908--12917}.
\newblock


\bibitem[Xu et~al\mbox{.}(2023)]%
        {xu2023causal}
\bibfield{author}{\bibinfo{person}{Shuyuan Xu}, \bibinfo{person}{Yingqiang Ge}, \bibinfo{person}{Yunqi Li}, \bibinfo{person}{Zuohui Fu}, \bibinfo{person}{Xu Chen}, {and} \bibinfo{person}{Yongfeng Zhang}.} \bibinfo{year}{2023}\natexlab{}.
\newblock \showarticletitle{Causal collaborative filtering}. In \bibinfo{booktitle}{\emph{Proceedings of the 46th {ACM} {SIGIR} International Conference on Theory of Information Retrieval (ICTIR'23)}}. \bibinfo{pages}{235--245}.
\newblock


\bibitem[Yang et~al\mbox{.}(2018)]%
        {yang2018mmcf}
\bibfield{author}{\bibinfo{person}{Hojin Yang}, \bibinfo{person}{Yoonki Jeong}, \bibinfo{person}{Minjin Choi}, {and} \bibinfo{person}{Jongwuk Lee}.} \bibinfo{year}{2018}\natexlab{}.
\newblock \showarticletitle{MMCF: Multimodal collaborative filtering for automatic playlist continuation}.
\newblock In \bibinfo{booktitle}{\emph{Proceedings of the 12th {ACM} Recommender Systems Challenge (RecSys'18)}}. \bibinfo{pages}{11:1--11:6}.
\newblock


\bibitem[Yang et~al\mbox{.}(2024)]%
        {yang2024distributional}
\bibfield{author}{\bibinfo{person}{Hao Yang}, \bibinfo{person}{Xian Wu}, \bibinfo{person}{Zhaopeng Qiu}, \bibinfo{person}{Yefeng Zheng}, {and} \bibinfo{person}{Xu Chen}.} \bibinfo{year}{2024}\natexlab{}.
\newblock \showarticletitle{Distributional fairness-aware recommendation}.
\newblock \bibinfo{journal}{\emph{ACM Transactions on Information Systems}} \bibinfo{volume}{42}, \bibinfo{number}{5} (\bibinfo{year}{2024}), \bibinfo{pages}{1--28}.
\newblock


\bibitem[Yang et~al\mbox{.}(2021)]%
        {yang2021enhanced}
\bibfield{author}{\bibinfo{person}{Yonghui Yang}, \bibinfo{person}{Le Wu}, \bibinfo{person}{Richang Hong}, \bibinfo{person}{Kun Zhang}, {and} \bibinfo{person}{Meng Wang}.} \bibinfo{year}{2021}\natexlab{}.
\newblock \showarticletitle{Enhanced graph learning for collaborative filtering via mutual information maximization}. In \bibinfo{booktitle}{\emph{Proceedings of the 44th International {ACM} {SIGIR} Conference on Research and Development in Information Retrieval (SIGIR'21)}}. \bibinfo{pages}{71--80}.
\newblock


\bibitem[Yao and Huang(2017)]%
        {FOCF}
\bibfield{author}{\bibinfo{person}{Sirui Yao} {and} \bibinfo{person}{Bert Huang}.} \bibinfo{year}{2017}\natexlab{}.
\newblock \showarticletitle{Beyond parity: Fairness objectives for collaborative filtering}. In \bibinfo{booktitle}{\emph{Proceedings of the 30th International Conference on Neural Information Processing Systems (NeurIPS'17)}}. \bibinfo{pages}{2921--2930}.
\newblock


\bibitem[Ye et~al\mbox{.}(2023)]%
        {ye2023towards}
\bibfield{author}{\bibinfo{person}{Haibo Ye}, \bibinfo{person}{Xinjie Li}, \bibinfo{person}{Yuan Yao}, {and} \bibinfo{person}{Hanghang Tong}.} \bibinfo{year}{2023}\natexlab{}.
\newblock \showarticletitle{Towards robust neural graph collaborative filtering via structure denoising and embedding perturbation}.
\newblock \bibinfo{journal}{\emph{ACM Transactions on Information Systems}} \bibinfo{volume}{41}, \bibinfo{number}{3} (\bibinfo{year}{2023}), \bibinfo{pages}{59:1--59:28}.
\newblock


\bibitem[Yu et~al\mbox{.}(2023)]%
        {MGCN}
\bibfield{author}{\bibinfo{person}{Penghang Yu}, \bibinfo{person}{Zhiyi Tan}, \bibinfo{person}{Guanming Lu}, {and} \bibinfo{person}{Bing-Kun Bao}.} \bibinfo{year}{2023}\natexlab{}.
\newblock \showarticletitle{Multi-view graph convolutional network for multimedia recommendation}. In \bibinfo{booktitle}{\emph{Proceedings of the 31st ACM International Conference on Multimedia (MM'23)}}. \bibinfo{pages}{6576--6585}.
\newblock


\bibitem[Yuan et~al\mbox{.}(2023)]%
        {MoRec}
\bibfield{author}{\bibinfo{person}{Zheng Yuan}, \bibinfo{person}{Fajie Yuan}, \bibinfo{person}{Yu Song}, \bibinfo{person}{Youhua Li}, \bibinfo{person}{Junchen Fu}, \bibinfo{person}{Fei Yang}, \bibinfo{person}{Yunzhu Pan}, {and} \bibinfo{person}{Yongxin Ni}.} \bibinfo{year}{2023}\natexlab{}.
\newblock \showarticletitle{Where to go next for recommender systems? ID- vs. modality-based recommender models revisited}. In \bibinfo{booktitle}{\emph{Proceedings of the 46th International {ACM} {SIGIR} Conference on Research and Development in Information Retrieval (SIGIR'23)}}. \bibinfo{pages}{2639--2649}.
\newblock


\bibitem[Zhang et~al\mbox{.}(2025)]%
        {HACD}
\bibfield{author}{\bibinfo{person}{Anran Zhang}, \bibinfo{person}{Xingfen Wang}, {and} \bibinfo{person}{Yuhan Zhao}.} \bibinfo{year}{2025}\natexlab{}.
\newblock \showarticletitle{HACD: Harnessing Attribute Semantics and Mesoscopic Structure for Community Detection}. In \bibinfo{booktitle}{\emph{Proceedings of the Eighteenth ACM International Conference on Web Search and Data Mining (WSDM'25)}}. \bibinfo{pages}{616--624}.
\newblock


\bibitem[Zhang et~al\mbox{.}(2021)]%
        {LATTICE}
\bibfield{author}{\bibinfo{person}{Jinghao Zhang}, \bibinfo{person}{Yanqiao Zhu}, \bibinfo{person}{Qiang Liu}, \bibinfo{person}{Shu Wu}, \bibinfo{person}{Shuhui Wang}, {and} \bibinfo{person}{Liang Wang}.} \bibinfo{year}{2021}\natexlab{}.
\newblock \showarticletitle{Mining latent structures for multimedia recommendation}. In \bibinfo{booktitle}{\emph{Proceedings of the 29th ACM International Conference on Multimedia (MM'21)}}. \bibinfo{pages}{3872--3880}.
\newblock


\bibitem[Zhang et~al\mbox{.}(2023)]%
        {zhang2023causal}
\bibfield{author}{\bibinfo{person}{Shengyu Zhang}, \bibinfo{person}{Ziqi Jiang}, \bibinfo{person}{Jiangchao Yao}, \bibinfo{person}{Fuli Feng}, \bibinfo{person}{Kun Kuang}, \bibinfo{person}{Zhou Zhao}, \bibinfo{person}{Shuo Li}, \bibinfo{person}{Hongxia Yang}, \bibinfo{person}{Tat-seng Chua}, {and} \bibinfo{person}{Fei Wu}.} \bibinfo{year}{2023}\natexlab{}.
\newblock \showarticletitle{Causal distillation for alleviating performance heterogeneity in recommender systems}.
\newblock \bibinfo{journal}{\emph{IEEE Transactions on Knowledge and Data Engineering}} \bibinfo{volume}{36}, \bibinfo{number}{2} (\bibinfo{year}{2023}), \bibinfo{pages}{459--474}.
\newblock


\bibitem[Zhang et~al\mbox{.}(2017)]%
        {zhang2017joint}
\bibfield{author}{\bibinfo{person}{Yongfeng Zhang}, \bibinfo{person}{Qingyao Ai}, \bibinfo{person}{Xu Chen}, {and} \bibinfo{person}{W~Bruce Croft}.} \bibinfo{year}{2017}\natexlab{}.
\newblock \showarticletitle{Joint representation learning for top-n recommendation with heterogeneous information source}. In \bibinfo{booktitle}{\emph{Proceedings of the 2017 {ACM} on Conference on Information and Knowledge Management (CIKM'17)}}. \bibinfo{pages}{1449--1458}.
\newblock


\bibitem[Zhao et~al\mbox{.}(2023c)]%
        {zhao2023fair}
\bibfield{author}{\bibinfo{person}{Chen Zhao}, \bibinfo{person}{Le Wu}, \bibinfo{person}{Pengyang Shao}, \bibinfo{person}{Kun Zhang}, \bibinfo{person}{Richang Hong}, {and} \bibinfo{person}{Meng Wang}.} \bibinfo{year}{2023}\natexlab{c}.
\newblock \showarticletitle{Fair representation learning for recommendation: A mutual information perspective}. In \bibinfo{booktitle}{\emph{Proceedings of the 37th AAAI Conference on Artificial Intelligence (AAAI'23)}}. \bibinfo{pages}{4911--4919}.
\newblock


\bibitem[Zhao et~al\mbox{.}(2023b)]%
        {zhao2023disentangled}
\bibfield{author}{\bibinfo{person}{Weiqi Zhao}, \bibinfo{person}{Dian Tang}, \bibinfo{person}{Xin Chen}, \bibinfo{person}{Dawei Lv}, \bibinfo{person}{Daoli Ou}, \bibinfo{person}{Biao Li}, \bibinfo{person}{Peng Jiang}, {and} \bibinfo{person}{Kun Gai}.} \bibinfo{year}{2023}\natexlab{b}.
\newblock \showarticletitle{Disentangled causal embedding with contrastive learning for recommender system}. In \bibinfo{booktitle}{\emph{Companion Proceedings of the ACM Web Conference (WWW'23)}}. \bibinfo{pages}{406--410}.
\newblock


\bibitem[Zhao et~al\mbox{.}(2025a)]%
        {ZCC25}
\bibfield{author}{\bibinfo{person}{Yuhan Zhao}, \bibinfo{person}{Rui Chen}, \bibinfo{person}{Li Chen}, \bibinfo{person}{Shuang Zhang}, \bibinfo{person}{Qilong Han}, {and} \bibinfo{person}{Hongtao Song}.} \bibinfo{year}{2025}\natexlab{a}.
\newblock \showarticletitle{From Pairwise to Ranking: Climbing the Ladder to Ideal Collaborative Filtering with Pseudo-Ranking}. In \bibinfo{booktitle}{\emph{Proceedings of the AAAI Conference on Artificial Intelligence (AAAI'25)}}, Vol.~\bibinfo{volume}{39}. \bibinfo{pages}{13392--13400}.
\newblock


\bibitem[Zhao et~al\mbox{.}(2024)]%
        {ZCH24}
\bibfield{author}{\bibinfo{person}{Yuhan Zhao}, \bibinfo{person}{Rui Chen}, \bibinfo{person}{Qilong Han}, \bibinfo{person}{Hongtao Song}, {and} \bibinfo{person}{Li Chen}.} \bibinfo{year}{2024}\natexlab{}.
\newblock \showarticletitle{Unlocking the Hidden Treasures: Enhancing Recommendations with Unlabeled Data}. In \bibinfo{booktitle}{\emph{Proceedings of the 18th ACM conference on recommender systems (RecSys'24)}}. \bibinfo{pages}{247--256}.
\newblock


\bibitem[Zhao et~al\mbox{.}(2023a)]%
        {Aug}
\bibfield{author}{\bibinfo{person}{Yuhan Zhao}, \bibinfo{person}{Rui Chen}, \bibinfo{person}{Riwei Lai}, \bibinfo{person}{Qilong Han}, \bibinfo{person}{Hongtao Song}, {and} \bibinfo{person}{Li Chen}.} \bibinfo{year}{2023}\natexlab{a}.
\newblock \showarticletitle{Augmented Negative Sampling for Collaborative Filtering}. In \bibinfo{booktitle}{\emph{Proceedings of the 17th ACM conference on recommender systems (RecSys'23)}}. \bibinfo{pages}{256–266}.
\newblock


\bibitem[Zhao et~al\mbox{.}(2025b)]%
        {Denoising}
\bibfield{author}{\bibinfo{person}{Yuhan Zhao}, \bibinfo{person}{Rui Chen}, \bibinfo{person}{Riwei Lai}, \bibinfo{person}{Qilong Han}, \bibinfo{person}{Hongtao Song}, {and} \bibinfo{person}{Li Chen}.} \bibinfo{year}{2025}\natexlab{b}.
\newblock \showarticletitle{Denoising and Augmented Negative Sampling for Collaborative Filtering}.
\newblock \bibinfo{journal}{\emph{ACM Transactions on Recommender Systems}} \bibinfo{volume}{3}, \bibinfo{number}{4} (\bibinfo{year}{2025}), \bibinfo{pages}{1--23}.
\newblock


\bibitem[Zhou et~al\mbox{.}(2023a)]%
        {DRAGON}
\bibfield{author}{\bibinfo{person}{Hongyu Zhou}, \bibinfo{person}{Xin Zhou}, {and} \bibinfo{person}{Zhiqi Shen}.} \bibinfo{year}{2023}\natexlab{a}.
\newblock \showarticletitle{Enhancing dyadic relations with homogeneous graphs for multimodal recommendation}. In \bibinfo{booktitle}{\emph{Proceedings of the 26th European Conference on Artificial Intelligence (ECAI'23)}}. \bibinfo{pages}{3123--3130}.
\newblock


\bibitem[Zhou et~al\mbox{.}(2023c)]%
        {zhou2023comprehensive}
\bibfield{author}{\bibinfo{person}{Hongyu Zhou}, \bibinfo{person}{Xin Zhou}, \bibinfo{person}{Zhiwei Zeng}, \bibinfo{person}{Lingzi Zhang}, {and} \bibinfo{person}{Zhiqi Shen}.} \bibinfo{year}{2023}\natexlab{c}.
\newblock \showarticletitle{A comprehensive survey on multimodal recommender systems: taxonomy, evaluation, and future directions}.
\newblock \bibinfo{journal}{\emph{arXiv preprint arXiv:2302.04473}} (\bibinfo{year}{2023}).
\newblock


\bibitem[Zhou and Shen(2023)]%
        {FREEDOM}
\bibfield{author}{\bibinfo{person}{Xin Zhou} {and} \bibinfo{person}{Zhiqi Shen}.} \bibinfo{year}{2023}\natexlab{}.
\newblock \showarticletitle{A tale of two graphs: Freezing and denoising graph structures for multimodal recommendation}. In \bibinfo{booktitle}{\emph{Proceedings of the 31st ACM International Conference on Multimedia (MM'23)}}. \bibinfo{pages}{935--943}.
\newblock


\bibitem[Zhou et~al\mbox{.}(2023b)]%
        {BM3}
\bibfield{author}{\bibinfo{person}{Xin Zhou}, \bibinfo{person}{Hongyu Zhou}, \bibinfo{person}{Yong Liu}, \bibinfo{person}{Zhiwei Zeng}, \bibinfo{person}{Chunyan Miao}, \bibinfo{person}{Pengwei Wang}, \bibinfo{person}{Yuan You}, {and} \bibinfo{person}{Feijun Jiang}.} \bibinfo{year}{2023}\natexlab{b}.
\newblock \showarticletitle{Bootstrap latent representations for multi-modal recommendation}. In \bibinfo{booktitle}{\emph{Proceedings of the {ACM} Web Conference (WWW'23)}}. \bibinfo{pages}{845--854}.
\newblock


\bibitem[Zhu et~al\mbox{.}(2024)]%
        {zhu2024adaptive}
\bibfield{author}{\bibinfo{person}{Xinyu Zhu}, \bibinfo{person}{Lilin Zhang}, {and} \bibinfo{person}{Ning Yang}.} \bibinfo{year}{2024}\natexlab{}.
\newblock \showarticletitle{Adaptive fair Representation learning for personalized fairness in recommendations via information alignment}. In \bibinfo{booktitle}{\emph{Proceedings of the 47th International {ACM} {SIGIR} Conference on Research and Development in Information Retrieval (SIGIR'24)}}. \bibinfo{pages}{427--436}.
\newblock


\end{thebibliography}
